\pdfoutput=1
\documentclass[preprint,12pt]{elsarticle}

\journal{Computer-Aided Design}
\usepackage[disable]{todonotes} %draft or disable

\usepackage{rotating}
\usepackage{graphicx}
\usepackage{subfig}
\usepackage{multicol}
\usepackage{algorithm}
\usepackage{algpseudocode}
\usepackage{amsmath}
\usepackage{amsfonts}
\usepackage{enumerate}
\usepackage{mathrsfs}
\usepackage{dsfont}
\usepackage{stackengine}

\DeclareMathOperator*{\argmin}{argmin}

\newcommand{\zR}{\mathds{R}}

\newcommand{\zmetric}{\textbf{M}}

\newcommand{\ztr}{\mathrm{tr}}
\newcommand{\zmaster}{E^M}

\newcommand{\zequilater}{E^{\triangle}}

\newcommand{\zphysical}{E^P}

\newcommand{\zequilatermap}{\zphi_{\triangle}}

\newcommand{\zphysicalmap}{\zphi_P}

\newcommand{\zJacobianequilater}{\textbf{D}\zequilatermap}

\newcommand{\zJacobianphysical}{\textbf{D}\zphysicalmap}

\newcommand{\zu}{\textbf{u}}

\newcommand{\zx}{\textbf{x}}
\newcommand{\zxi}{\boldsymbol{\xi}}

\newcommand{\zp}{\textbf{p}}

\newcommand{\zphi}{\boldsymbol{\phi}}

\newcommand{\zDline}{\textbf{D}_{\text{line}}}
\newcommand{\zDcross}{\textbf{D}_{\text{cross}}}
\newcommand{\zmapline}{\psi_{\text{line}}}
\newcommand{\zmapcross}{\psi_{\text{cross}}}
\graphicspath{{./figures/}}

\begin{document}
	
	\algnewcommand\algorithmicswitch{\textbf{switch}}
	\algnewcommand\algorithmiccase{\textbf{case}}
	\algdef{SE}[SWITCH]{Switch}{EndSwitch}[1]{\algorithmicswitch\ #1\ \algorithmicdo}{\algorithmicend\ \algorithmicswitch}%
	\algdef{SE}[CASE]{Case}{EndCase}[1]{\algorithmiccase\ #1}{\algorithmicend\ \algorithmiccase}%
	\algtext*{EndSwitch}%
	\algtext*{EndCase}%
	
%	\verso{Guillermo Aparicio-Estrems \textit{etal}}
	
	\begin{frontmatter}
		\title{A globalized and preconditioned Newton-CG solver for metric-aware curved high-order mesh optimization}
		
		\author{Guillermo Aparicio-Estrems}
		\ead{guillermo.aparicio@bsc.es}
		\author{Abel Gargallo-Peir\'{o}}
		\ead{abel.gargallo@bsc.es}
		\author{Xevi Roca\corref{mycorrespondingauthor}}
		\cortext[mycorrespondingauthor]{Corresponding author}
		\ead{xevi.roca@bsc.es}
		\address{Barcelona Supercomputing Center, Carrer de Jordi Girona, 29, 31, Barcelona 08034, Spain}

		\begin{abstract}
			We present a specific-purpose globalized and preconditioned Newton-CG solver to minimize a metric-aware curved high-order mesh distortion. The solver is specially devised to optimize curved high-order meshes for high polynomial degrees with a target metric featuring non-uniform sizing, high stretching ratios, and curved alignment --- exactly the features that stiffen the optimization problem. To this end, we consider two ingredients: a specific-purpose globalization and a specific-purpose Jacobi-$\text{iLDL}^{\text{T}}(0)$ preconditioning with varying accuracy and curvature tolerances (dynamic forcing terms) for the CG method. These improvements are critical in stiff problems because, without them, the large number of non-linear and linear iterations makes curved optimization impractical. First, to enhance the global convergence of the non-linear solver, the globalization strategy modifies Newton’s direction to a feasible step. In particular, our specific-purpose globalization strategy memorizes the length of the feasible step (step-length continuation) between the optimization iterations while ensuring sufficient decrease and progress. Second, to compute Newton's direction in second-order optimization problems, we consider a conjugate-gradient iterative solver with specific-purpose preconditioning and dynamic forcing terms. To account for the metric stretching and alignment, the preconditioner uses specific orderings for the mesh nodes and the degrees of freedom. We also present a preconditioner switch between Jacobi and $\text{iLDL}^{\text{T}}(0)$ preconditioners to control the numerical ill-conditioning of the preconditioner. In addition, the dynamic forcing terms determine the required accuracy for the Newton direction approximation. Specifically, they control the residual tolerance and enforce sufficient positive curvature for the conjugate-gradients method. Finally, to analyze the performance of our method, the results compare the specific-purpose solver with standard optimization methods. For this, we measure the matrix-vector products indicating the solver computational cost and the line-search iterations indicating the total amount of objective function evaluations. When we combine the globalization and the linear solver ingredients, we conclude that the specific-purpose Newton-CG solver reduces the total number of matrix-vector products by one order of magnitude. Moreover, the number of non-linear and line-search iterations is mainly smaller but of similar magnitude.
		\end{abstract}
		
		\begin{keyword}
		%\texttt{elsarticle.cls}\sep \LaTeX\sep Elsevier \sep template
		%\MSC[2010] 00-01\sep  99-00
		optimization, $r$-adaption, curved high-order meshes
%		49M15  	Newton-type methods
		\end{keyword}
		
	\end{frontmatter}
	
	%\linenumbers
	
	\section{Introduction}\label{sec:intro}
	To enhance accuracy in problems where the solution presents pronounced curved features, the community of unstructured high-order methods has started to curve not only the boundary but also the interior \cite{knupp2021adaptive,dobrev2019target,dobrev2021hr,sanjaya2016improving,rochery2021p2,zhangthesis,coupez:BasisFrameworkHighOrderAnisotropicMeshAdaptation,marcon2019mesh,zahr2020implicit,ekelschot2019parallel} of unstructured high-order meshes.
	These methods aim to match the pronounced curved features of the solution by exploiting the capability of high-order elements to represent curved shapes.
%	These methods aim to match the pronounced curved features of the solution by exploiting the non-constant Jacobian of curved high-order elements.
	To this end, these methods modify the high-order mesh topology and coordinates \cite{dobrev2021hr,rochery2021p2,zhangthesis,ekelschot2019parallel} or only the coordinates \cite{knupp2021adaptive,dobrev2019target,sanjaya2016improving,coupez:BasisFrameworkHighOrderAnisotropicMeshAdaptation,marcona2017variational,zahr2020implicit}. In both families, the modification of the mesh coordinates is a crucial ingredient.
	
	In a broad sense, the modification of the mesh coordinates can be understood as a curved high-order $r$-adaption to the solution features. This curved high-order $r$-adaption is driven by an objective function accounting for a pointwise varying target \cite{dobrev2019target,dobrev2018towards,dobrev2020simulation,CAMIER2023111808,aparicio2018defining,aparicio2022metricinterpolation,aparicio2023combining}. The target can be a deformation matrix \cite{dobrev2019target,dobrev2018towards,dobrev2020simulation,CAMIER2023111808} or a metric \cite{aparicio2018defining,aparicio2022metricinterpolation,aparicio2023combining} that encodes the curved geometric features of the solution, features such as the local stretching, and alignment.
	
	Regarding the optimization of the objective function, we can iteratively modify the coordinates of either all the free nodes (all nodes) or one free node (one node) per non-linear iteration by using either gradient-based (first-order) or Hessian-based (second-order) optimization methods. For linear elements, there are several studies on the performance of local \cite{diachin2006comparison,diachin2004comparison,sastry2009comparison,diachin2004comparison} first and second-order optimization methods. One common conclusion is that when highly optimized and accurate meshes are required, especially in isotropic meshes featuring high gradations of the element size, a specific-purpose all-nodes globalized Newton method \cite{steihaug1983conjugate} outperforms local optimization methods \cite{diachin2006comparison,diachin2004comparison,sastry2012performance}. Following these conclusions, we focus on devising a specific-purpose all-nodes globalized Newton method for metric-aware optimization of curved high-order meshes.
	
%	To provide a basic context, we describe the main steps of a standard optimization algorithm \cite{nocedal2006numerical}. First, we start computing a descent direction. Specifically, for the Newton method, we obtain this descent direction by solving a linear system of equations, the Newton equation. Here play an important role the forcing terms and the preconditioner to control the computational cost indicated by the total amount of matrix-vector products. Second, from a globalization strategy, we transform the descent direction into a feasible step. In particular, the line-search strategy considers a linear approximation of the objective function to predict the decrease and feasibility of the step. Here play an important role the step-length continuation and the model of the objective function. Third, we apply the feasible step to the current point. Finally, we iterate until convergence or a stopping condition is achieved.

	To provide a basic context, we describe the main steps of an optimization algorithm. First, we start computing a descent direction. Specifically, for the Newton method, we obtain this descent direction by solving a linear system of equations. When this system is solved with an iterative method, we need to provide the required solution tolerance. Moreover, for the conjugate gradient method, it is standard to provide a curvature tolerance threshold that ensures that the curvature of the conjugate gradient is sufficiently positive along the direction of the approximated linear solution. From now on, when the term curvature complements conjugate gradient, tolerance, or direction, we are referring to a scalar that corresponds to the curvature of a conjugate gradient step with respect to the matrix of the linear system. This matrix is the Hessian of the conjugate gradients problem, which is also the Hessian of the objective function of the current mesh. The accuracy and curvature tolerances can be static or dynamic. The dynamic tolerances, determined by dynamic forcing terms, allow solving the system only up to the required accuracy for the current optimization stage. Hence, these forcing terms save unrequired steps of the iterative linear solver. Also to accelerate the solution of the linear systems, it is standard to use a pre-conditioner that can be based on diagonal matrices or incomplete matrix factorizations. Second, using a globalization strategy, we transform the descent direction into a feasible step. In particular, the line-search strategy considers a linear approximation of the objective function to predict the decrease and feasibility of a scaled step. Standard approaches re-initiate the step-length scaling factor to be one, yet it might be advantageous to continue the step-lengths between successive iterations of the non-linear solver. Third, we use the step-length to obtain a feasible step. Finally, we iterate until the non-linear tolerance or a stopping criterion is achieved.
	
	To enhance global convergence of Newton's direction, globalization strategies modify Newton’s direction to a feasible step. Standard globalization methods are divided into those using either trust-region (TR) \cite{nocedal2006numerical,conn2000trust,Bulteau1985} or line-search (LS) globalization \cite{nocedal2006numerical}. On the one hand, trust-region methods consistently deal with negative-curvature steps and direction candidates on subspaces. To this end, standard TR methods enable a step-length continuation by only evaluating the objective function, and they do it by promoting not only a sufficient decrease but also a sufficient progress criterion. For this, standard TR methods consider a predictor model, comparing the non-linear behavior of the objective function with a quadratic model in terms of the step size \cite{conn2000trust}. However, it is unclear how to choose the initial trust-region radius in terms of the current mesh size. On the other hand, we prefer the simplicity of a backtracking line-search (BLS) strategy for a first implementation trial. Specifically, a standard BLS globalization considers the Newton direction reduced by a step-length factor using a sufficient decrease criterion \cite{nocedal2006numerical}.
	
	To compute Newton's direction in large second-order optimization problems, it is standard to use an inexact Newton method with a conjugate gradient (CG) method \cite{diachin2006comparison,sastry2009comparison,sastry2012performance}, using constant residual tolerance, and Jacobi preconditioning \cite{bertaccini2018iterative}.
	
	The preference for the CG method is based on three factors. First, the CG method is specific for symmetric and positive-definite matrices. This design is relevant near a minimum, where the symmetric Hessian of the objective function is also positive-definite. Second, its short-recurrence property allows computing a solution without requiring additional memory. Third, its negative-curvature termination condition is helpful in line-search strategies \cite{nocedal2006numerical}.
	
	In iterative linear solvers, it is standard to set a constant tolerance threshold for the residual norm as a stopping criterion to control the accuracy. Furthermore, specifically for the CG method, one can consider a curvature tolerance threshold as a stopping criterion. The choice of these tolerance parameters impacts the accuracy and number of iterations of the iterative method and hence, on the evolution and computational cost of the nonlinear solver.
	
	For a given constant tolerance, preconditioning techniques reduce the total number of matrix-vector products while preserving a comparable number of non-linear iterations. The total number of sparse matrix-vector products indicates the computational cost of inexact Newton solvers \cite{bertaccini2018iterative}. In Newton-CG methods, this number corresponds to the total number of CG iterations. 
	
	Unfortunately, for metric-aware curved high-order mesh optimization, standard globalized and preconditioned Newton-CG solvers have robustness and efficiency issues. In curved high-order metric-aware mesh optimization, we observe that these issues are triggered by non-uniform sizing, stretching ratios, and curved alignment. When more remarkable these characteristics are, more difficult the convergence with a general-purpose optimization solver. First, in each non-linear step, highly non-uniform mesh gradation stiffens the validity of the mesh deformations and the corresponding linear systems. Second, for high stretching ratios, the deformations in some directions are locally stiffer than in other directions. Third, curved alignment requires curved high-order elements. For these elements, when higher is the order, stiffer is the corresponding linear system.
	
%	The previous mesh characteristics challenge the global convergence of the non-linear solver and the solution of the corresponding linear systems.
	Because the previous mesh characteristics stiffen the optimization problem, they challenge the global convergence of the non-linear solver and the solution of the corresponding linear systems. Specifically, progressing towards convergence, standard solvers might present three main issues:
	
	\begin{itemize}
		\item They might need additional backtracking line-search iterations because they do not continuously ensure sufficient decrease and progress. A standard BLS globalization reduces the Newton direction by a step-length factor using a sufficient decrease criterion \cite{nocedal2006numerical}. However, the step length is restarted at each non-linear iteration, impeding its continuous evolution during the optimization. Moreover, BLS does not promote sufficient progress.
		
		\item They might accumulate additional iterations of the linear solver because they use constant linear solver tolerance. Constant tolerances do not correctly predict the accuracy of a descent direction for a highly non-linear and non-convex objective function. Thus, they might require excessive precision far from the optima or feature insufficient accuracy to promote quadratic convergence near an optimum \cite{eisenstat1996choosing,Dembo1983}.
		
		\item They might obtain inaccurate steps because the preconditioner is inaccurate or numerically singular. Jacobi preconditioners favor a low computational cost instead of an accurate approximation of the Hessian matrix. This loss of accuracy in solving Newton's equation compromises the computational cost of the solver near an optimum, where quadratic convergence must be prioritized. Incomplete Cholesky factorization favors accurate Hessian approximation instead of low computational cost. However, it might lead to singular preconditioning when the Hessian is numerically singular.
	\end{itemize}
	
	It is critical to devise a solver to alleviate these issues because, without such a solver, it might be impossible to demonstrate the potential advantages of curved high-order optimization for high polynomial degrees, especially when the target metric features non-uniform size, high stretching, and curved alignment. The implementation of the resulting high-order mesh optimization solver can be later accelerated using fast GPU implementations \cite{CAMIER2023111808}.
	
	We aim to alleviate the issues of standard solvers for metric-aware curved high-order mesh optimization. To this end, we propose a specific-purpose globalized and preconditioned Newton-CG solver. To devise the solver, we propose three main contributions:
	
	\begin{itemize}
		\item To continuously ensure sufficient decrease and progress for the LS globalization, we uniquely combine a step-length predictor featuring not only reduction but also amplification of the step length. This line search features memory and continuation of the step length while favoring the quadratic convergence of the Newton method.
		
		\item To reduce the number of iterations of the linear solver, we propose new dynamic forcing sequences that control the residual tolerance and sufficient positive curvature. Specifically, we propose a new forcing sequence for the residual. This sequence is suited to limit the number of CG iterations at the beginning of the optimization process and allow the necessary CG iterations to obtain a quadratic convergence rate near an optimum. To emulate steps with sufficient positive curvature, we also propose to define the normalized curvature of a given direction and a new dynamic forcing sequence for the curvature of the CG step. We define this sequence to limit CG-iterations when the Hessian is near to positive semidefinite without breaking the quadratic convergence rate near an optimum.
		
		\item To avoid numerically singular linear pre-conditioning, we propose three ingredients for our pre-conditioner. The first ingredient is a novel predictor that switches between the Jacobi pre-conditioner and the \emph{root-free incomplete Cholesky factorization} ($\text{iLDL}^\text{T}(0)$) with zero levels of fill-in \cite{bertaccini2018iterative}. This switch uses a parameter indicating the acceptable numerical ill-conditioning of the factorization. The second is a new inequality accounting for the curvature of the resulting direction computed from the CG method. If the direction violates the curvature inequality, we consider that the computation of the used pre-conditioner is numerically unstable. The third ingredient reorders the unknowns used to compute the factorization. Several results presented in the literature indicate that the ordering of a matrix impacts the numerical instability of its factorization \cite{bertaccini2018iterative}. To control this instability, we propose to use an ordering that tries to minimize the discarded fill of the incomplete factorization \cite{d1992ordering,persson2008newton}. We also propose to reorder the mesh nodes according to the first nonzero eigenvalue of a metric-aware Laplacian spectral problem with Neumann boundary conditions.

	\end{itemize}
	
	Finally, to measure the performance of our specific-purpose solver in metric-aware curved high-order mesh optimization, we compare it with a standard solver. For the solver ingredients, we also compare the standard and specific-purpose approaches. To perform these comparisons, we measure the number of iterations for the non-linear loop and backtracking line-search globalization. In addition, we compare the total number of matrix-vector products. The results allow us to describe the influence of each ingredient on the proposed specific-purpose non-linear solver. We conclude that the previous contributions are key to optimize curved high-order meshes for polynomial degrees with a target metric featuring non-uniform sizing, high stretching ratios, and curved alignment.
	
	The remainder of this paper is organized as follows.
	In Section \ref{sec:problemStatement}, we introduce the $r$-adaption problem, the distortion minimization formulation, and an optimization overview.
	In Section \ref{sec:updating}, we present the standard and specific-purpose line-search globalizations for Newton's method.
	In Section \ref{sec:updatinglinearsystem}, we present the standard and specific-purpose linear solvers for the inexact Newton method.
	In Section \ref{sec:results}, we present a set of examples to compare both the standard and specific-purpose implementations.
	In Section \ref{sec:discussion}, we present a discussion on different aspects related to our methods and results.
	Finally, in Section \ref{sec:conclusions}, we present the main conclusions and sum up the future work to develop.
	\section{The problem: $r$-adaption, formulation, and optimization overview}\label{sec:problemStatement}
	
	We aim to propose a robust specific-purpose solver for the piece-wise polynomial mesh $r$-adaption problem.
	In this adaption problem, the input is a domain, equipped with a metric, and meshed with a piece-wise polynomial mesh, see Section \ref{sec:modelCase} for a model case.
	We want to relocate the node coordinates of the input mesh, without modifying the topology, to obtain an output mesh that matches the stretching and alignment prescribed by the given metric.
	To this end, we can minimize the mesh distortion measure proposed in \cite{aparicio2018defining}, with the corresponding free node coordinates as design variables, see Section \ref{sec:minimizationProblem}.
	Unfortunately, we have observed that the existent optimization solvers equipped with standard globalization strategies, see an overview in Section \ref{sec:optimizationOverview}, might fail to drive an initial mesh to a local distortion minimum when the initial mesh highly mismatches the stretching and alignment of the given metric, especially when higher are the polynomial degrees, stretching ratios, and curved shape of the alignment features. We seek a new robust and globalized minimization solver that overcomes these issues. 
	
	%	\subsection{Piece-wise polynomial $r$-adaption: model case}\label{sec:modelCase}
	\subsection{Curved high-order $r$-adaption: model case}\label{sec:modelCase}
	\begin{figure}[t]
		\centering
		\hspace{-0.35cm}
		\begin{tabular}{cc}
			\subfloat[]{\label{fig:color}		\includegraphics[width=0.4\textwidth]{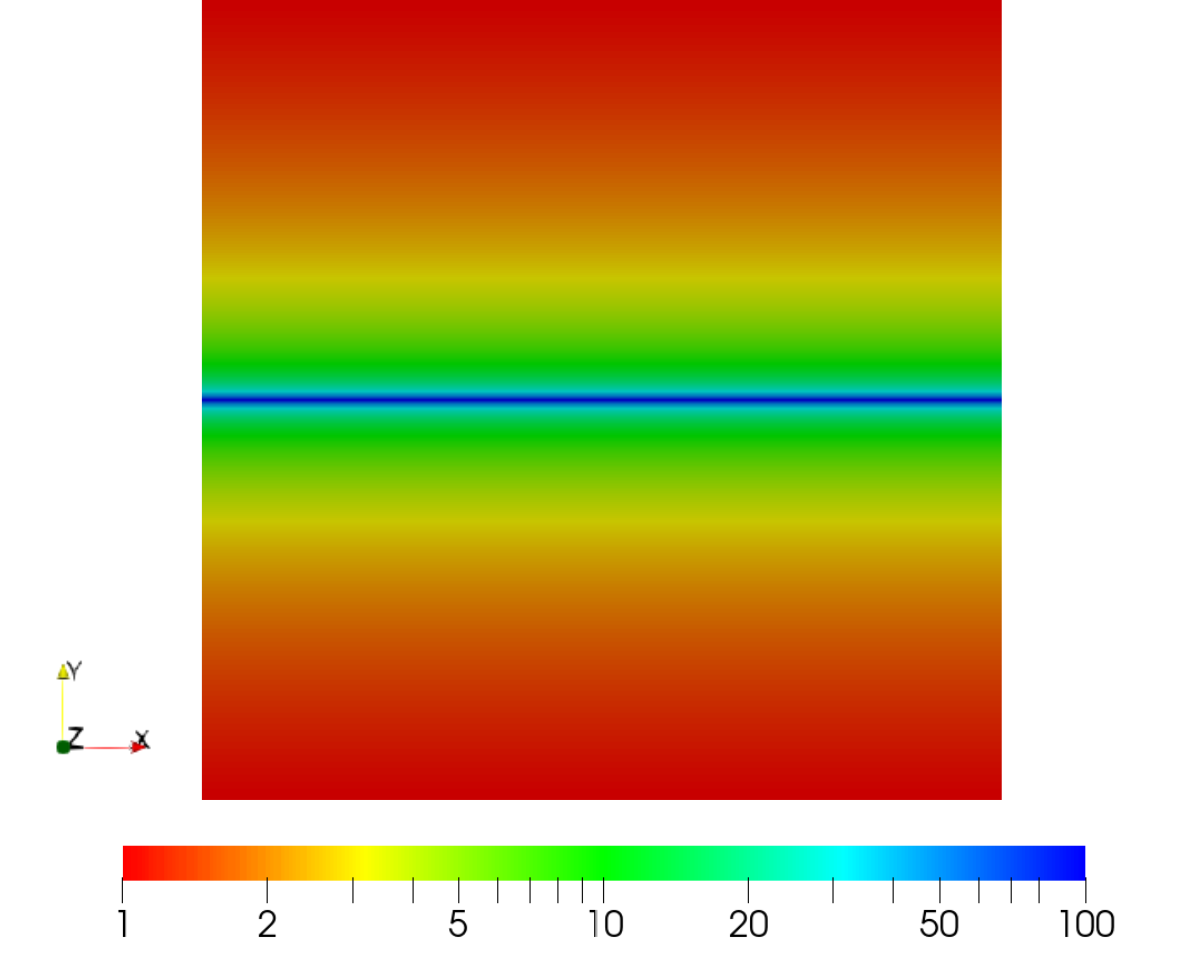}}
			&
			\subfloat[]{\label{fig:ratio}		\includegraphics[width=0.4\textwidth]{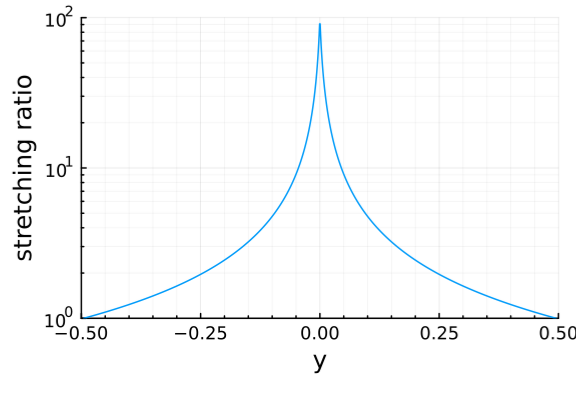}}
%			\\
%			\subfloat[]{\includegraphics[width=0.3\textwidth]{metrifig/bar1_100_0.png}}
%			&
%			\\
		\end{tabular}
		\caption{Unit square equipped with a metric matching a shear layer. Stretching ratio in logarithmic scale: (a) over the domain and (b) along the $y$-direction.}
		%\caption{Unit square equipped with a metric matching a shear layer: stretching ratio in logarithmic scale.}
		\label{fig:metric}
	\end{figure}
	To illustrate the $r$-adaption problem and to test the globalized minimization solvers considered through this work, we use a model case. In this model case, we consider the quadrilateral domain $\Omega=[-0.5,0.5]^2$, equipped with a metric matching a shear layer, and meshed with isotropic straight-sided triangular meshes of different polynomial degree but with the same resolution. This is achieved by subdividing lower-order elements at equispaced element nodes.
	
	The shear layer aligns with the $x$-axis, requires a constant unit element size along the $x$-direction, and a non-constant element size along the $y$-direction. This vertical element size grows linearly with the distance to the $x$-axis, with a factor $\gamma= 2$, and starts with the minimal value $h_{\min}=10^{-2}$. Thus, as illustrated in Figure \ref{fig:metric}, between $y=-0.5$ and $y=0.5$ the stretching ratio blends from $1:100$ to $1:1$. To match the shear layer, we define the metric as:
	\begin{equation*}
	\zDline = \left(
	\begin{array}{cc}
	1 & 0\\
	0 & 1/h(y)^2
	\end{array}
	\right),
	\end{equation*}
	where
	\begin{equation}\label{eq:htest}
		h(x)= h_{\min} + \gamma |x|.
	\end{equation}
	
	\begin{figure}[t]
		\centering
		\hspace{-0.35cm}
		\begin{tabular}{cccc}
			\subfloat[]{\label{fig:p1_0}
				\includegraphics[width=0.2\textwidth]{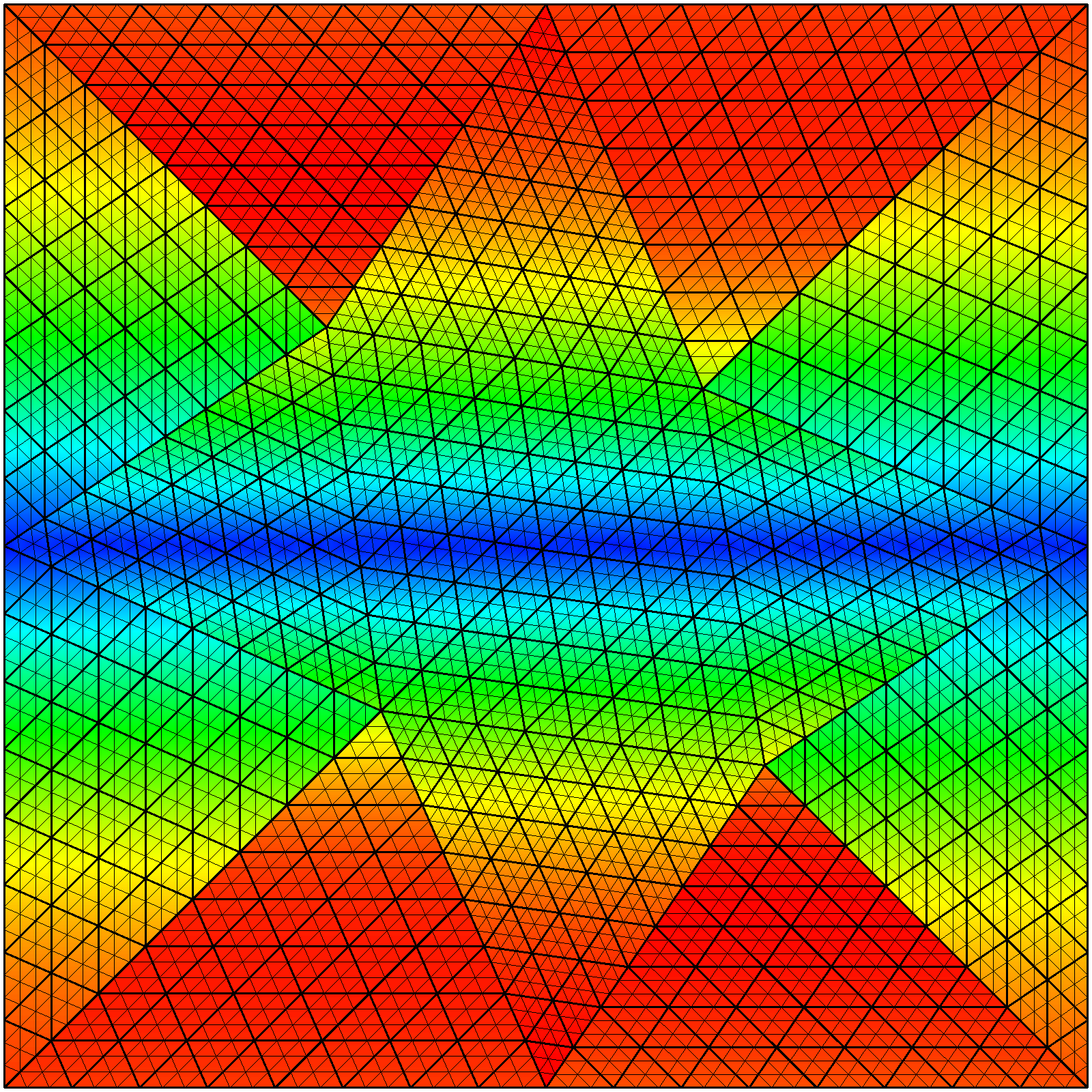}}
			&
			\subfloat[]{\label{fig:p2_0}
				\includegraphics[width=0.2\textwidth]{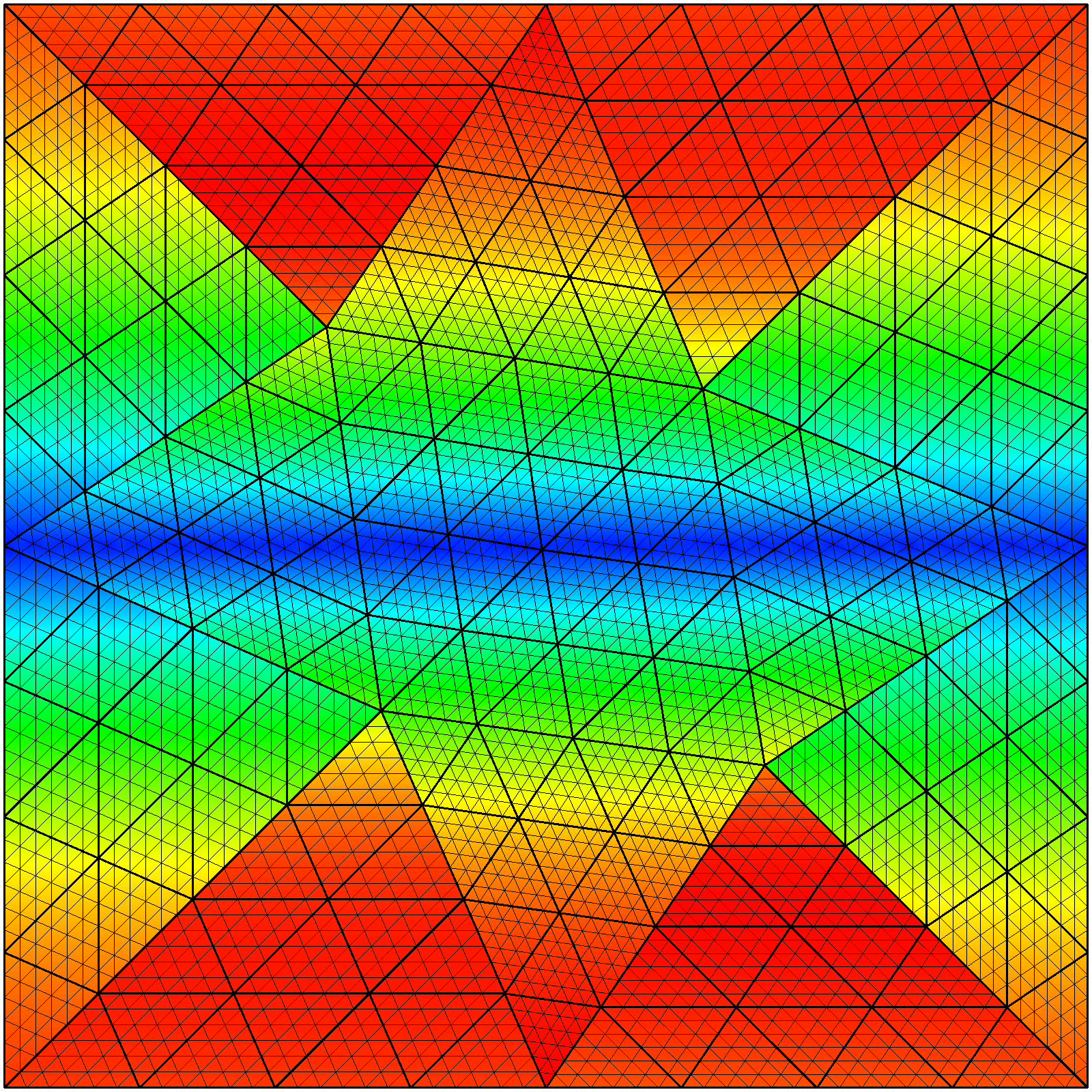}}
			&
			\subfloat[]{\label{fig:p4_0}
				\includegraphics[width=0.2\textwidth]{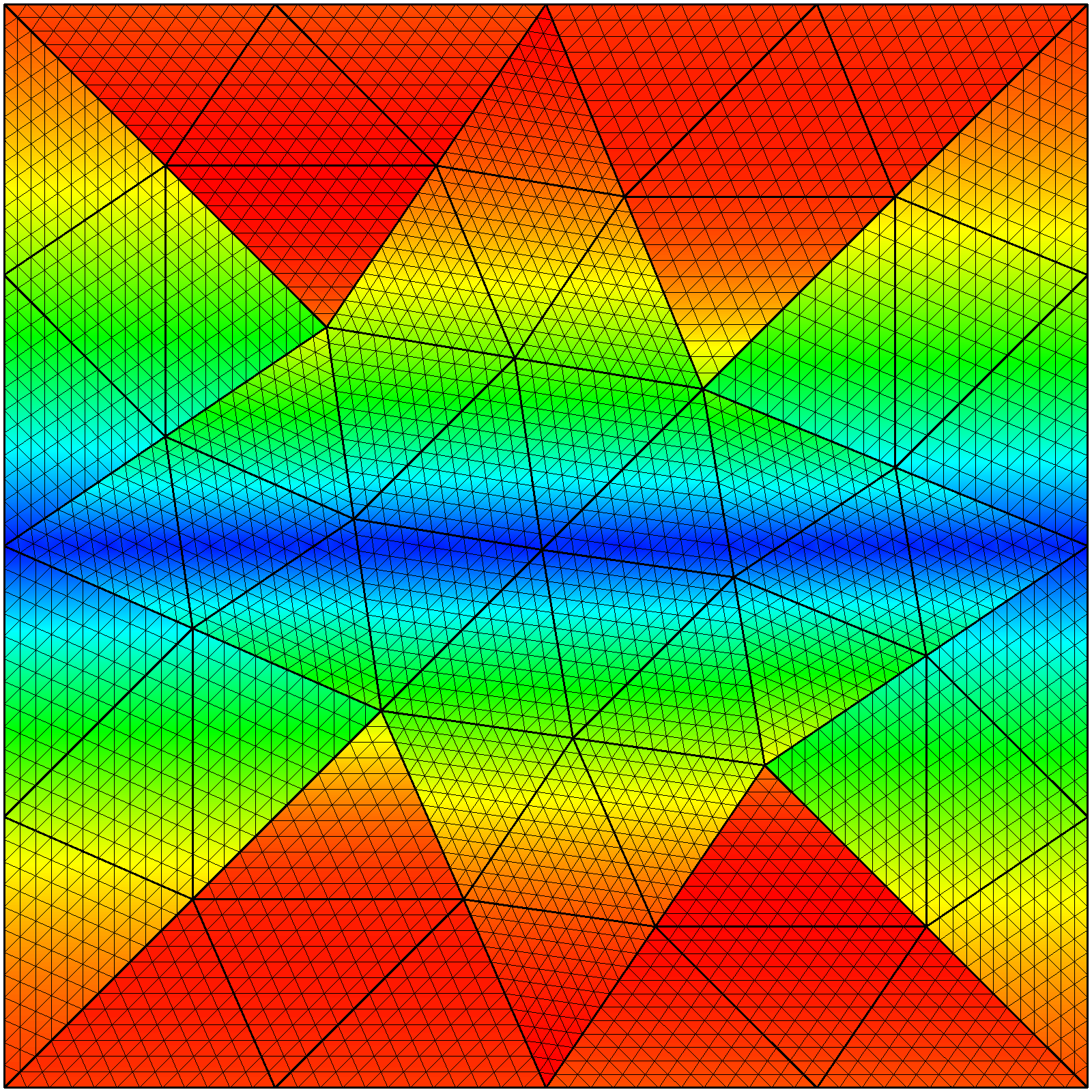}}
			&
			\subfloat[]{\label{fig:p8_0}
				\includegraphics[width=0.2\textwidth]{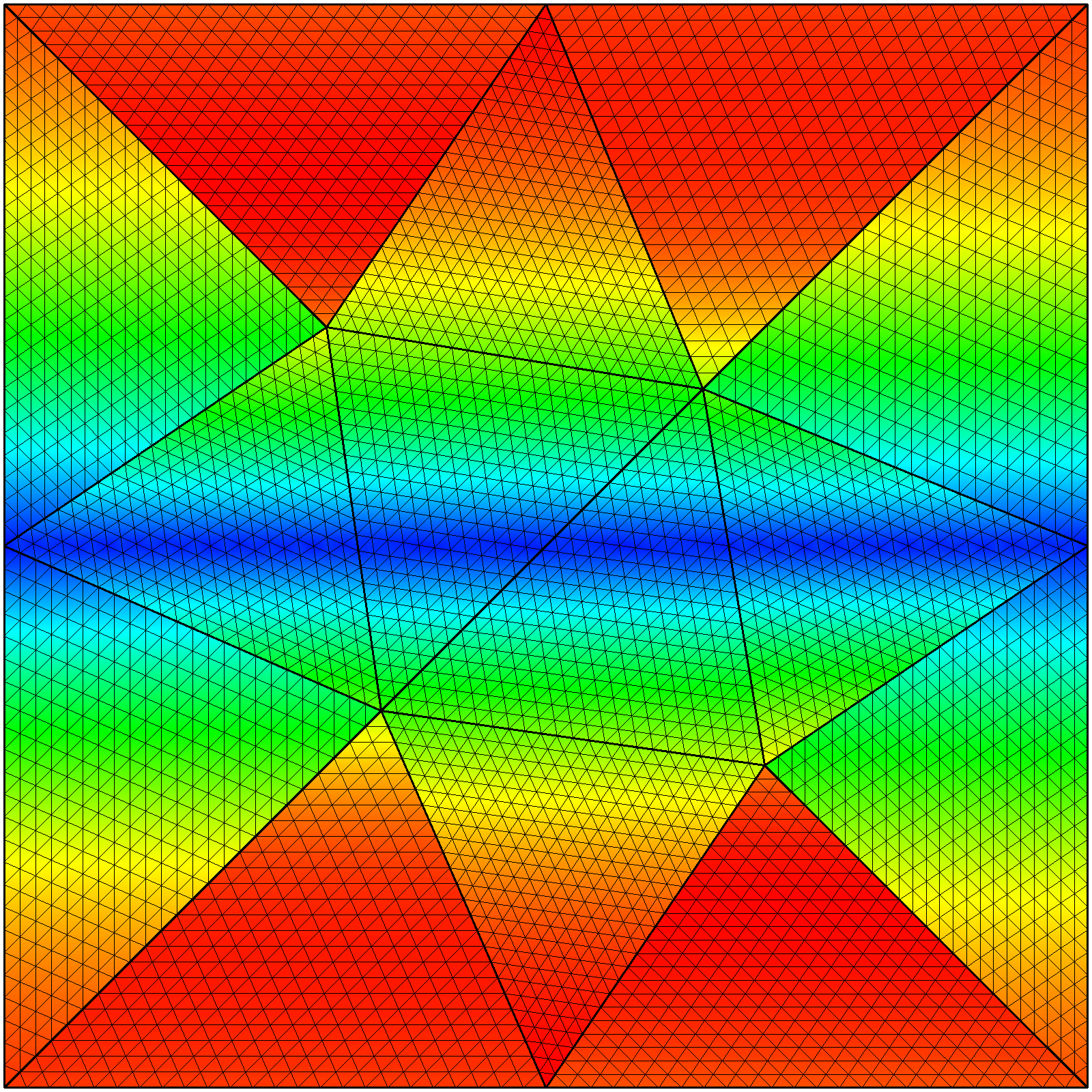}}
			\\
			\subfloat[]{\label{fig:p1_1}
				\includegraphics[width=0.2\textwidth]{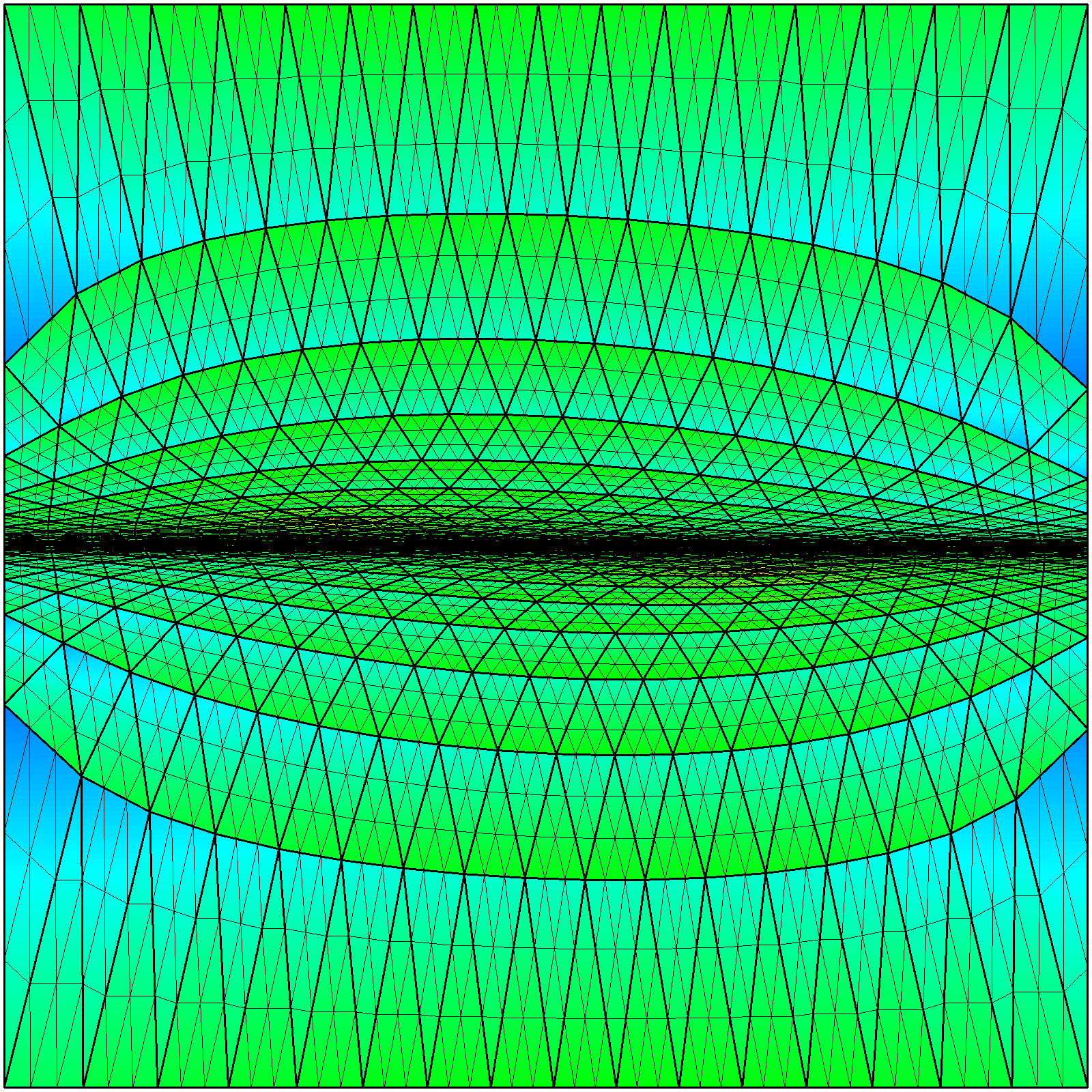}}
			&
			\subfloat[]{\label{fig:p2_1}
				\includegraphics[width=0.2\textwidth]{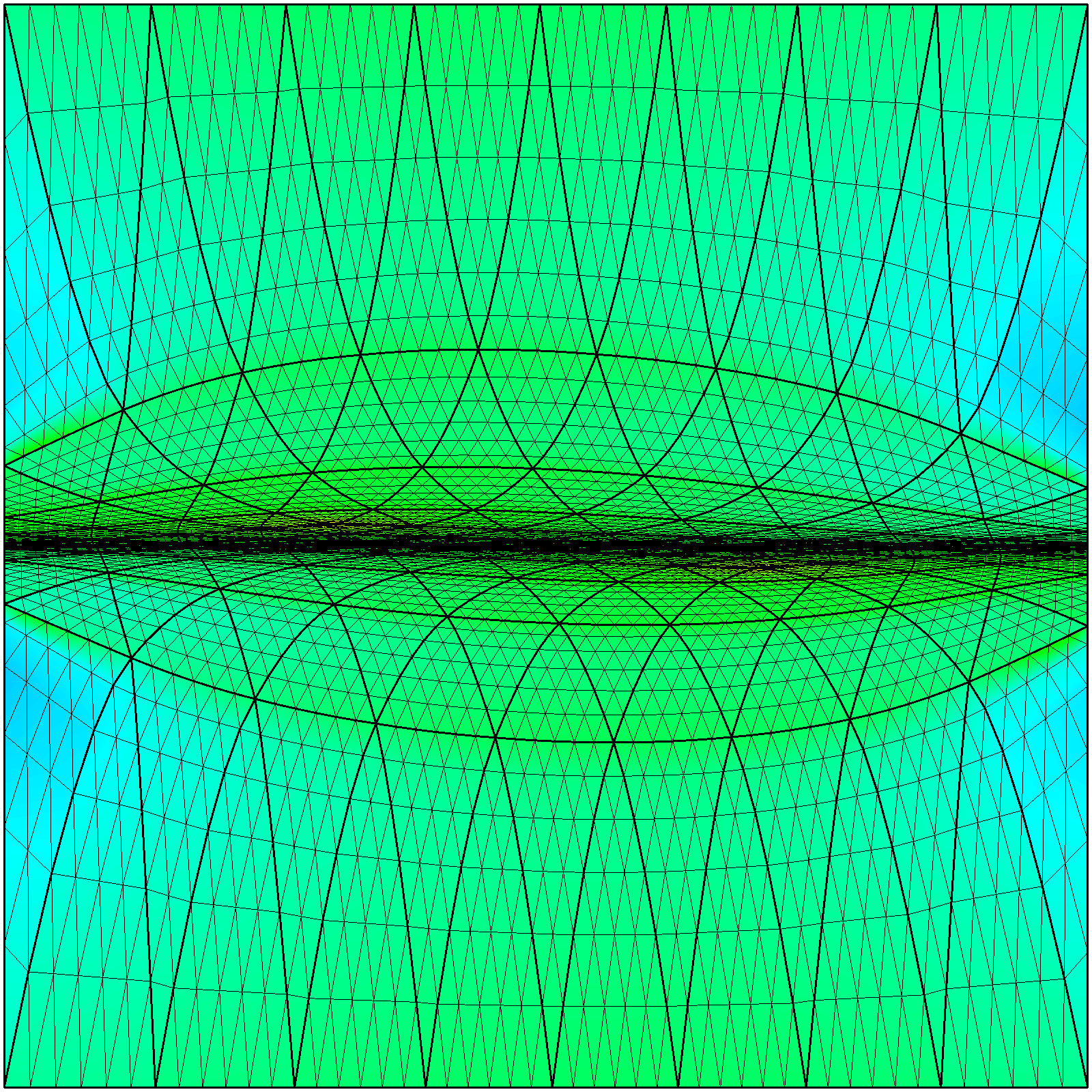}}
			&
			\subfloat[]{\label{fig:p4_1}
				\includegraphics[width=0.2\textwidth]{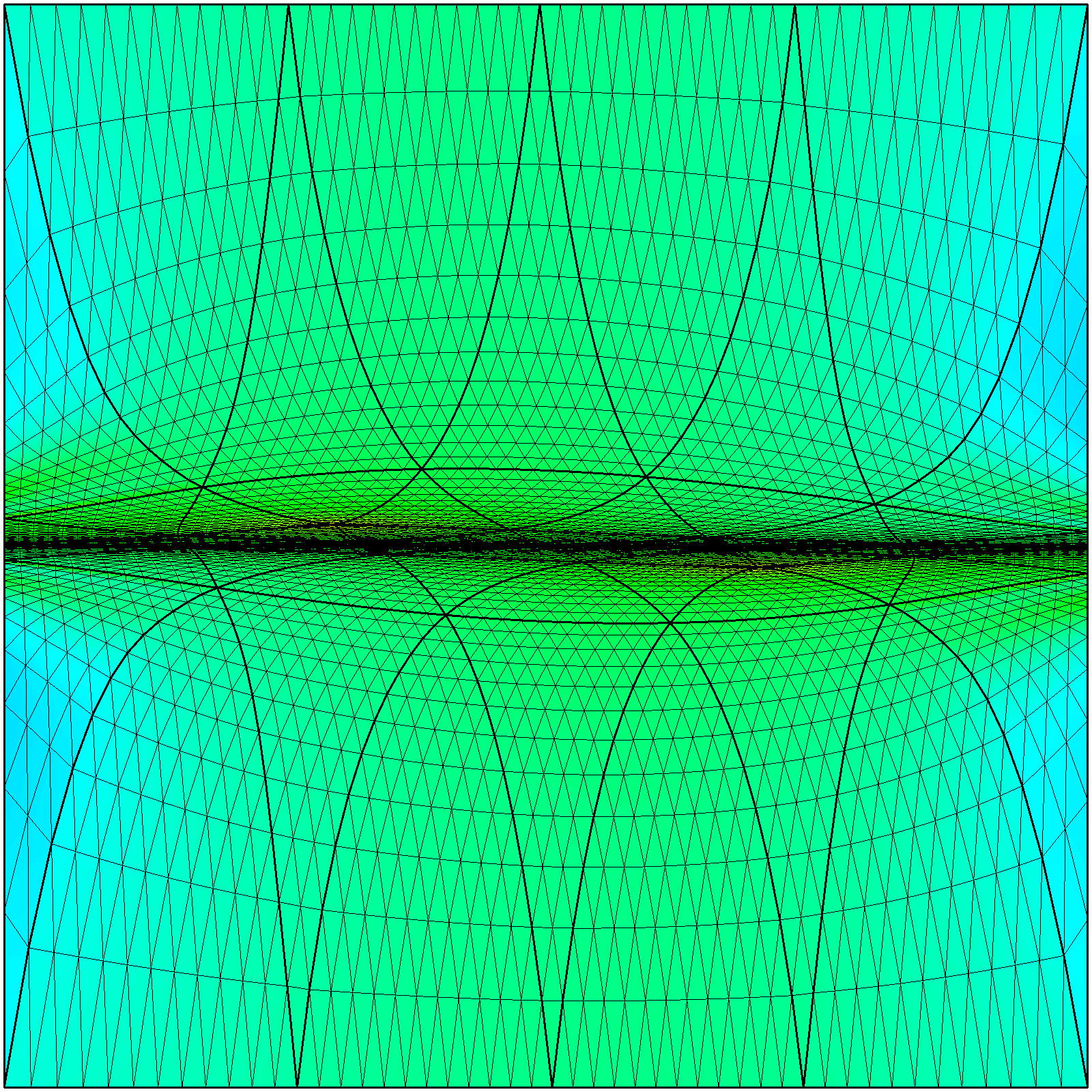}}
			&
			\subfloat[]{\label{fig:p8_1}
				\includegraphics[width=0.2\textwidth]{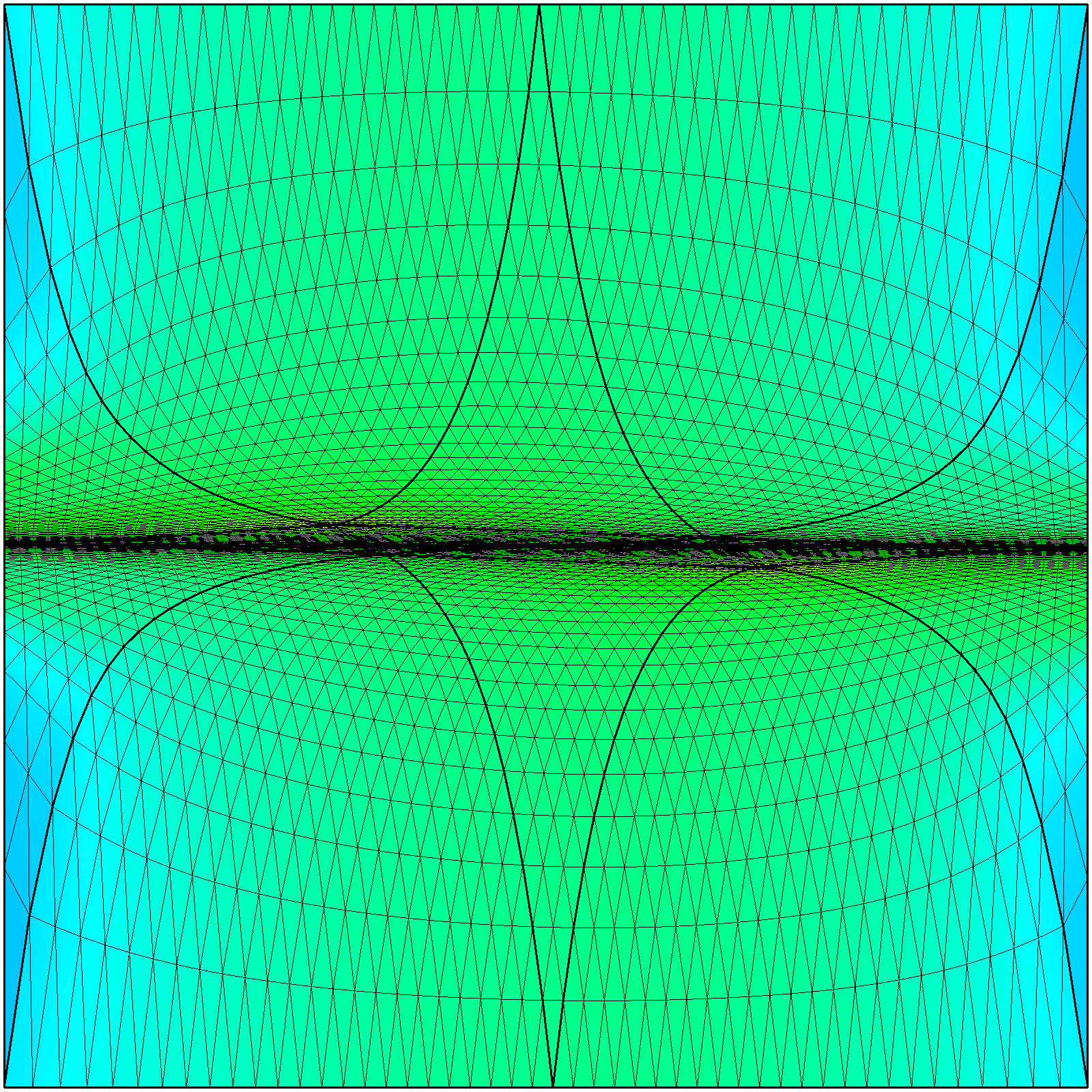}}
			\\
		\end{tabular}
		\\
		\includegraphics[width=0.3\textwidth]{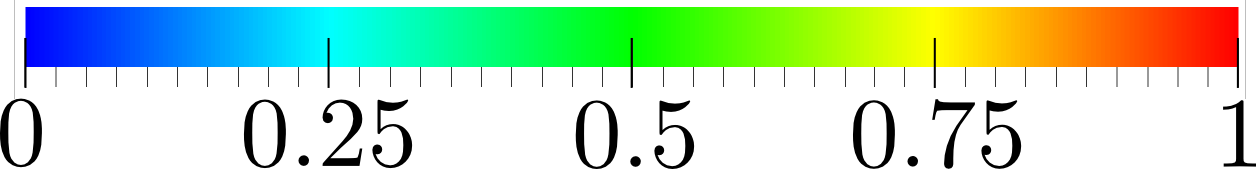} 
		\caption{Pointwise quality measure for triangular meshes of polynomial degree 1, 2, 4, and 8 in columns. Initial straight-sided isotropic meshes and optimized meshes from initial meshes in rows. These element vertices are for a visualization purpose, they are not the high-order degrees of freedom.}
		\label{fig:ex}
	\end{figure}
	
	The meshes are of polynomial degree 1, 2, 4, and 8, and since they have the same resolution, they are composed of the same number of nodes, 481 nodes, but a different number of elements, 896, 224, 56, and 14 elements, respectively. In Figures \ref{fig:ex}\subref{fig:p1_0}, \ref{fig:ex}\subref{fig:p2_0}, \ref{fig:ex}\subref{fig:p4_0}, and \ref{fig:ex}\subref{fig:p8_0} we show these meshes colored according to the pointwise stretching and alignment quality measure, proposed in \cite{aparicio2018defining} which will be detailed in Section \ref{sec:minimizationProblem}. Points in blue color have low quality and points with red color have high quality. As we observe, the elements lying in the region of highest stretching ratio have less quality than the elements lying in the isotropic region. This is because the generated meshes are almost isotropic and, when we equip them with the metric $\zDline$, the mesh quality measures a high deviation between the pointwise stretching and alignment of the mesh and the one of the metric near the region $y = 0$.
	
	The node coordinates of the isotropic mesh may be far from the configuration satisfying the stretching and alignment of the metric. Furthermore, the stretching and alignment of the metric might be impossible to be fulfilled depending on the initial generated mesh. In our case, we look for an optimal configuration, which may not be unique, that approximates the stretching and alignment of the metric.
	
	%	To obtain an optimal configuration, we minimize the distortion measure proposed by changing the coordinates of all the mesh nodes and preserving their connectivity.
	%	The coordinates of the inner nodes, and the one-dimensional coordinates of the inner nodes of the boundary segments, are the design variables. Thus, the inner nodes are free to move, the vertex nodes are fixed, while the rest of boundary nodes are enforced to slide along the boundary segments.
	To obtain an optimal configuration, we minimize the distortion measure proposed by changing the coordinates of all the mesh nodes and preserving their connectivity.
	This can be done by considering all mesh node coordinates targeting a representation of the boundary \cite{aparicio2023combining} or by restricting the boundary mesh nodes to slide over the geometric boundary \cite{aparicio2018defining}.
	%\cite{aparicio2023combining}
	Herein, we consider that the coordinates of the inner nodes, and the one-dimensional coordinates of the inner nodes of the boundary segments, are the design variables. Thus, the inner nodes are free to move, the vertex nodes are fixed, while the rest of boundary nodes are enforced to slide along the boundary segments.
	%	As it is detailed in \cite{aparicio2023combining}, the metric-aware distortion measure can also be minimized by considering all the mesh node coordinates as free variables targeting an implicit geometric model.
	
	The optimized meshes are illustrated in Figures \ref{fig:ex}\subref{fig:p1_1}, \ref{fig:ex}\subref{fig:p2_1}, \ref{fig:ex}\subref{fig:p4_1}, and \ref{fig:ex}\subref{fig:p8_1}. We observe that the elements away from the anisotropic region are enlarged vertically whereas the elements lying in the anisotropic region are compressed. In the optimized mesh, the minimum quality is improved and the standard deviation of the element qualities is reduced when compared with the initial configuration.
	
	\subsection{The minimization formulation: metric-aware distortion measure and free nodes}
	\label{sec:minimizationProblem}
	To match the stretching and alignment of a given metric, we relocate the nodes by minimizing the mesh distortion proposed in \cite{aparicio2018defining} with the corresponding free node coordinates as design variables. Following we summarize the definitions of the metric-aware pointwise, element, and mesh distortion, and  we then state the minimization problem.  
	%	To relocate the node coordinates to match the stretching and alignment of a given metric, we can minimize the mesh distortion measure proposed in \cite{aparicio2018defining} with the corresponding free node coordinates as design variables. Following we summarize the definitions of the metric-aware pointwise, element, and mesh distortion, and  we then state the minimization problem.  
	
\begin{figure}[t]
		\centering
		\includegraphics[width=1.0\textwidth]{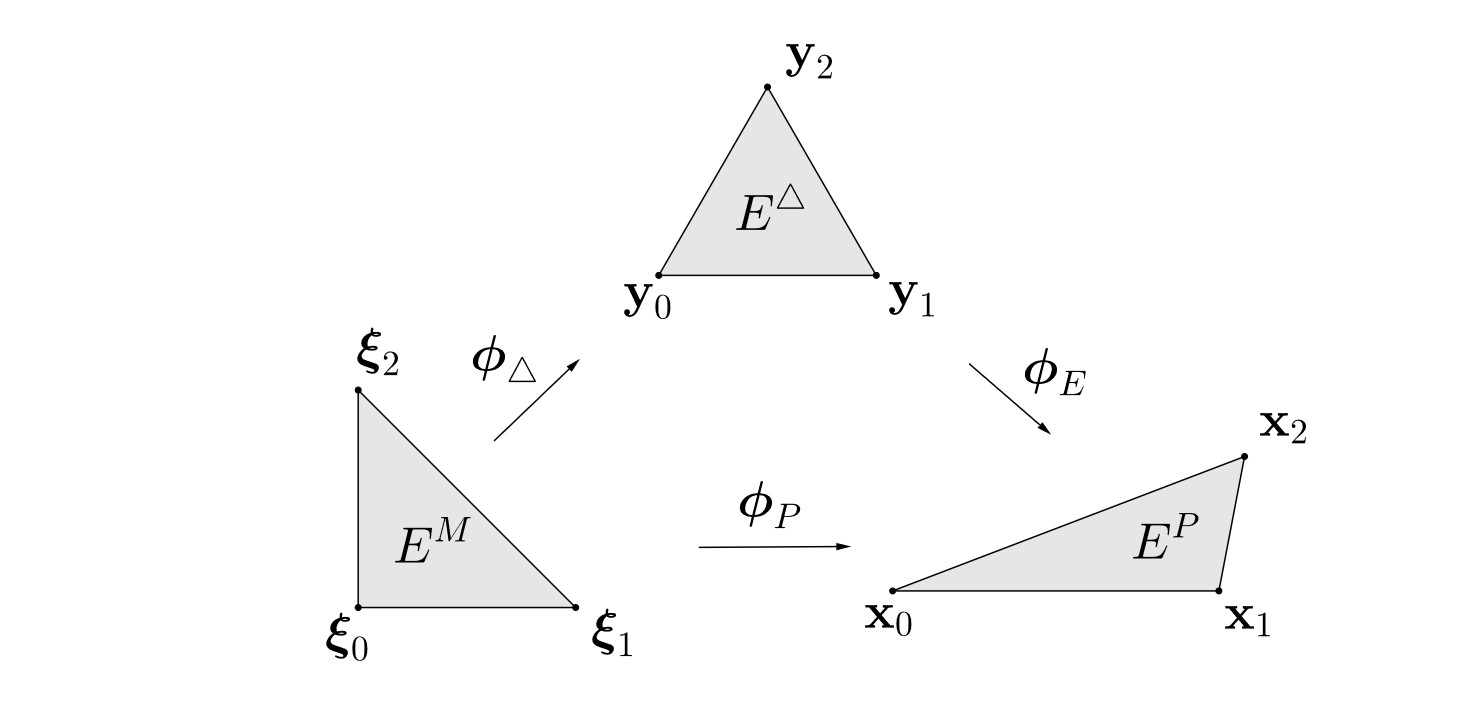}
		\caption{Mappings between the master, the ideal, and the physical elements in the linear case.}
		\label{fig:mappingRefIdealPhysical}
	\end{figure}

	%	\cite{aparicio2018defining}
	To define and compute the distortion of a piece-wise polynomial mesh $\mathcal{M}$ that approximates a domain $\Omega\subset\mathds{R}^d$ equipped with an input metric $\zmetric$, we need mappings between three elements: the master, the equilateral, and the physical, see Figure \ref{fig:mappingRefIdealPhysical} for 2D simplices.. The master $\zmaster\subset \mathds{R}^d$ is the element from which the iso-parametric mapping is defined. The equilateral (regular) element $\zequilater$ is characterized by the element having unitary edge lengths. The physical element $\zphysical\in \mathcal{M}$ is the element to be measured. The respective mappings $\zequilatermap:\zmaster\rightarrow\zequilater,\quad \zphysicalmap:\zmaster\rightarrow\zphysical$ between the equilateral and the physical elements through the master element are obtained. The mapping $\zequilatermap(\zxi)$ between the master element and the equilateral element depends only on a parameter $\zxi\in \zmaster$ while the mapping $\zphysicalmap(\zxi;\zx_e)$ between the master and the physical element $e\in \mathcal{M}$ depends both on the parameter $\zxi$ and the corresponding physical element nodes $\zx_e$.
	
	Then, we define the pointwise distortion measure $\eta$ of the physical element $e\in \mathcal{M}$ at a point $\zu$ as
	\begin{equation}\label{eq:point-wise}
	\eta(\zu;\zx_e) := \frac{\ztr\left(\textbf{A}(\zu;\zx_e)^\mathrm{T}\ \zmetric(\zphysicalmap(\zequilatermap^{-1}(\zu);\zx_e))\ \textbf{A}(\zu;\zx_e)\right)}{d\ \sigma_0^{2/d}},
	\end{equation}
	where $\textbf{A}(\zu;\zx_e) := \zJacobianphysical(\zequilatermap^{-1}(\zu);\zx_e)\ \left(\zJacobianequilater(\zu)\right)^{-1},$ for $\zu\in \zequilater$, and where
	\begin{equation*}
	\sigma_0 := \frac{1}{2}(\sigma + |\sigma|),\quad \sigma := \det\left(\textbf{A}(\zu;\zx_e)\right)\sqrt{\det \zmetric(\zphysicalmap(\zequilatermap^{-1}(\zu);\zx_e))}.
	\end{equation*}
	Here, $\sigma_0$ corresponds to the regularized determinant.
	
	We define the elemental distortion measure $\eta_e$ of the physical element $e\in \mathcal{M}$ as the $L^1$ mean over the equilateral element $\zequilater$
	\begin{equation}\label{eq:elementwise}
	\eta_e := \frac{\big\| \eta\left(\ \cdot\ ; \zx_e \right)\big\|_{L^1(\zequilater)}}{\| 1 \|_{L^1(\zequilater)}},
	\end{equation}
	where the $L^p$-norm is defined by
	\begin{equation*}
		\|f\|_{L^p(\zequilater)} := \left(\int_{\zequilater}|f|^p\right)^{1/p}.
	\end{equation*}
	We will use the reciprocal of the elemental distortion measure of Equation \ref{eq:elementwise} to statistically quantify the mesh element qualities.
	
	Then, for the mesh $\mathcal{M}$ with nodes $\zx_i\in\mathds{R}^d,\ i = 1,...,k$, and equipped with an input metric $\zmetric$, we define the functional that measures the distortion by
	\begin{equation}\label{eq:objective}
	F(\zx_1,...,\zx_k) := \sum_{e\in\mathcal{M}} \big\| \eta\left(\ \cdot\ ; \zx_{e,1},...,\zx_{e,n_p} \right)\big\|_{L^2(\zequilater)}^2,
	\end{equation}
	where we denote the coordinates of the $n_p$ element nodes by $\zx_e = \left(\zx_{e,1},...,\zx_{e,n_p}\right)$, and each pair $(e,j)$ in $\zx_{e,j}$ identifies the local $j$-th node of element $e$ with their global mesh number $i$. That is, for nodal high-order elements is equivalent to determining the configuration of the nodes of the high-order mesh. Moreover, the element contribution to the objective function only depends on the nodes of that element.
	
	For the optimization of the function $F$, each interior node is able to move in $\mathds{R}^d$ and only the normal coordinates of the mesh nodes of the boundary are fixed. Hence, the variables are composed of all the coordinates of the interior nodes and the tangential coordinates of the boundary nodes. We denote the vector containing all the $n$ variable coordinates by $x\in \mathds{R}^n$, and since the other coordinates are fixed we can define $f(x):=F(\zx_1,...,\zx_k)$ with $f:\mathds{R}^n\rightarrow \mathds{R}$. Then, the optimization of the mesh distortion leads to an optimal mesh $\mathcal{M}^*$, where the nodes set $(\zx_1^*,...,\zx_k^*)$ is determined by including the fixed node coordinates to the optimal solution $x^*$. Note that this problem corresponds to an unconstrained minimization problem, and thus, we can solve it using the standard minimization and globalization techniques over-viewed in the following section.
	
	If the domain boundaries are not flat we can consider all coordinates as degrees of freedom. In this case, we can account for the geometric deviation between the mesh boundary and the CAD model with an implicit representation, see \cite{aparicio2023combining} for the details.
	
	Although inverted elements and the form of functional $F$ might have a significant impact on the convergence of the solver towards the optimal mesh, we expect that improvements in the optimization solver tested on initially valid meshes and for our specific functional might also be beneficial to other meshes and functionals. Although there might be other alternative metric-aware functionals, the chosen one is a metric-aware generalization of a shape distortion proven to be successful for isotropic optimization of straight-edged and curved meshes. The chosen distortion only ensures to lead to a local optima of the shape distortion according to the prescribed metric stretching and alignment. That is, this mesh distortion functional does not guarantee a unique optimal mesh.
	
	Because we are detailing a specific-purpose solver, we do not detail how to obtain the input metric $\textbf{M}$. Nevertheless, it is possible to exploit existent high-order goal-oriented \cite{yano2012,fidkowski2011review} and interpolation-oriented \cite{loseille2011continuous,coulaud:VeryHighOrderAnisotropic} error estimators that provide a metric as an output. In practice, these output metrics can be interpolated on a background mesh with the method detailed in \cite{aparicio2022metricinterpolation,aparicio2023combining}.
	\subsection{Optimization overview}\label{sec:optimizationOverview}
	We have casted our adaption problem to an unconstrained minimization problem. To solve the problem, we first recall essential unconstrained optimization concepts, conditions, and notation, to finally detail an optimization algorithm.
	
	Let us consider the unconstrained minimization of a non-linear smooth function $f: \mathds{R}^n\rightarrow \mathds{R}$:
	\begin{equation*}\label{eq:min}
	x^* = \argmin\limits_{x\in\mathds{R}^n} f(x), 
	\end{equation*}
	with gradient and Hessian denoted by $\nabla f$ and $\text{H} f$, respectively.
	
	To decide which points are candidates or local minimizers, we can derive first and second-order conditions from $f$ \cite{nocedal2006numerical}. To derive these conditions, we consider a point $x\in \mathds{R}^n$ and a sufficiently small step $s\in \mathds{R}^n$ to obtain two local approximations of $f(x+s)$ from Taylor's expansion. These two approximations lead to the first and second-order conditions of the minimization problem, respectively. On the one hand, an approximation of first order in $s$, \emph{linear model}, can be computed as
	\begin{equation}\label{eq:linearmodel}
	f(x + s) \approx f(x) + s^\text{T}\nabla f(x).
	\end{equation} 
	The linear model leads to the first-order necessary conditions. That is, if $x^*$ is a local minimizer of $f$ then
	\begin{equation*}\label{eq:firstOrderConditions}
	\nabla f(x^*) = 0.
	\end{equation*} 
	We refer to $x^*$ as a \emph{stationary point} if it fulfills the latter condition. On the other hand, a second-order approximation in $s$,  \emph{quadratic model}, can be computed as
	\begin{equation}\label{eq:quadraticmodel}
	f(x + s) \approx f(x) + s^\text{T}\nabla f(x) + \frac{1}{2}s^\text{T}\text{H}f(x) s.
	\end{equation}
	The quadratic model leads to the second-order sufficient conditions. That is, if 
	\begin{equation*}\label{eq:secondOrderConditions}
	\nabla f(x^*) = 0\quad \text{and} \quad \text{H}f(x^*)\quad \text{is positive definite},
	\end{equation*}
	then $x^*$ is a strict local minimizer of $f$, see a proof in \cite{nocedal2006numerical}. Note that second-order sufficient conditions are not necessary. For instance, there are functions with strict local minimizers where the Hessian matrix vanishes.
	
	Accordingly, to minimize $f$, given an initial point $x_0$, we seek a sequence of non-linear iterates $\{ x_k\}$ that has to converge to a stationary point $x^*$,
	\begin{equation*}
		\lim\limits_{x_k \rightarrow x^*} \| \nabla f(x_k) \| = 0,
	\end{equation*}
	expecting to find a local minimizer. We terminate the sequence when either no more progress can be made or when it seems that a solution point has been approximated with sufficient accuracy, \emph{e.g.}, when the residual $\| \nabla f(x_k) \|$ is smaller than a fixed tolerance. In practice, the sequence is obtained by iteratively computing, from the current point $x_{k}$, a step $s_{k}$ that determines a next point $x_{k+1} = x_k + s_{k}$ with a lower value of $f$, that is, $f(x_{k+1}) < f(x_{k})$. To ensure a sufficient decrease of the objective function and the convergence to either a stationary point or even to a local minimizer, it is standard to compute the step $s_k$ using a \emph{globalization} strategy, see Section \ref{sec:updating}.
	
	We consider globalization strategies that start from a given search direction $p_k$. This search direction can be obtained either from the linear or the quadratic model, and ideally it should lead to a decrease of the objective function. To this end, the direction is required to be a \emph{descent direction}, that is,
	\begin{equation}\label{eq:descent}
	p_{k}^\text{T}\nabla f(x_{k}) < 0.
	\end{equation}
	% The \emph{steepest descent} direction is the  search direction that locally produces the greatest decrease in the linear model (Equation \eqref{eq:linearmodel}) and is given by 
	%
	The search direction that locally produces the greatest decrease in the linear model, Equation \eqref{eq:linearmodel}, is the \emph{steepest descent} direction given by
	\begin{equation*}\label{eq:steepestdescent}
	p_{k} = -\nabla f(x_{k}).
	\end{equation*}
	However, for non-linear functions, the steepest-descent direction might not provide a sufficient decrease of $f$.
	For instance, this is the case near those minimum where the function is locally quadratic.
	
	In this region, we can derive from the quadratic model, Equation \eqref{eq:quadraticmodel}, a direction with a quadratic rate of local convergence, the \emph{Newton direction} \cite{nocedal2006numerical}. This direction satisfies the \emph{Newton Equation} given by the linear system of equations
	\begin{equation}\label{eq:Newton}
	\text{H} f(x_{k})p_{k} = -\nabla f(x_{k}).
	\end{equation}
	The corresponding Newton direction is a descent direction, see Equation \eqref{eq:descent}, whenever the Hessian is positive definite. 
	
	To enforce the search direction fulfills the descent property, we might need to switch to the opposite of the Newton direction. This is so since when the Hessian is non-singular but non-positive definite, the Newton direction is defined but it might violate the descent condition in Equation \eqref{eq:descent}. In this case, a practical choice for $p_{k}$ is the Newton direction times the sign of its scalar product with the steepest-descent direction. That is,
	\begin{equation*}
		p_{k} = -\text{sign}(\tilde{p}_k^\text{T}\nabla f(x_{k})) \tilde{p}_k,
	\end{equation*}
	where $\tilde{p}_k$ satisfies Equation \eqref{eq:Newton}. We call this direction the \emph{signed Newton direction}.	
	\begin{algorithm}[t!]
		\caption{Second-order optimization}\label{alg:optim1}
		\textbf{Input:} $f,\ \nabla f,\ \text{H}f,\ x,\ \text{solver},\ \text{preconditioner}$\\
		\textbf{Output:} $x^*$
%		\begin{multicols}{2} {
				\begin{algorithmic}[1]
					\Procedure{OptimizeFunction}{}
					\State stop $\gets$ false\label{lst:line:stop0}
					\State $k \gets 0$\label{lst:line:setk1}
					\State $\alpha \gets 1$\label{lst:line:setalpha1}
					\State $s \gets \textbf{0}$\label{lst:line:sets1}
					\State $g \gets \nabla f(x),\ H \gets \text{H}f(x)$\label{lst:line:gHeval10}
					\State $g_0\gets g,\ H_0\gets H$\label{lst:line:gHeval11}
					\State $\eta \gets 0.5,\ \tau \gets 0.01$\label{lst:line:setfseq}
					\While {stop is false}\label{lst:line:nonlinloop1}
					\State $p \gets \text{NewtonDirection}(g,H,x,s,\text{solver},\text{preconditioner},\eta,\tau)$\label{lst:line:newtondirection1}
	%				\Statex $\text{preconditioner},\eta,\tau)$
					\State $s,\ \alpha \gets \text{GlobalizationLS}(x,p,\alpha,f,g,H)$\label{lst:line:glob1}
					\If{$k = 0$}
					\State $s_0\gets s$\label{lst:line:sets10}
					\EndIf
					\State $x \gets x + s$\label{lst:line:pointupdate1}
					\State $g \gets \nabla f(x),\ H \gets \text{H}f(x)$\label{lst:line:gHeval1}
					\State $\eta,\ \tau \gets \text{ForcingSequences}\left( x,s_0,g_0,H_0,s,g,H\right)$\label{lst:line:fseqfun}
					\State stop $\gets$ StoppingCondition$(\nabla f,x,s,k)$\label{lst:line:stop1}
					\State $k \gets k + 1$\label{lst:line:iter1}
					\EndWhile
					\State $x^* \gets x$\label{lst:line:sol1}
					\EndProcedure
				\end{algorithmic}
%		}\end{multicols}
	\end{algorithm}
	
	%	The non-linear second-order optimization method is presented in Algorithm \ref{alg:optim1}.
	A general second-order optimization solver, incorporating the previous concepts, obtains a step $s_k$ by applying a globalization strategy to a numerical approximation of the Newton direction $p_k$, see Algorithm \ref{alg:optim1}. This algorithm corresponds to the general scheme for the standard solvers we aim to modify to obtain our specific-purpose solver. The inputs are the objective function, its gradient, its Hessian, an initial guess, and the linear solver choice: direct or iterative. Note that, for the iterative solver, we need to detail the choice of a preconditioner. The output is a configuration expected to be at least locally optimal up to an input tolerance. First, we setup the parameters, Lines \ref{lst:line:stop0}-\ref{lst:line:setfseq}. In particular, we set the stopping criterion, Line \ref{lst:line:stop0},
	the initial non-linear iteration, Line \ref{lst:line:setk1}, the initial step length, Line \ref{lst:line:setalpha1}, and the initial step, Line \ref{lst:line:sets1}. In addition, we set the initial values for the dynamic estimators, Line \ref{lst:line:setfseq}, considered in the specific-purpose optimization solvers. Second, we perform the non-linear iteration, Line \ref{lst:line:nonlinloop1}. Specifically, we compute the exact/inexact approximation of the Newton direction, Line \ref{lst:line:newtondirection1}. Then, we obtain the corresponding step via a line-search globalization, Line \ref{lst:line:glob1}. Following, we obtain the new point by applying the step, Line \ref{lst:line:pointupdate1}. Next, we evaluate the gradient and the Hessian of the objective function, Line \ref{lst:line:gHeval1}. They are used in the next non-linear iteration and, for the specific-purpose case, to update the dynamic estimators, Line \ref{lst:line:fseqfun}. Then, we check the stopping criterion, Line \ref{lst:line:stop1}.	Finally, we upgrade the current non-linear iteration, Line \ref{lst:line:iter1}. Once the loop stopped we set the output point as the obtained one, Line \ref{lst:line:sol1}.
	We remark that the globalized solver aims to ensure convergence from remote starting points, but not necessarily convergence to a global minimum.
	\section{Line-search globalization: standard and specific-purpose strategies}\label{sec:updating}
	
	%%Header: \checkmark \textbf{Section overview, contribution, references}\\
	
	To propose a robust specific-purpose optimization solver for the $r$-adaption problem, Section \ref{sec:modelCase}, a specific-purpose globalization strategy is critical. To obtain such a strategy, we propose to improve the standard line-search globalization. To this end, in Section \ref{sec:standard}, we first review the standard backtracking line-search strategy \cite{nocedal2006numerical}. Then, in Section \ref{sec:final}, we detail the proposed modification.
	Our contribution is to propose a linear predictor model and a new procedure for the computation of the step length.
	%	Our contribution is to propose a linear predictor model (predictor based in linear model) and a new procedure for the computation of the step length. %\textsl{}At the end of each section, the corresponding algorithm scheme for the globalization strategy is presented.
	
	\subsection{Standard backtracking line search: sufficient decrease}\label{sec:standard}
	
	%Change: It is a specific line-search strategy that reduces the step-length to ensure a sufficient decrease of the objective function. To this end, it requires to determine the value of several constants. To enforce convergence, both the algorithm and the value of the involved constants are provided in the literature \cite{nocedal2006numerical}. Furthermore, the convergence properties are well-known for different search directions. Specifically, it seems that Newton's direction properties might be more attractive for challenging target metrics.
	
	%\checkmark \textbf{Overview BLS}\\
	
	The \emph{backtracking} line-search (BLS) strategy is a systematic approach to promote global convergence in a non-linear solver. It is a specific line-search globalization strategy. These strategies minimize a function over a sequence of search paths reducing a multi-variable problem into a one-dimensional problem. A basic line-search strategy consists in computing a suitable \emph{step length} $\alpha_{k}$ for a given descent direction $p_{k}$, see Equation \eqref{eq:descent}, and determining the step $s_k$ as:
	\begin{equation*}\label{eq:LSstep}
	s_k = \alpha_k p_k.
	\end{equation*}
	In order to obtain a sufficient decrease of $f$, such step length should satisfy the \emph{Armijo condition} \cite{nocedal2006numerical}
	\begin{equation}\label{eq:Armijo}
	f(x_{k}+s_k) < f(x_{k}) + cs_k^\text{T} \nabla f(x_{k}),
	\end{equation}
	where $c$ is a constant in $(0,1)$. In its most basic form, a backtracking line-search strategy proceeds by reducing the step length until the Armijo condition is satisfied.
	
	%\checkmark \textbf{Internal iteration} \\
	
	% %Header: \textbf{Halving, sufficient decrease cmin}\\
	To enforce that successive reduction of the step length leads to a sufficient decrease, Equation \eqref{eq:Armijo}, it is preferred to use a small constant like $c = 10^{-4}$, see \cite{nocedal2006numerical}. The constant $c$ of the Armijo condition controls the balance between the decrease of the objective function and the step-length condition. A large constant admits only those step lengths providing a large decrease of the objective function. Accordingly, the desired decrease might not be achieved by reducing the step length monotonically, and hence, one needs an advanced search strategy to set a valid step length. On the contrary, a small constant admits the step lengths providing a small decrease, and thus, we can use the simple successive reduction strategy to set the step length.
	
	%The Armijo condition, Equation \eqref{eq:Armijo}, is also an indicator of which step-lengths provide a reasonable decrease of the objective function.
	
	% imposes a sufficient decrease of the objective function necessary to control the convergence to a minimizer in the convex region. It serves also as an indicator of which step-lengths provide a reasonable decrease of the objective function. When a step-length violates the Armijo condition it is standard to reduce it. Moreover, it is expected that, after some halving iterations, the step-length satisfies the Armijo condition. The constant $c_{\min} := c$ of the Armijo condition is designed to control the relation between the decrease of the objective function and the step-length, a small constant admits the step-lengths providing a small decrease and a big constant admits only those step-lengths providing a big decrease of the objective function. One drawback of choosing a big constant is that the desired decrease may not be achieved by reducing monotonically the step-length, it requires an advanced search of the step-length. For this reason, one chooses a small constant like $c_{\min} = 10^{-4}$ \cite{nocedal2006numerical}.
	
	\begin{algorithm}[t]
		\caption{Standard BLS}\label{alg:LSinitial}
		\textbf{Input:} $x_k,\ p_k,\ \alpha_k,\ f,\ g_k$\\
		\textbf{Output:} $ s_{k},\ \alpha_{k+1}$\\
		\textbf{Set:} $c = 10^{-4}$, $\gamma = 2$, $\alpha_{\min} = 2^{-20}$
		\begin{algorithmic}[1]
			\Procedure{StandardGlobalizationBLS}{}
			\State $s_k \gets \alpha_k p_k$
			\While{$f(x_k + s_k) > f(x_k) + c s^T_k g_k$ and $\alpha_k > \alpha_{\min}$} \label{lst:line:Armijo}
			\State $\alpha_k \gets \alpha_k/\gamma$
			\State $s_k \gets \alpha_k p_k$
			\EndWhile
			%			\State $x_{k+1} \gets x_k + s_k$\label{lst:line:LS1pointupdate}
			\State $\alpha_{k+1} \gets 1$\label{lst:line:LS1steplength}
			\EndProcedure
		\end{algorithmic}
	\end{algorithm}
	
	In Algorithm \ref{alg:LSinitial}, we detail a standard BLS strategy with constants $c = 10^{-4}$, $\gamma = 2$, and $\alpha_{\min} = 2^{-20}$ such as presented in \cite{nocedal2006numerical}. The algorithm inputs are: the point $x_k$, the descent direction $p_k$, the step length $\alpha_k$, the objective function $f$, and the value of the gradient of $f$ at $x_k$, $g_k:=\nabla f(x_k)$. The algorithm outputs are: the new point $x_{k+1}$, the step $s_k$, and the next initial value of the step length $\alpha_{k+1}$. The step length $\alpha_k$ is divided by a factor $\gamma > 1$ iteratively until it satisfies the Armijo condition, Equation \eqref{eq:Armijo}, and while the factor $\alpha > \alpha_{\min}$, Line \ref{lst:line:Armijo}. Finally, the standard BLS strategy restarts the next initial value of the step length to one, Line \ref{lst:line:LS1steplength}.
	
	\subsection{Specific-purpose line search: prediction and continuation of the step length}\label{sec:final}
	
	%%Header: \checkmark \textbf{Motivation, section overview}\\
	To propose a line-search strategy that promotes sufficient decrease and progress, we detail two main ingredients. First, we consider a predictor that indicates if a step length is either large or small. Second, taking into account the predictor, we propose to promote sufficient decrease and progress by either reducing or amplifying the step length. Finally, we combine these ingredients to propose a line-search algorithm featuring memory and continuation of the step length while favoring quadratic convergence of Newton method.
	
	\subsubsection{Step-length predictor: indicating large and small step length}
	
	%%Header: \checkmark \textbf{Predictor BLS (contribution)} \\
	As in the standard strategy presented in Algorithm \ref{alg:LSinitial}, we consider a step length determined by the Armijo condition. However, instead of using the standard inequality presented in Equation \eqref{eq:Armijo}, we propose to use the linear model of the objective function
	\begin{equation*}\label{eq:linearmodelpredictor}
	\phi(s;x) := f(x) + s^{\text{T}}\nabla f(x).
	\end{equation*}
	Note that the step $s$ is a descent direction, see Equation \eqref{eq:descent}, if and only if $\phi(s;x) < \phi(0;x)$.

	Analogously to the standard trust-region formulation presented in \cite{conn2000trust}, for each step $s$ and for the model $\phi$, we can define a predictor given by 
	%Equation \eqref{eq:pred}
	\begin{equation*}\label{eq:predictor}
	\rho(s;x) := \frac{f(x) - f(x + s)}{\phi(0;x) - \phi(s;x)},
	\end{equation*}
	where the model $\phi$ is linear in our line-search strategy, while it is quadratic in trust-region strategies.
	
	The predictor serves as an indicator of the quality of the step length of a descent direction. For a given descent direction, $\phi(s;x) < \phi(0;x)$, the predictor can be either non-positive or positive. When $\rho(s;x) \leq 0$ it indicates that the step does not provide a decrease of the objective function, that is $f(x + s) \geq f(x)$. When $\rho(s;x) > 0$, there is a decrease in the objective function, and thus, $\rho$ indicates the quality of the step length. On the one hand, a low value of the predictor, $\rho(s;x) \approx 0$, indicates a step length far away from those neighborhoods where the function behaves as the linear model. This negligible decrease indicates a large step length. On the other hand, a high value of the predictor, $\rho(s;x) \approx 1$, means that the objective function behaves as the linear model. This linear behavior indicates a small step length. In addition, $\rho(s;x) > 1$ indicates that the objective function does not behave as the linear model. In this situation, we obtain a higher decrease of the objective function than the one expected by the linear model.
	
	\subsubsection{Promoting sufficient decrease and progress: reducing and amplifying step length}
	
	We propose to control the step length according to the value of the predictor. We aim to promote a step length that provides a sufficient decrease of the objective function and that is sufficiently large so that the objective function is not in the linear regime. Heuristically, if we reduce the step length we expect to increase the value of the predictor. On the contrary, if we amplify the step length we expect to decrease the value of the predictor.
	
	We can control the sufficient decrease of the objective function in terms of the predictor. This is so since the Armijo condition of Equation \eqref{eq:Armijo} is equivalent to the bound $\rho(s;x) > c_{\min}$. In particular, for a descent step $s$, the condition $f(x+s) < f(x) + c_{\min}s^T\nabla f(x)$ is equivalent to
	\begin{equation}\label{eq:PredArmijo}
	\rho(s;x) = \frac{f(x) - f(x + s)}{-s^T \nabla f(x)} > c_{\min}.
	\end{equation} 
	
	Even if the decrease of the objective function is reasonable, the step might be too short. The successive reduction of the step length might not ensure reasonable progress. To address this issue, it is standard to use line-search globalizations accounting for the Wolfe conditions  \cite{nocedal2006numerical}. Herein, we propose an alternative adequate to our problem. In contrast to the existent line-search strategies for the Wolfe conditions, our methodology does not require additional evaluations of the gradient of the objective function at the line-search iterations.
	
	To promote sufficient progress, we propose to amplify the step length iteratively that is, $\alpha_{k} \gets \gamma \alpha_{k}$. The amplification of the step length leads to a greater decrease of the objective function. However, it might also reduce the value of the predictor.  To avoid an excessive reduction of the predictor value, which might violate the Armijo condition of Equation \eqref{eq:PredArmijo}, we propose a stopping criterion for the amplifying iterations.
	
	Our criterion stops the amplifying iterations whenever the predictor indicates that the step-length quality exceeds a threshold. Specifically, we stop when $\rho(\gamma s_k;x_k) < c_{\max}$ for a given constant $c_{\max}$. By choosing $c_{\max} \geq c_{\min}$ we ensure that the Armijo condition is satisfied for the step $s_k$ at each amplifying iteration. It may happen that amplifying the step length does not decrease the objective function monotonically, not fulfilling the goal of the line-search iteration. To address this issue in the amplifying iteration, we propose to add the condition $f(x_k + \gamma s_k) < f(x_k + s_k)$. This condition enforces to decrease the objective function.
	
	\subsubsection{Specific-purpose line-search algorithm}
	
	\begin{algorithm}[p!]
		\caption{Specific-purpose LS}\label{alg:LSfinal}
		\textbf{Input:} $x_k,\ p_k,\ \alpha_k,\ f,\ g_k$\\
		\textbf{Output:} $s_k,\ \alpha_{k+1}$\\
		\textbf{Set:} $c_{\min} = 10^{-4}$ $,\ c_{\max} = 0.25,\ \gamma = 2$,$\ \alpha_{\min} = 2^{-20}$
%		\begin{multicols}{2} {
				\begin{algorithmic}[1]
					\Procedure{Specific-purposeGlobalizationLS}{}
					%		\State $ \phi(s;x_k) \gets f(x_k) + s^T Z_k^T\nabla f(x_k) + \frac{1}{2} s^T Z_k^T \text{H} f(x_k)Z_k s $
					\State $s_k \gets \alpha_k p_k$ \label{lst:line:LS1initialsetup}
					\State $\phi(0;x_k) \gets f(x_k)$ \label{lst:line:LS1modelsetup}
					\State $\phi(s_k;x_k) \gets f(x_k) + s^{\text{T}}_k g_k$ \label{lst:line:LS1model2setup}
					\State $\rho(s_k;x_k) \gets \frac{f(x_k) - f(x_k + s_k)}{\phi(0;x_k) - \phi(s_k;x_k)}$ \label{lst:line:LS1endsetup}
					
					\If{$\rho(s_k;x_k) < c_{\min}$}\label{lst:line:LSintermediatepredictorless}
					\While{$\rho(s_k;x_k) < c_{\min}$ and $\alpha_k > \alpha_{\min}$}\label{lst:line:LS1halvingloop}
					\State $\alpha_k \gets \alpha_k/\gamma$
					\State $s_k \gets \alpha_k p_k$
					\EndWhile\label{lst:line:LSendintermediatepredictorless}
					%		\If{$\rho(s_k;x_k) \geq c_{\max}$}
					\Else\label{lst:line:LSintermediatepredictorgreater}
					\While{$\rho(\gamma s_k;x_k) > c_{\max} \text{ and } f(x_k + \gamma s_k) < f(x_k + s_k)$}\label{lst:line:doublingcriterionLS}
					\State $\alpha_k \gets \gamma \alpha_k$
					\State $s_k \gets \alpha_k p_k$
					\EndWhile
					\EndIf\label{lst:line:LS1endloop}
					%			\State $x_{k+1} \gets x_k + s_k$\label{lst:line:stepfinal}
					\If{$\rho(s_k;x_k) < c_{\max}$}\label{lst:line:updatein2LS}
					\State $\alpha_{k+1} \gets \alpha_k /\gamma$ %assegurar que el pujes del cmin, doubling assegura que no et passes
					\Else
					\State $\alpha_{k+1} \gets \alpha_k$
					\EndIf\label{lst:line:LS1endupdate}
					\EndProcedure
				\end{algorithmic}
%			} \end{multicols}
	\end{algorithm}
	
	%%Header: \textbf{Intencio algorithm}\\
	The main objective of the specific-purpose LS, Algorithm \ref{alg:LSfinal}, is to perform a continuation of the step length. This continuation is expected to generate a smooth sequence of non-linear iterations. The inputs and the outputs of Algorithm \ref{alg:LSfinal} are the same as the ones of Algorithm \ref{alg:LSinitial}. The constants $c_{\min} = 10^{-4},\ \gamma = 2,$ and $\alpha_{\min} = 2^{-20}$ correspond to standard values \cite{nocedal2006numerical}. In addition, we propose to set the new constant $c_{\max} = 0.25$ to favor quadratic convergence of Newton method near the optimum without additional line-search iterations, see the reasoning in  \ref{sec:set_c_max}. The algorithm starts, Lines \ref{lst:line:LS1initialsetup}-\ref{lst:line:LS1endsetup}, setting up the main variables and functions: step, current model, model for the step, and predictor.
	
	The algorithm continues by deciding to either reduce or amplify the step length, Lines \ref{lst:line:LSintermediatepredictorless}-\ref{lst:line:LS1endloop}. If the sufficient-decrease condition is violated, we decide to reduce the step length, Line \ref{lst:line:LSintermediatepredictorless}. Otherwise, we decide to amplify the step length, Line \ref{lst:line:LSintermediatepredictorgreater}. Then, we proceed to the line-search iteration. The reduction iterations are the ones of the standard BLS, Lines \ref{lst:line:LSintermediatepredictorless}-\ref{lst:line:LSendintermediatepredictorless}. In contrast, we improve the standard BLS, Lines \ref{lst:line:LSintermediatepredictorgreater}-\ref{lst:line:LS1endupdate}, by enlarging and updating the step length $\alpha_k$ and the step $s_k$, respectively.
	
	First, we decide to amplify the step length while the sufficient-progress condition is violated and the additional decrease of the objective function is fulfilled, Line \ref{lst:line:doublingcriterionLS}. We remark that if no amplifying iterations are performed, the input step length for the current direction is preserved. Finally, we update the step length, Lines \ref{lst:line:updatein2LS}-\ref{lst:line:LS1endupdate}. These instructions provide a step-length memory to the specific-purpose strategy instead of restarting with a step length equal to one in the standard strategy, Line \ref{lst:line:LS1steplength} of Algorithm \ref{alg:LSinitial}. In particular, we only update the step length by reducing it whenever it has not sufficient quality, Line \ref{lst:line:updatein2LS}. We consider this update to prevent an additional reduction iteration at the next non-linear iteration, as it is proposed for trust-region methods \cite{conn2000trust}.
	
	Note that, in the reduction iteration of Line \ref{lst:line:LS1steplength} in Algorithm \ref{alg:LSfinal}, if $\alpha \leq \alpha_{\min}$, the globalization strategy will end. The optimization is successful only if the residual satisfies a given tolerance and fails otherwise.
	
	\section{Newton-CG solvers: standard and specific-purpose methods}\label{sec:updatinglinearsystem}
	After proposing the specific-purpose globalization in Section \ref{sec:updating}, we aim to improve the performance of the non-linear optimization method.
	For this, in this section, we present the standard inexact Newton method, with the standard residual and curvature tolerances and the standard preconditioner. Then, we present the specific-purpose inexact Newton method, with specific-purpose residual and curvature tolerances and a specific-purpose preconditioner.
	\subsection{Standard Newton-CG method}\label{sec:inexactNewton0}
	Next, we present the standard features of the inexact Newton method. These are the residual and curvature forcing sequences and the preconditioner. Then, we combine them to obtain a numerical approximation of the Newton direction.
	\subsubsection{Existing residual and curvature forcing sequences}\label{sec:forcingseq0}
	In an inexact Newton optimization process, the residual and curvature tolerances of the CG method are given by the so-called forcing sequences and forcing terms \cite{eisenstat1996choosing,Dembo1983}. In this section, we present an existing choice of these two estimators. On the one hand, residual forcing terms are presented in \cite{eisenstat1996choosing,Dembo1983,NASH1990219}. They are proposed to avoid \emph{oversolving} the linear system of Newton Equation \eqref{eq:Newton}. On the other hand, a standard constant curvature forcing term is presented for the CG method in \cite{Dembo1983}. It is proposed to limit the total amount of CG iterations.
	
	The role of the forcing sequences is to control the numerical approximation of the Newton direction during the optimization process. This is because, at the starting iterations, it is sufficient to compute an inaccurate direction. Then, in the convergence region, we are interested in promoting quadratic convergence with an accurate direction. In particular, the Newton direction.
	
	The first estimator is the forcing sequence for the residual $r_k$ of the iterative method. Specifically, it is denoted by $\eta$ and it is used as a stopping criterion for the iterative method through the following expression
	\begin{equation*}
	\|r_k\| < \eta \|\nabla f(x_{k})\|.
	\end{equation*}
	In practice, it is standard to set $\eta = 10^{-9}$, in order to achieve a desirable accuracy, see Algorithm \ref{alg:fseq0}. Herein, we define the residual $r_k$ of the linear solver as $r\left(d;x_k\right) = -\text{H}f\left(x_k\right) d - \nabla f\left(x_k\right)$, where $d$ corresponds to the iterative CG-step.
	
	In contrast, dynamic forcing sequences $\{\eta_k\}$ for the residual $r_k$ have been proposed in the literature \cite{eisenstat1996choosing,Dembo1983,NASH1990219}. Specifically, the stopping criterion for the iterative method is now given by
	\begin{equation}\label{eq:fseq}
	\|r_k\| < \eta_{k} \|\nabla f(x_{k})\|.
	\end{equation}
	The choice of $\eta_k$ have been reported to be critical to the efficiency of the inexact Newton method \cite{eisenstat1996choosing}.
	
	Referring to curvature forcing sequences, a constant estimator for the sufficient positive curvature of the CG-step $d_k$ is presented in the literature \cite{Dembo1983}. Specifically, it is denoted by $\epsilon$ and it is used as a stopping criterion for the iterative method by the following expression
	\begin{equation}\label{eq:fseqcurv}
	d_k^{\text{T}} \text{H}f(x_k) d_k < \epsilon\ d_k^{\text{T}}d_k,
	\end{equation}
	where $d_k^{\text{T}} \text{H}f(x_k) d_k$ corresponds to the curvature of the direction $d_k$ according to the objective function $f$, defined as the second-order variation of $f$.
%	Note that, in the literature, $d_k^{\text{T}} \text{H}f(x_k) d_k$ stands for the curvature of the direction $d_k$ according to the objective function $f$ \cite{nocedal2006numerical}.

	It is standard to set $\epsilon = 0$ in Equation \eqref{eq:fseqcurv} to avoid negative curvature directions in the next CG iterations, see Algorithm \ref{alg:fseq0}. In the specific-purpose solver, we will distinguish between curvature forcing term, $\epsilon$ and curvature forcing sequence, $\tau$, being $\tau = \epsilon$ in the standard solver.
	
	\begin{algorithm}[t]
		\caption{Standard Forcing Sequences}\label{alg:fseq0}
		\textbf{Input:} $x,\ s_0,\ g_0,\ H_0,\ s,\ g,\ H$\\
		\textbf{Output:} $\eta,\ \tau$
		\begin{algorithmic}[1]
			\Procedure{ForcingSequences}{}
			\State $\eta \gets 10^{-9},\ \tau \gets 0$
			\EndProcedure
		\end{algorithmic}
	\end{algorithm}
	\subsubsection{Standard Preconditioner}\label{sec:preconditioner0}
	In addition to the forcing sequences, the use of a preconditioner constitutes an important ingredient to improve the efficiency and accuracy of the CG method. For instance, when the initial guess is far from a minimizer, the diagonal preconditioner is a cheap but sufficient approximation of the Hessian to obtain a desirable inexact approximation of the Newton direction. In contrast, accurate preconditioners based on incomplete decomposition are sensitive to the magnitude entries of the Hessian matrix, because the use of pivots is prone to numerical instabilities. Incomplete decompositions become advantageous in the convergence region because we can exploit the numerical stability of the Hessian matrix and the quadratic convergence of an accurate Newton direction.
	\subsubsection{Standard Numerical Approximation of Newton Direction}\label{sec:inexactNewtonalg0}
	The standard inexact Newton method is summarized in terms of the standard forcing sequences and the standard preconditioner presented in Sections \ref{sec:forcingseq0} and \ref{sec:preconditioner0}, respectively. This procedure is used to determine the descent direction, Line \ref{lst:line:newtondirection1}, for the optimization method, Algorithm \ref{alg:optim1}.
	
	\begin{algorithm}[t!]
		\caption{Standard Numerical Approximation of Newton Direction}\label{alg:Newton0}
		%		\textbf{Input:} $g,\ H,\ \sigma,\ x,\ \text{preconditioner},\ \text{solver}$\\
		\textbf{Input:} $g,\ H,\ \sigma,\ x,\ \text{solver},\ \text{preconditioner},\ \eta,\ \tau$\\
		\textbf{Output:} $p$
		%		\textbf{Set:} $\eta = 10^{-9},\ \epsilon = 0$
%		\begin{multicols}{2} {
				\begin{algorithmic}[1]
					\Procedure{NewtonDirection}{}
					\Switch{solver}
					\Case{direct}\label{lst:line:directsolver0}
					\State $p \gets - H \backslash g$\label{lst:line:solve0}
					\EndCase
					\Case{iterative}\label{lst:line:iterative0}
					\State $\text{preconfun}(r) \gets \text{diag}(H)\backslash r$\label{lst:line:setprecon0}
					\State $\epsilon \gets \tau$
					\State $p \gets \text{CG}(H,-g,\textbf{0},\text{preconfun},n,\eta, \epsilon)$\label{lst:line:directioncg0}
					\EndCase
					\EndSwitch
					\State $p \gets \text{sign}(-g^{\text{T}}p)\ p$\label{lst:line:correctsign0}
					\EndProcedure
				\end{algorithmic}
%		}\end{multicols}
	\end{algorithm}
	
	%The inputs are the evaluated gradient and Hessian, a permutation of the $n$ degrees for freedom, the current point, the solver choice, and the preconditioner.
	In Algorithm \ref{alg:Newton0}, we present the numerical approximation of the Newton direction.
%	The inputs are the gradient $g = \nabla f(x)$, the Hessian $H = \text{H}f(x)$, the MDF ordering of the $n$ unknowns for the initial Hessian $\sigma = \text{MDF}\left(\text{H}f(x_0)\right)$, the current point $x$, the solver type (iterative), and the preconditioner function.
	The inputs are the gradient $g = \nabla f(x)$, the Hessian $H = \text{H}f(x)$, the chosen ordering $\sigma$ of the $n$ unknowns for the initial Hessian $\text{H}f(x_0)$, the current point $x$, the solver type (iterative), and the preconditioner function.
	It is standard to set the parameters $\eta = 10^{-9}$ and $\tau = 0$, see Algorithm \ref{alg:fseq0}. The output is a descent direction. First, we decide which solver is used to compute the Newton approximation: direct for an exact Newton approximation, Line \ref{lst:line:directsolver0}, and iterative for an inexact Newton approximation, Line \ref{lst:line:iterative0}. The exact Newton approximation is computed using a complete factorization, Line \ref{lst:line:solve0}. In particular, we use the backslash operator to denote the solution of the linear system by using a sparse $\text{LDL}^\text{T}$ factorization. To compute the inexact Newton approximation, we consider the diagonal of the Hessian as a preconditioner, Line \ref{lst:line:setprecon0}. Then we apply the preconditioned CG algorithm with null initial guess, Line \ref{lst:line:directioncg0}. Note that $n$ corresponds to the number of degrees of freedom. Finally, we obtain a descent direction by correcting its sign according to the steepest-descent direction, Line \ref{lst:line:correctsign0}.
	\subsection{Specific-purpose Newton-CG method}\label{sec:inexactNewton}
	In what follows we present the specific-purpose Newton-CG method. For this, we first detail the specific-purpose residual and curvature forcing sequences and then specific-purpose preconditioner. Finally, we combine them to obtain a specific-purpose numerical approximation of the Newton direction.
	\subsubsection{Specific-purpose residual and curvature forcing sequences}\label{sec:forcingseq}
	The main disadvantage of the standard forcing sequences is the failure of accuracy prediction for inexact Newton approximations. On the one hand, constant forcing sequences keep the accuracy fixed. This is unpractical because the additional accuracy required near an optimum may require, at the same time, an unnecessary computational cost at the first iterations, far from that optimum. On the other hand, the dynamic forcing sequences for the residual presented in Section \ref{sec:forcingseq0} predict the accuracy in terms of a scaled variation between the objective function and the linear model \cite{eisenstat1996choosing}. We have observed that, even if they predict a better accuracy than the fixed sequences, they do not predict a desirable accuracy in our specific problem.
	
	Next, we clarify why standard forcing sequences do not exploit the characteristics of our problem. The standard forcing sequences account for the current step, usually a numerical approximation of the Newton direction, but they do not account for the steepest descent direction. In our approach, we restrict the Hessian on the space spanned by these two directions. Accordingly, when the steepest descent is a better descent direction the forcing term accounts for this, and similarly when the Newton direction is a better descent direction. The impact of this choice becomes relevant when switching from an initial condition far away from a minimizer, to the convergence region. This is because at starting iterations the steepest descent may provide a higher decrease of the objective function. In contrast, in the convergence iterations, it is the Newton direction the one providing an optimal decrease of the objective function. It is for this reason that we advocate for forcing sequences considering both directions, the steepest descent and the numerical approximation of the Newton direction.
	
	\begin{algorithm}[t]
		\caption{Specific-purpose Forcing Sequences}\label{alg:fseq}
		\textbf{Input:} $x_k,\ s_0,\ g_0,\ H_0,\ s_k,\ g_k,\ H_k$\\
		\textbf{Output:} $\eta_k,\ \tau_k$\\
		\textbf{Set:} $\eta_{\max} = 0.5,\ \tau_{\max} = 0.01$
		\begin{multicols}{2} {
				\begin{algorithmic}[1]
					\Procedure{ForcingSequences}{}
					\State{$Z_0 \gets \text{GramSchmidt}\left(-g_0,s_0\right)$}
					\State{$\tilde{g}_0 \gets Z_0^\text{T} g_0,\ \tilde{H}_0 \gets Z_0^\text{T} H_0  Z_0$}
					\State{$\tilde{q}_0 \gets - \tilde{H}_0 \backslash \tilde{g}_0$}
					\State{$\tilde{\kappa}_0 \gets \tilde{q}_0^\text{T} \tilde{H}_0 \tilde{q}_0 / \tilde{q}_0^\text{T}\tilde{q}_0$}
					
					\State{$Z_k \gets \text{GramSchmidt}\left( -g_k,s_k \right)$}
					\State{$\tilde{g}_k \gets Z_k^\text{T} g_k,\quad \tilde{H}_k \gets Z_k^\text{T} H_k  Z_k$}
					\State{$\tilde{q}_k \gets - \tilde{H}_k \backslash \tilde{g}_k$}
					\State{$\tilde{\kappa}_k \gets \tilde{q}_k^\text{T} \tilde{H}_k \tilde{q}_k / \tilde{q}_k^\text{T}\tilde{q}_k$}
					\State{$\eta_k \gets \frac{\| \tilde{q}_k \|}{\|s_0\|},\quad \tau_k \gets \frac{|\tilde{\kappa}_k|}{|\tilde{\kappa}_0|}$}
					\State{$\eta_k \gets \min\left(\eta_k,\eta_{\max}\right)$}
					\State{$\tau_k \gets \min\left(\tau_k,\tau_{\max},\eta_k\right)$}
					\EndProcedure
				\end{algorithmic}
			}	\end{multicols}
	\end{algorithm}
	Next, we present the specific-purpose dynamic forcing sequences for the CG method. For this, we use two additional inexact approximations of the Newton direction: the \emph{restricted Newton direction} (based in a subspace restriction concept presented in \cite{Bulteau1985}) and the \emph{incomplete Newton direction}. Finally, the residual and curvature forcing terms are obtained in terms of these approximations and the corresponding forcing sequences. We summarize the presented procedure in Algorithm \ref{alg:fseq}.
	
	Our restricted Newton direction is given by the Newton Equation
	\begin{equation}\label{eq:newtonLinear}
	H_{k} q_{k} = -g_{k},
	\end{equation}
	restricted to the subspace $W_k:= \text{span} \{-g_k ,s_{k-1}\}$ generated by the steepest-descent direction $-g_k$ and the last step $s_{k-1}=x_{k} - x_{k-1}$, and where $g_{k} := \nabla f(x_{k})$ and $\ H_{k} := \text{H}f(x_{k})$. In particular, we consider the Gram-Schmidt orthonormalization procedure to the ordered basis $\{-g_k ,s_{k-1}\}$. This results in an orthonormal basis $Z_k$ of the subspace $W_k$, where the columns of $Z_k$ are the vectors forming the basis. From this basis, we define the projection of the gradient and the Hessian onto the subspace $W_k$ as
	\begin{equation*}
	\tilde{g}_{k} := Z_{k}^\text{T}g_{k},\quad \tilde{H}_k := Z_k^\text{T} H_{k} Z_k.
	\end{equation*}
	Then, in the restricted form, Equation \eqref{eq:newtonLinear} reduces to the two-dimensional linear system
	\begin{equation*}\label{eq:restrictedNewtonLinear}
	\tilde{H}_{k} \tilde{q}_{k} = - \tilde{g}_{k},
	\end{equation*}
	and the restricted Newton direction is given by the pre-projected direction $Z_k\tilde{q}_k$. Then, we define the forcing sequences by
	\begin{equation}\label{eq:forcingseqs}
	\eta_k := \frac{\| \tilde{q}_{k} \|}{\|s_0\|},\quad \tau_k := \frac{|\tilde{\kappa}_k|}{|\tilde{\kappa}_0|},
	\end{equation}
	where $\tilde{\kappa}_k := \kappa\left( Z_k\tilde{q}_k;x_k \right)$ is the normalized curvature of the restricted Newton direction, see  \ref{sec:NormalizedCurvature}.
	The sequence $\eta_k$ is used for the stopping criterion presented in Equation \eqref{eq:fseq}. In addition, the sequence $\tau_k$ is used for the stopping criterion presented in Equation \eqref{eq:fseqcurv}. Specifically, we set the curvature forcing term $\epsilon_k = \tau_k|\kappa_k|$, where $\kappa_k$ is the curvature of the Newton direction. To compute $\kappa_k$, we observe that, from Equation \eqref{eq:newtonLinear}, we have
	\begin{equation}\label{eq:curvaturesimplification}
	\kappa_k := \frac{q_k^{\text{T}}   H_{k}   q_k}{q_k^{\text{T}} q_k} = \frac{q_k^{\text{T}}   \left( - g_{k}\right)}{q_k^{\text{T}} q_k}.
	\end{equation}
	Instead of computing a solution $q_k$ of Equation \eqref{eq:newtonLinear}, we compute an incomplete approximation $\hat{q}_k$ of $q_k$ using the chosen preconditioner, denoted as $M_k$, and defining $\hat{q}_k$ by
	\begin{equation*}\label{eq:preconditionedNewton}
	M_{k} \hat{q}_{k} = -g_{k}.
	\end{equation*}
	We call $\hat{q}_k$ the incomplete Newton direction.
	Then, using Equation \eqref{eq:curvaturesimplification} and the equation presented above, we approximate the $\kappa_k$ as follows
	\begin{equation}\label{eq:approxNewtoncurv}
	\kappa_k \approx \mu := \frac{\hat{q}_k^{\text{T}}   \left( - g_{k}\right)}{\hat{q}_k^{\text{T}}  \hat{q}_k}.
	%\hat{\kappa}_k :=
	\end{equation}
	Finally, we approximate the curvature forcing term as follows $\epsilon_k \approx \tau_k|\mu|$.
	%\hat{\kappa}_k
	
	%Header: \textbf{Safeguards}\\
	It is standard to apply safeguards to the forcing sequences \cite{Dembo1983,eisenstat1996choosing}. Similarly, we observe that by choosing a safeguard for the forcing sequences presented in Equation \eqref{eq:forcingseqs}, we can improve the inexact Newton method. One for the residual forcing sequence $\eta_k$ given by $\eta_k \gets \min(\eta_k,\eta_{\max})$, we set $\eta_{\max} = 0.5$. The other for the curvature forcing sequence $\tau_k$ given by $\tau_k \gets \min(\tau_{k},\tau_{\max},\eta_k)$ with $\tau_{\max} = 0.01$. This value is set to avoid an excessive influence of the curvature forcing sequence at the initial non-linear iterations.
	
	For our optimization problem, we propose a new forcing sequence for the residual which is suited to limit the number CG-iterations at the beginning of the optimization process and allowing the necessary CG-iterations to obtain a quadratic convergence rate near an optimum. On the other hand, we propose to define the normalized curvature of a given direction and a new dynamic forcing sequence for the curvature of the CG-step to emulate CG-steps with sufficient positive curvature. We define this sequence to limit CG-iterations when the Hessian is near to positive semi-definite without breaking the quadratic convergence rate near an optimum.
	
	In our problem, the main advantage of the specific-purpose forcing sequences is to efficiently predict a desirable accuracy of the inexact Newton approximation at each stage of the optimization process that is, far and near an optimum. This is because they are based in a cheap but faithful approximation of the Newton direction. Specifically, this approximation is obtained by restricting the Newton equation in a subspace spanned by the steepest-descent direction and the step of the last non-linear iteration. Consequently, the forcing sequences predict a decrease of accuracy at the first iterations, obtaining steps approximating the steepest-descent direction. In addition, they predict an increase of accuracy near an optimum, obtaining steps approximating the Newton direction, in order to preserve second-order convergence.
	\subsubsection{Specific-purpose preconditioner}\label{sec:preconditioner}
	In addition to the forcing sequences presented before, the choice of the preconditioner impacts on the efficiency of the iterative method. We remark that a more accurate preconditioner, sensitive to the magnitude of the entries of the Hessian matrix, can be numerically unstable for an ill-conditioned matrix. For this reason, we propose three procedures to reduce both the numerical instabilities and its potential impact in the non-linear optimization process. In this section, we define the preconditioner and then, we present its numerical instability issues together with the three procedures to mitigate them. Then, we present the linear solver obtained from the modifications presented in this section and in Section \ref{sec:inexactNewton0}.
	
	In what follows, we present the specific-purpose preconditioner for the CG method. In addition, we control the numerical instability issues by applying three different procedures: a switch criterion between two preconditioners, a curvature inequality limitation, and an ordering that minimizes the discarded fill of the factorization.
	
	The first procedure consists in switching between the Jacobi preconditioner and the \emph{root-free incomplete Cholesky factorization} ($\text{iLDL}^\text{T}(0)$) with zero levels of fill-in \cite{bertaccini2018iterative}. This switch uses a parameter indicating the numerical instability of the factorization.
	
	The second one is based on an inequality of the curvature of the resulting direction computed from the CG method. If the direction violates the curvature inequality, we consider that the computation of the used preconditioner is numerically unstable.
	
	Finally, the third condition consists in the ordering of the unknowns used to compute the factorization. Several results presented in the literature indicate that the ordering of a matrix has an impact on the numerical instability of its factorization \cite{bertaccini2018iterative}. To control this instability we propose to use an ordering that tries to minimize the discarded fill of the incomplete factorization.
	
	When the initial guess is far from a minimizer, the minimization meets different configurations of the objective function. These configurations can be determined in terms of the Hessian. Roughly speaking, the Hessian starts at a highly indefinite configuration where the positive and negative eigenvalues have large magnitudes. Then, the magnitude of the negative eigenvalues become smaller and the Hessian tends to be nearly singular. After this, the Hessian is positive definite and nearly singular, with small positive eigenvalues. Finally, in the convergence region, the Hessian is positive definite with no small positive eigenvalues. Between these configurations some oscillations may occur, exceptionally switching between an indefinite configuration to a positive definite one. These Hessian configurations are approximately represented in the preconditioner.
	
	Accordingly, we propose to use the preconditioner to detect the Hessian configurations. Specifically, we expect that the diagonal matrix $D$ of a Hessian decomposition indicates when the factorization is indefinite, positive definite, and numerically singular. To this end, in addition to the Jacobi preconditioner, Section \ref{sec:preconditioner0}, we consider the \emph{root-free incomplete Cholesky factorization} $\text{iLDL}^\text{T}(0)$ \cite{bertaccini2018iterative}. Note that the $(0)$ specification indicates a zero level of fill-in.
%	In what follows, we use the notation $\text{iLDL}^\text{T}(0)$ and $\text{iLDL}^\text{T}(0)$.
	
	When applied to the CG method, the $\text{iLDL}^\text{T}(0)$ preconditioner provides an accurate approximation of the Newton direction. This is especially useful for points near a minimizer, where the Newton direction needs to be solved with a high level of accuracy to preserve the quadratic convergence. However, when the initial guess is far from a minimizer, the $\text{iLDL}^\text{T}(0)$ preconditioner may provide low-quality directions interfering with the evolution of the optimization process. Finally, we have observed that when the negative values of the matrix $D$ tend to cluster, the factorization tends to be more numerically stable.
	%	Hence, to detect at which configurations of the objective function we must use the $\text{iLDL}^\text{T}(0)$ preconditioner.
	
	We propose to use the Jacobi preconditioner whenever the $\text{iLDL}^\text{T}(0)$ factorization is supposed to provide low quality directions. %	Header: \checkmark \textbf{Definition of preconditioner}\\
	Specifically, we first obtain the $\text{iLDL}^\text{T}(0)$ preconditioner from an incomplete $\text{LU}$ factorization with zero levels of fill-in ($\text{iLU}(0)$), as 
	%
	%	\begin{equation*}
	%	M_k := \frac{1}{4}\left( L + D\backslash U^{\text{T}} \right) D \left( L + D \backslash U^{\text{T}} \right)^{\text{T}},\ D = \text{diag}(U),
	%	\end{equation*}
	\begin{equation*}
	M_k :=  \tilde{L} D \tilde{L}^{\text{T}},
	\end{equation*}
	where the $\tilde{L}$ and $D$ factors are given by
	\begin{equation*}
	\tilde{L} = \frac{1}{2}\left(L + D\backslash U^{\text{T}}\right),\quad D = \text{diag}(U).
	\end{equation*}
	%	where $L$ and $U$ are the corresponding factors of the $\text{iLU}(0)$ factorization of $H_k=\text{H}f(x_k)$. 
	%
	%	Header: \textbf{ILDL with CG:+ -}\\
	To guess the factorization quality, we consider the negative value with smallest magnitude, $d_{\min}$, and the negative value with largest magnitude, $d_{\max}$, of the diagonal matrix $D$. Then, we use the $\text{iLDL}^\text{T}(0)$ factorization whenever the quantities $d_{\min}$ and $d_{\max}$ are similar. This condition corresponds to check if their ratio is smaller than some fixed quantity. In particular, we consider that the quantities $d_{\min}$ and $d_{\max}$ are similar when the following condition is satisfied
	\begin{equation}\label{eq:ildlcriterion}
	\frac{d_{\max}}{d_{\min}} < \delta,
	\end{equation}
	for $\delta := 10$. We assume that we are near an optimum when the matrix $D$ has no negative values and, in such case, we use the $\text{iLDL}^\text{T}(0)$ factorization. On the contrary, when $d_{\max}/d_{\min} \geq \delta$, we will use the Jacobi preconditioner. The larger the parameter $\delta$, more $\text{iLDL}^\text{T}(0)$ factorizations are used instead of the Jacobi preconditioner. For ill-conditioned problems, this may cause some instability issues breaking the continuity of the optimization process by choosing consecutive steps with nearly opposite directions.
	
	%	Header: \textbf{Curvature limitation inequality}\\
	In addition to the numerical instabilities described before, we have observed that the $\text{iLDL}^\text{T}(0)$ preconditioner can provide low quality directions $p_k$. That is, directions with a low value of the predictor $\rho(\alpha_k p_k;x_k)$ and requiring too many reductions of the length step $\alpha_k$. To avoid such directions, we use the Jacobi preconditioner whenever the CG method with the $\text{iLDL}^\text{T}(0)$ preconditioner stopped because a CG-step of negative curvature is encountered and the CG solution $p_k$ violates the limited curvature inequality
	\begin{equation}\label{eq:limitedcurv}
	\kappa \left(p_k; x_k\right)10^{-2} \leq \tau_{k}|\kappa_k|,
	\end{equation}
	where $\kappa_k$ is approximated as in Equation \eqref{eq:approxNewtoncurv} and $\tau_{k}$ is presented in Equation \eqref{eq:forcingseqs}. In both cases, these quantities are computed using a diagonal preconditioner, $M_k = \text{diag}(H_k)$.
	
	%	Header: \textbf{Ordering for the incomplete factorization}
	
	We have observed that when an $\text{iLU}$ type preconditioner is used (including $\text{iCHOL}$ and $\text{iLDL}^{\text{T}}$ preconditioners) the ordering of the unknowns has a major effect on the convergence of the conjugate gradients iterative method. In our case, where at a given non-linear iteration the mesh may contain highly stretched and curved elements, it is crucial to compute a high-quality preconditioner to ensure convergence of the conjugate gradients method. Furthermore, we are interested in orderings that can take into account in an automatic way both the principal directions of the anisotropy and the ordering of the elements instead of the individual unknowns, especially for high-order elements.
	
	%	Header:	\textbf{MDF}\\
	For anisotropic problems \cite{d1992ordering} and high-order elements \cite{persson2008newton}, the minimum discarded fill (MDF) method provides good convergence results. We only compute the MDF ordering at the beginning of the optimization process that is, for the initial Hessian $H_0 = \text{H}f(x_0)$. We use the computed permutation when the $\text{iLDL}^{\text{T}}$ factorization of the Hessian $H_k = \text{H}f(x_k)$ is computed at the non-linear iteration $k$ and when the corresponding linear system of equations in Equation \eqref{eq:newtonLinear} is solved, in Lines \ref{lst:line:CGprecon1} and \ref{lst:line:CGprecon2} of Algorithm \ref{alg:CG}, see  \ref{sec:CG}. We remark that this ordering is not used for the matrix-vector products.

	\begin{algorithm}[t!]
		\caption{Preconditioner}\label{alg:ildl}
		\textbf{Input:} $H,\ \sigma,\ \text{preconditioner}$\\
		\textbf{Output:} $\text{preconfun}$\\
		\textbf{Set:} $\delta = 10$
%		\begin{multicols}{2} {
				\begin{algorithmic}[1]
					\Procedure{Factorize}{}
					\Switch{preconditioner}\label{lst:line:switch}
					\Case{Jacobi}\label{lst:line:diag}
					\State $M \gets \text{diag}(H)$
					\EndCase
					\Case{iLDL}\label{lst:line:ildl}
					\State $H^{\sigma} \gets H(\sigma,\sigma)$\label{lst:line:mdf}
					\State $[L,U] \gets \text{iLU}(H^{\sigma},0)$\label{lst:line:ilu}
					\State $D \gets \text{diag}(U)$\label{lst:line:setdiag}
					\State $\tilde{L} \gets 0.5\left(L + D\backslash U\right)$\label{lst:line:iluildl}
					\State $d_{\max} \gets \max\limits_{1 \leq i\leq n,\ D_{ii} < 0}|D_{ii}|$\label{lst:line:switch0}
					\State $d_{\min} \gets \min\limits_{1 \leq i\leq n,\ D_{ii} < 0}|D_{ii}|$
					\If{$\frac{d_{\max}}{d_{\min}} \geq \delta$}\label{lst:line:preconswitch}
					\State $M \gets \text{diag}(H)$
					\State $\text{preconditioner} \gets \text{Jacobi}$
					\Else
					\State $P = \text{Id}(:,\sigma)$\label{lst:line:permat}
					\State $M \gets P^{T}\tilde{L}D\tilde{L}^{\text{T}}P$\label{lst:line:factorization}
					\State $\text{preconditioner} \gets \text{iLDL}^{\text{T}}\text{(0)}$
					\EndIf\label{lst:line:switch1}
					\EndCase
					\EndSwitch
					\State $\text{preconfun}(r) = M \backslash r$
					\EndProcedure
				\end{algorithmic}
%			}	\end{multicols}
	\end{algorithm}
	In Algorithm \ref{alg:ildl}, we detail the factorization of the Hessian. The inputs are the evaluated Hessian $H$, the MDF permutation $\sigma$, and the preconditioner choice. First, in Line \ref{lst:line:switch}, we switch between the Jacobi and the $\text{iLDL}^\text{T}(0)$ preconditioner. When the $\text{iLDL}^\text{T}(0)$ preconditioner is chosen, we first apply the permutation to the Hessian, Line \ref{lst:line:mdf}. Then, we compute the $\text{iLDL}^\text{T}(0)$ in terms of the $\text{iLU}(0)$ preconditioner, Lines \ref{lst:line:ilu}-\ref{lst:line:iluildl}. Note that, it is standard to describe the $\text{iLDL}^\text{T}(0)$ factorization of the Hessian $H$, Line \ref{lst:line:factorization}, in terms of the matrix representation $P$ of the permutation $\sigma$, Line \ref{lst:line:permat}, where $\text{Id}(:,\sigma)$ denotes the identity matrix $\text{Id}$ with columns arranged according to $\sigma$ \cite{bertaccini2018iterative}. Finally, we apply the switching criterion in Lines \ref{lst:line:switch0}-\ref{lst:line:switch1}. The output is the preconditioner function.
	
	We propose to use as a preconditioner an incomplete, symmetric, and root-free factorization. Firstly, we have chosen an incomplete factorization as a matter of performance and storage. It is well known that computing a complete factorization of a sparse matrix produces, in general, almost dense triangular factors \cite{bertaccini2018iterative}, leading to a more expensive matrix-vector products (if required in the factorization) and requiring more memory to store the matrix. Secondly, since the CG method requires symmetric matrices, the Cholesky factorization is more appropriate than other factorizations, such as LU. Finally, contrary to the standard Cholesky factorization, its root-free version can be computed at each non-linear iteration of the optimization process. This is because the existence of the root-free factorization does not depend on the matrix $H_k$ being positive definite or indefinite \cite{bertaccini2018iterative,kershaw1978incomplete}.
	\subsubsection{Specific-purpose Numerical Approximation of Newton Direction}\label{sec:updatedNewton}
	The specific-purpose inexact Newton method is summarized in terms of the specific-purpose forcing sequences and the specific-purpose preconditioner presented in Sections \ref{sec:forcingseq} and \ref{sec:preconditioner}, respectively. This procedure is used to determine the descent direction, Line \ref{lst:line:newtondirection1}, for the optimization method, Algorithm \ref{alg:optim1}.
	
	\begin{algorithm}[t!]
		\caption{Specific-purpose Numerical Approximation of Newton Direction with standard preconditioner}\label{alg:Newton1}
		\textbf{Input:} $g,\ H,\ \sigma,\ x,\ \text{solver},\ \text{preconditioner},\ \eta,\ \tau$\\
		\textbf{Output:} $p$
%		\begin{multicols}{2} {
				\begin{algorithmic}[1]
					\Procedure{NewtonDirection}{}
					\Switch{solver}
					\Case{direct}
					\State $p \gets - H \backslash g$\label{lst:line:solve1}
					\EndCase
					\Case{iterative}\label{lst:line:iterative1}
					
					%				\State $\text{preconfun}(r) = \text{diag}(H)\backslash r$\label{lst:line:Newton1precon}
	%				\State $\text{preconfun} \gets$
	%				\Statex $\text{Factorize}(H,\sigma,\text{preconditioner})$\label{lst:line:setprecon1}
					\State $\text{preconfun}\gets\text{Factorize}(H,\sigma,\text{preconditioner})$\label{lst:line:setprecon1}
					\State $\hat{q} \gets \text{preconfun}(-g)$\label{lst:line:curvft1}
					\State $\kappa \gets \frac{\hat{q}^{\text{T}}(-g)}{\hat{q}^{\text{T}}\hat{q}}$
					\State $\epsilon \gets \tau |\kappa|$\label{lst:line:curvft2}
					\State $p \gets \text{CG}(H,-g,\textbf{0},\text{preconfun},n,\eta ,\epsilon)$\label{lst:line:cgpoint1}
					\EndCase
					\EndSwitch
					\State $p \gets \text{sign}(-g^{\text{T}}p)\ p$
					\EndProcedure
				\end{algorithmic}
%			} \end{multicols}
	\end{algorithm}
	%	Header: \textbf{Algorithm inexact Newton 1}\\
	In Algorithm \ref{alg:Newton1}, we summarize the updates of the inexact Newton method, presented in this section and in Section \ref{sec:inexactNewton0}. The inputs are the gradient $g = \nabla f(x)$, the Hessian $H = \text{H}f(x)$, the MDF ordering of the $n$ unknowns for the initial Hessian $\sigma = \text{MDF}\left(\text{H}f(x_0)\right)$, the current point $x$, the solver type (iterative), the preconditioner function, and the value of the residual and curvature forcing sequences at the current non-linear iteration $\eta$ and $\tau$ respectively, see Equation \eqref{eq:forcingseqs}. First, in Line \ref{lst:line:setprecon1}, we compute the preconditioner of the permuted matrix $H^\sigma$, see Line \ref{lst:line:mdf} of Algorithm \ref{alg:ildl}, which is the $\text{iLDL}^\text{T}(0)$ factorization or the $\text{Jacobi}$ preconditioner depending on the criterion presented in Equation \eqref{eq:ildlcriterion}. Then, in Lines \ref{lst:line:curvft1}-\ref{lst:line:curvft2}, we compute the curvature forcing term from the forcing sequence and, next, in Line \ref{lst:line:cgpoint1}, we compute the CG direction. The output of the algorithm is the descent direction $p$.
	
	\begin{algorithm}[t!]
		\caption{Specific-purpose Numerical Approximation of Newton Direction with $\text{iLDL}^\text{T}(0)$ preconditioner }\label{alg:Newton2}
		\textbf{Input:} $g,\ H,\ \sigma,\ x,\ \text{solver},\ \text{preconditioner},\ \eta,\ \tau$\\
		\textbf{Output:} $p$
%		\begin{multicols}{2} {
				\begin{algorithmic}[1]
					\Procedure{NewtonDirection}{}
					\Switch{solver}
					\Case{direct}
					\State $p \gets - H \backslash g$\label{lst:line:solve2}
					\EndCase
					\Case{iterative}\label{lst:line:iterative2}
					
					\State$\text{preconfun},\text{preconditioner}\gets\text{Factorize}(H,\sigma,\text{preconditioner})$\label{lst:line:setprecon2}
	%				\Statex$\text{Factorize}(H,\sigma,\text{preconditioner})$\label{lst:line:setprecon2}
					\State $\hat{q} \gets \text{preconfun}(-g)$\label{lst:line:curvft0}
					\State $\kappa \gets \frac{\hat{q}^{\text{T}}(-g)}{\hat{q}^{\text{T}}\hat{q}}$
					\State $\epsilon \gets \tau |\kappa|$\label{lst:line:curvft}
					\State $p \gets \text{CG}(H,-g,\textbf{0},\text{preconfun},n,\eta ,\epsilon)$\label{lst:line:cgpoint}
					\If{$p^{\text{T}}Hp < 0$ and $\text{preconditioner} = \text{iLDL}^{\text{T}}(0)$}\label{lst:line:illconditioned}
					\State $M \gets \text{diag}(H)$
					\State $\text{preconfun}(r) = M\backslash r$
					\State $\hat{q} \gets \text{preconfun}(-g)$
					\State $\kappa \gets \frac{\hat{q}^{\text{T}}(-g)}{\hat{q}^{\text{T}}\hat{q}}$
					\State $\epsilon \gets \tau |\kappa|$
					\If{$|p^{\text{T}}Hp| > 10^{2}\epsilon\ p^{\text{T}}p$}\label{lst:line:illconditionedcond}
					\State $p \gets \text{CG}(H,-g,\textbf{0},\text{preconfun},n,\eta ,\epsilon)$\label{lst:line:cgpoint2}
					\EndIf
					\EndIf\label{lst:line:illconditioned1}
					\EndCase
					\EndSwitch
					\State $p \gets \text{sign}(-g^{\text{T}}p)\ p$
					\EndProcedure
				\end{algorithmic}
%			}\end{multicols}
	\end{algorithm}
	%	Header: \textbf{Algorithm inexact Newton 2}\\
	In addition, in Algorithm \ref{alg:Newton2}, we incorporate the curvature safeguard, see Equation \eqref{eq:limitedcurv}. Specifically, in Lines \ref{lst:line:illconditioned}-\ref{lst:line:illconditioned1}, we apply the curvature limitation criterion. We first check, in Line \ref{lst:line:illconditioned}, if the CG direction has negative curvature. In such case, we update the curvature forcing term in terms of the Jacobi preconditioner. Finally, in Line \ref{lst:line:illconditionedcond}, if the direction violates the limited curvature inequality, we compute the CG point using the diagonal preconditioner.
	\section{Results}\label{sec:results}	
	In this section, we compare both optimization solvers: specific-purpose versus standard. To do it so, we first present the implementation details, in Section \ref{sec:implementation}. Then, in Section \ref{sec:domainsmetrics}, we propose a set of $r$-adaption tests where the initial guess is far from a minimizer. In particular, to devise and test our solver, we only consider domains with flat boundaries. For results on regions having curved boundaries, see details on \cite{aparicio2023combining}.
	Following, in Sections \ref{sec:resultsglob} and \ref{sec:resultsinexactNewton}, we compare the specific-purpose versus the standard globalizations and linear solvers for the model case presented in Section \ref{sec:modelCase}. Finally, in Section \ref{sec:optimizationresults}, we compare the optimization solvers for all the $r$-adaption tests. They are compared in terms of the non-linear iterations, line-search iterations, and matrix-vector products. In addition, we compare both optimization solvers for an initial guess near to an optimal configuration, Section \ref{sec:MMG}. This is the case of a previously $h$-adapted mesh according to the test metric.
	
	Because our goal is to optimize the mesh distortion, instead of including mathematical proofs of mesh validity, we detail how we numerically enforce the positiveness of the element Jacobians.
	Specifically, we use a numerical valid-to-valid approach that uses four ingredients.
	First, because we want numerically valid results, we enforce mesh validity on the integration points.
	Second, to initialize the optimization, we start from a numerically valid mesh. Third, to penalize inverted elements, we modify the pointwise distortion to be infinity for non-positive Jacobians.
	Specifically, we regularize the element Jacobians to be zero for non-positive Jacobians, so their reciprocals are infinite.
	Note that these reciprocals appear in the distortion expression, and thus, they determine the infinite distortion value.
	Fourth, to enforce numerically valid mesh displacements, we equip Newton's method with a line-search, see Section \ref{sec:updating}.
	Specifically, if the mesh optimization update is invalid in any integration point, the objective function, Equation \eqref{eq:objective}, is infinite.
	In that case, the step is divided by two until it leads to a valid mesh update.
	\subsection{Implementation}\label{sec:implementation}
	As a proof of concept, a mesh optimizer is developed in Julia 1.4.2 \cite{bezanson2017julia}. The in-house implementations include the evaluation of the high-order mesh distortion, the numerical optimizers, and the MDF ordering for the degrees of freedom. These implementations use the Julia base and standard libraries as well as the following external packages: \texttt{Arpack.jl} v0.5.0, \texttt{Einsum.jl} v0.4.1, \texttt{ILUZero.jl} v0.1.0, and \texttt{TensorOperations.jl} v3.1.0. In addition, we use specific functions to solve sparse linear systems. First, we use the Julia internal CHOLMOD package from SuiteSparse as a direct solver, see Line \ref{lst:line:solve0} of Algorithms \ref{alg:Newton0}, \ref{alg:Newton1}, and \ref{alg:Newton2}. Specifically, we solve the linear system by computing a sparse $\text{LDL}^\text{T}$ factorization. Second, we use the preconditioned Conjugate Gradients (CG) algorithm \cite{saad:IterativeMethods} as an iterative method, see Line \ref{lst:line:directioncg0} of Algorithm \ref{alg:Newton0}, Line \ref{lst:line:cgpoint1} of Algorithm \ref{alg:Newton1}, and Lines \ref{lst:line:cgpoint} and \ref{lst:line:cgpoint2} of Algorithm \ref{alg:Newton2}. Third, we compute the $\text{iLU}(0)$ factorization with the ILUZero.jl package, see Line \ref{lst:line:ilu} of Algorithm \ref{alg:ildl}.
	
	%Header: \textbf{Implementation + stopping conditions}\\
	The Julia prototyping code is sequential, it corresponds to the implementation of the method presented in this work and to the method presented in \cite{aparicio2018defining}. In all the examples, the optimization is reduced to find a minimum of a non-linear unconstrained multi-variable function. The ordering of the mesh nodes and of the degrees of freedom is detailed in  \ref{sec:ordering}. The stopping condition is set to reach an absolute root mean square residual, that is $\| \nabla f(x)\|_{\ell^2}/\sqrt{n}$ for $x\in\mathds{R}^n$, smaller than $10^{-4}$.
	Moreover, the examples for the standard and the specific-purpose cases are both equally converged. Finally, each optimization process has been performed in a node featuring two Intel Xeon Platinum 8160 CPU with 24 cores, each at 2.10 GHz, and 96 GB of RAM memory.
	
	Even if the minimization is unconstrained, not all points are valid. In our $r$-adaption problem, the mesh is considered invalid if an element’s mapping Jacobian becomes non-positive at a quadrature point. For this reason, we regularize the objective function to ensure infinite values for inverted configurations. Whenever the Newton’s update provides an inverted configuration, the objective function becomes infinity and thus, the backtracking line-search shortens the update until a valid configuration is reached.
	%
	%	Each optimization process is performed in a single node of the MareNostrum4 super-computer located at the Barcelona Supercomputing Center. The node contains two Intel Xeon Platinum 8160 CPU with 24 cores, each at 2.10 GHz, and 96 GB of RAM memory.
	%	For this reason, we avoid invalid points by regularizing the objective function. This ensures infinite values for inverted configurations. Whenever the globalization update provides an inverted configuration, the objective function becomes infinity and thus, the backtracking line-search shortens the update until a valid configuration is reached.
	\subsection{Examples setup: domains and metrics}\label{sec:domainsmetrics}
	%Header: \textbf{Domain and metric}\\
	We consider the quadrilateral domain $\Omega=[-0.5,0.5]^2$ for the two-dimensional examples and the hexahedral domain $\Omega=[-0.5,0.5]^3$ for the three-dimensional ones. Each domain is equipped with a metric matching a shear layer. In particular, our target metric $\zmetric$ is characterized by a shear layer metric with a diagonal matrix $\textbf{D}$ and a deformation map $\varphi$ by the following expression
	\begin{equation}\label{eq:metricDeformation}
	\zmetric = \nabla \varphi^{\text{T}}\ \textbf{D} \ \nabla \varphi,
	\end{equation}
	where $\textbf{D}$ is a shear layer metric, and $\varphi$ is a deformation map used to align the stretching with a given manifold. The constructions of both $\textbf{D}$ and $\varphi$ are detailed in  \ref{sec:testmetrics}.
	
	The anisotropy of the metric $\zmetric$ can be described by two quantities: the anisotropic ratio and the anisotropic quotient \cite{loseille:AnisotropicAdaptiveSimulations}. On the one hand, the anisotropic ratio is defined by the maximum local elongation. Specifically, at a physical point $\zp\in\Omega\subset \zR^d$ it is given by
	\begin{equation}\label{eq:ratio}
	\text{ratio}\left(\textbf{p}\right) := \sqrt{\frac{\max\limits_{i = 1,...,d} \lambda_i\left(\zp\right)}{\min\limits_{i = 1,...,d} \lambda_i\left(\zp\right)}} > 1,
	\end{equation}
	where $\lambda_i\left(\zp\right) > 0, \ i = 1,...,d$ are the eigenvalues of $\zmetric\left(\zp\right)\in\zR^{d\times d}$. The maximum anisotropic ratio attained in $\Omega$ is denoted by $\text{ratio}_{\max} = \max\limits_{\zp\in\Omega} \text{ratio}\left(\zp\right)$.
	%	Note that here we define the anisotropic ratio as the inverse of the one defined in \cite{loseille:AnisotropicAdaptiveSimulations}.
	
	On the other hand, the anisotropic quotient represents the overall anisotropic ratio. Specifically, at a physical point $\zp\in\Omega\subset \zR^d$, the anisotropic quotient is given by
	\begin{equation}\label{eq:quotient}
	\text{quo}\left(\zp\right) := \frac{\sqrt{\det\left(\zmetric\left(\zp\right)\right)}}{\left(\min\limits_{i = 1,...,d} \lambda_i\left(\zp\right)\right)^{d/2}} > 1.
	\end{equation}
	The maximum anisotropic quotient attained in $\Omega$ is denoted by $\text{quo}_{\max} = \max\limits_{\zp\in\Omega} \text{quo}\left(\zp\right)$. 
	\begin{figure}[t!]
		\centering
		
		\hspace{-0.35cm}
		\footnotesize
		\begin{tabular}{ccc}
			\subfloat{\includegraphics[width=0.25\textwidth]{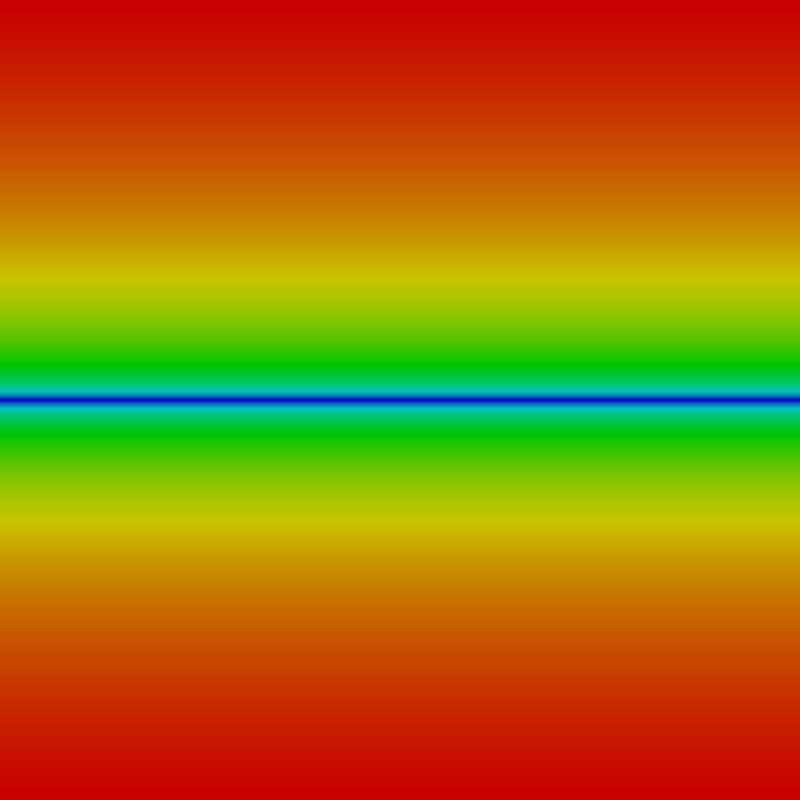}}
			&
			\subfloat{\includegraphics[width=0.25\textwidth]{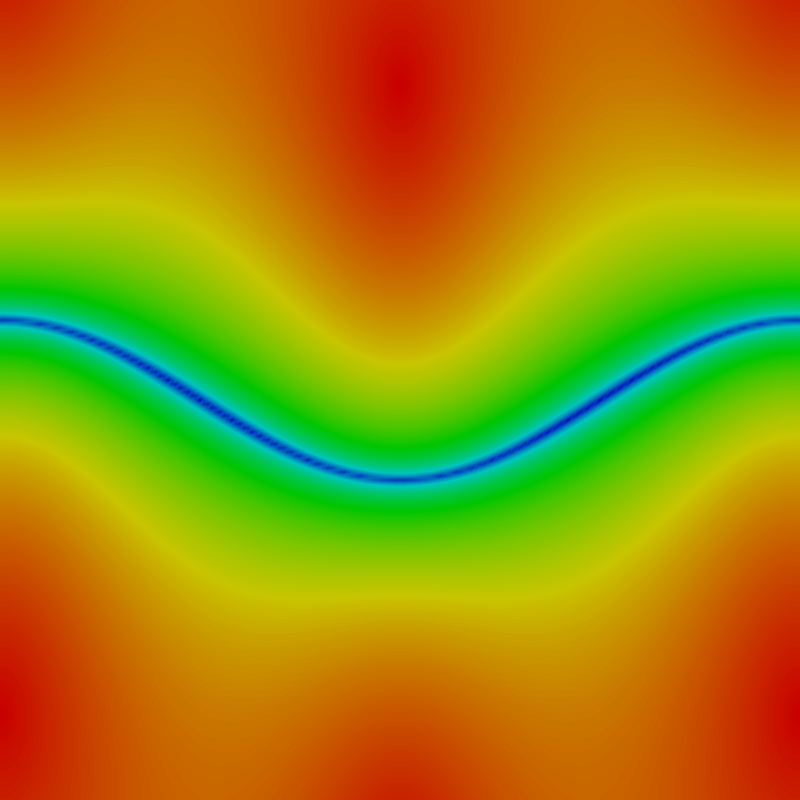}}
			&
			\subfloat{\includegraphics[width=0.25\textwidth]{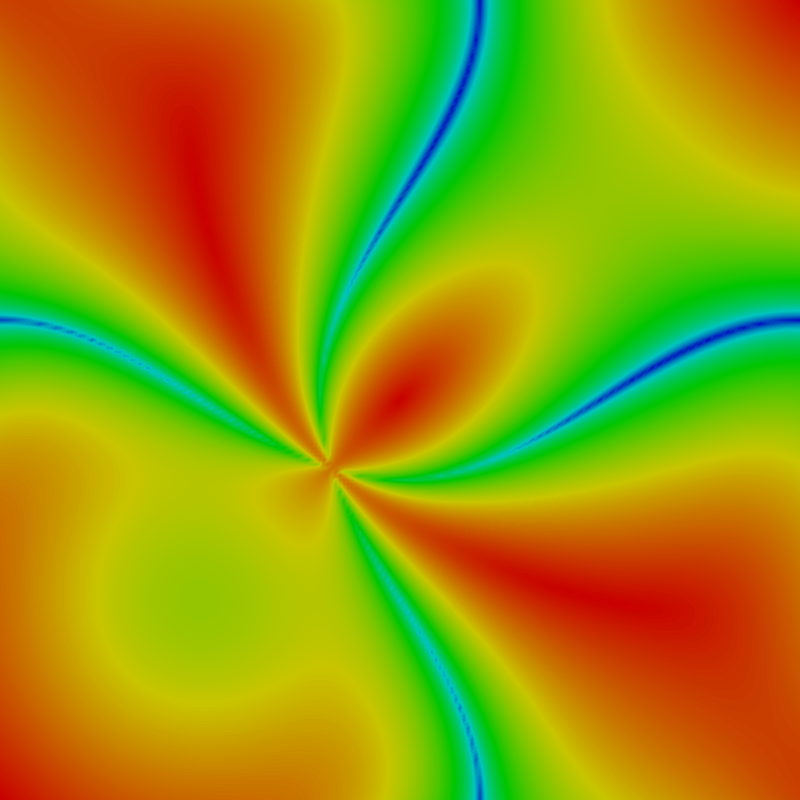}}
			\\
			\stackunder[2pt]{\subfloat{\label{fig:line}
					\includegraphics[width=0.3\textwidth]{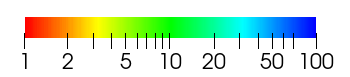}}}{(a)}
			&
			\stackunder[2pt]{\subfloat{\label{fig:curve}
					\includegraphics[width=0.3\textwidth]{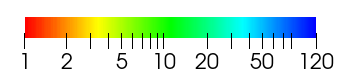}}}{(b)}
			&
			\stackunder[2pt]{\subfloat{\label{fig:curveIntersection}
					\includegraphics[width=0.3\textwidth]{bar1_120}}}{(c)}
			\\
			\subfloat{\includegraphics[width=0.25\textwidth]{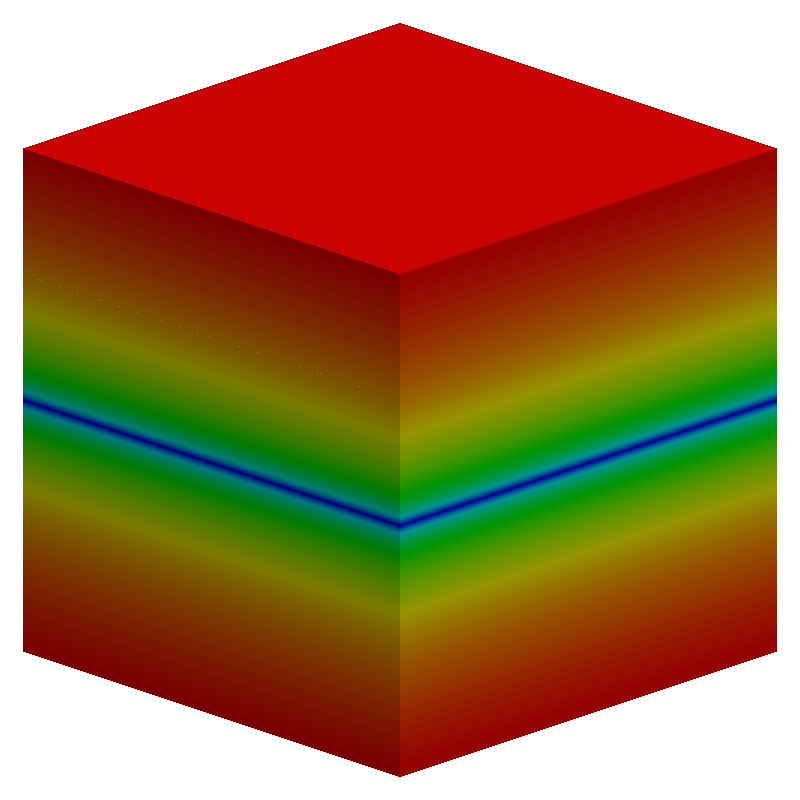}}
			&
			\subfloat{\includegraphics[width=0.25\textwidth]{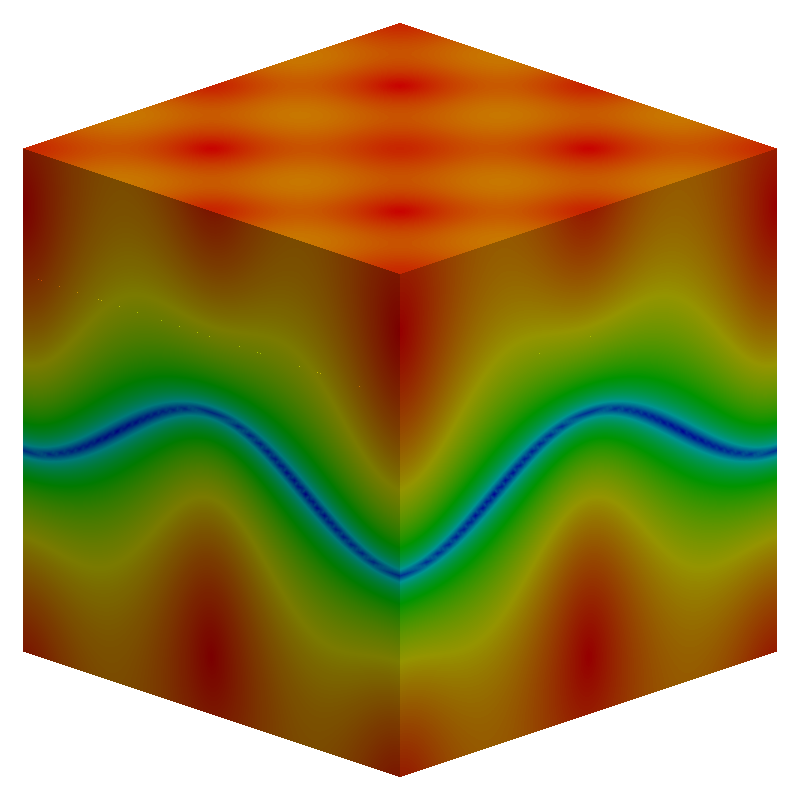}}
			&
			\subfloat{\includegraphics[width=0.25\textwidth]{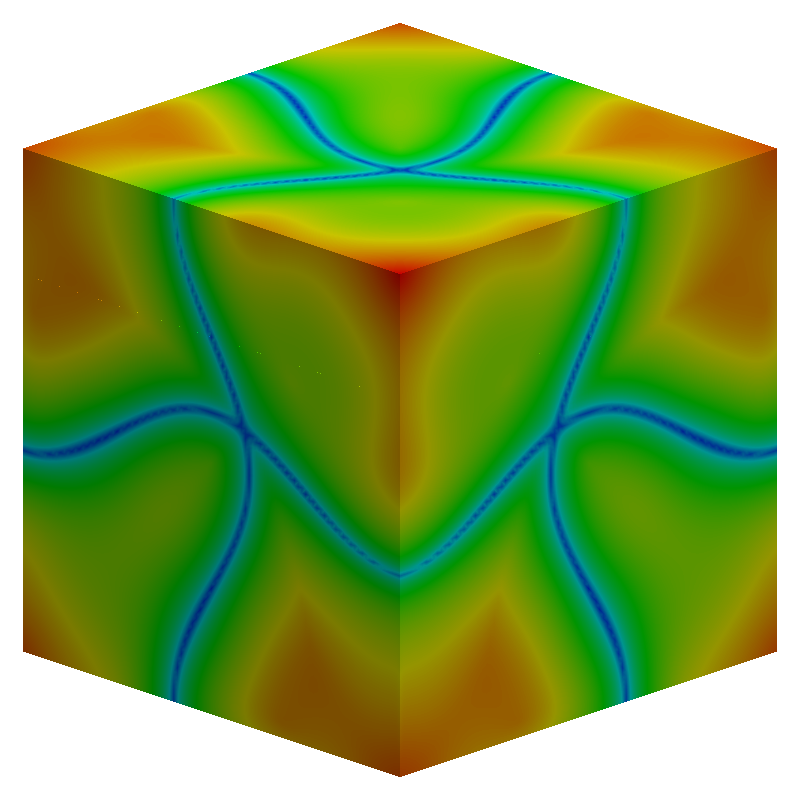}}
			\\
			\subfloat{\includegraphics[width=0.25\textwidth]{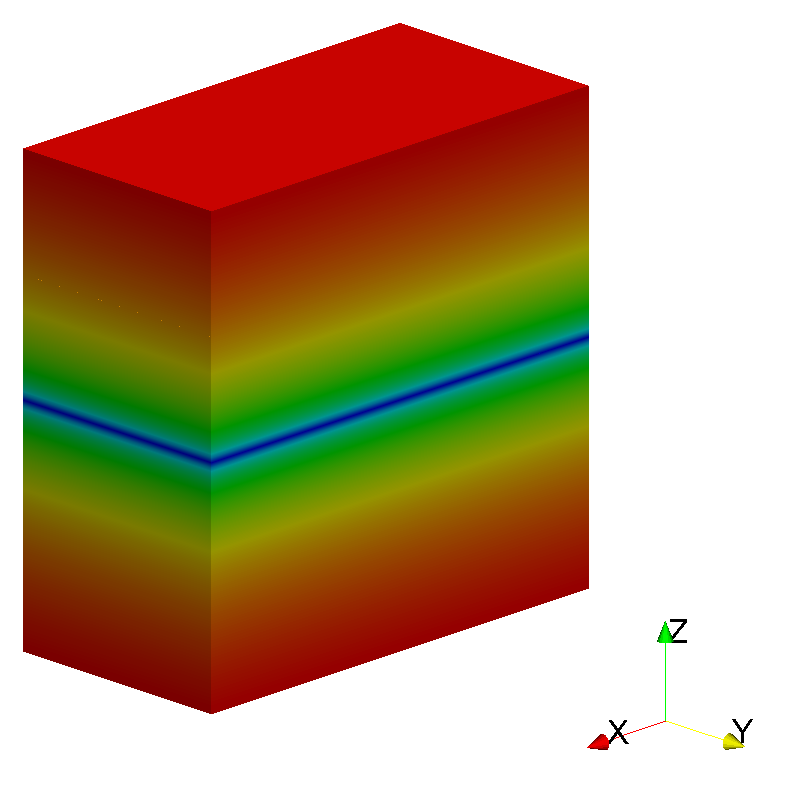}}
			&
			\subfloat{\includegraphics[width=0.25\textwidth]{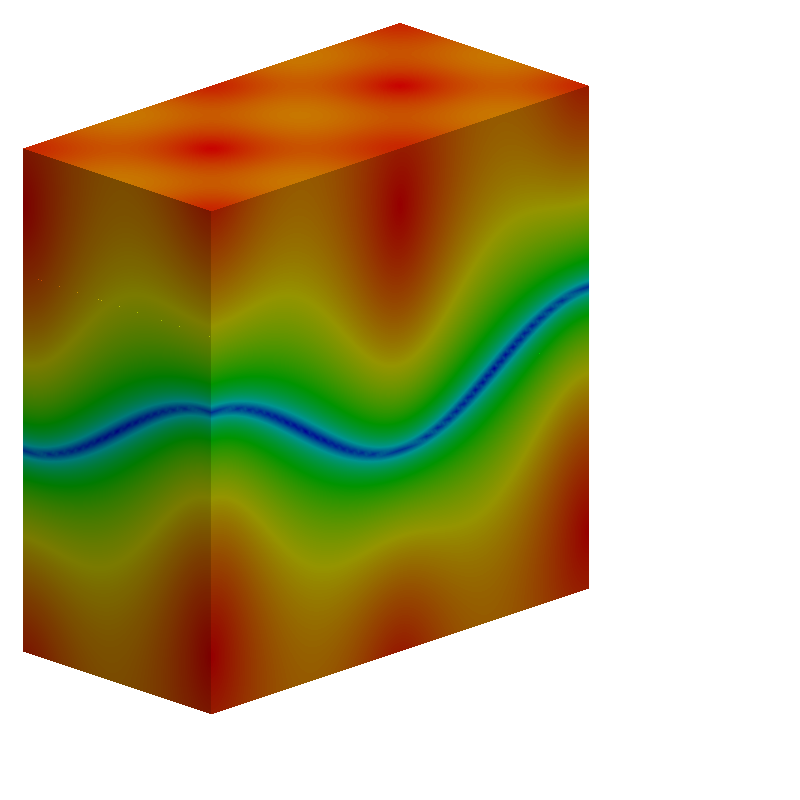}}
			&
			\subfloat{\includegraphics[width=0.25\textwidth]{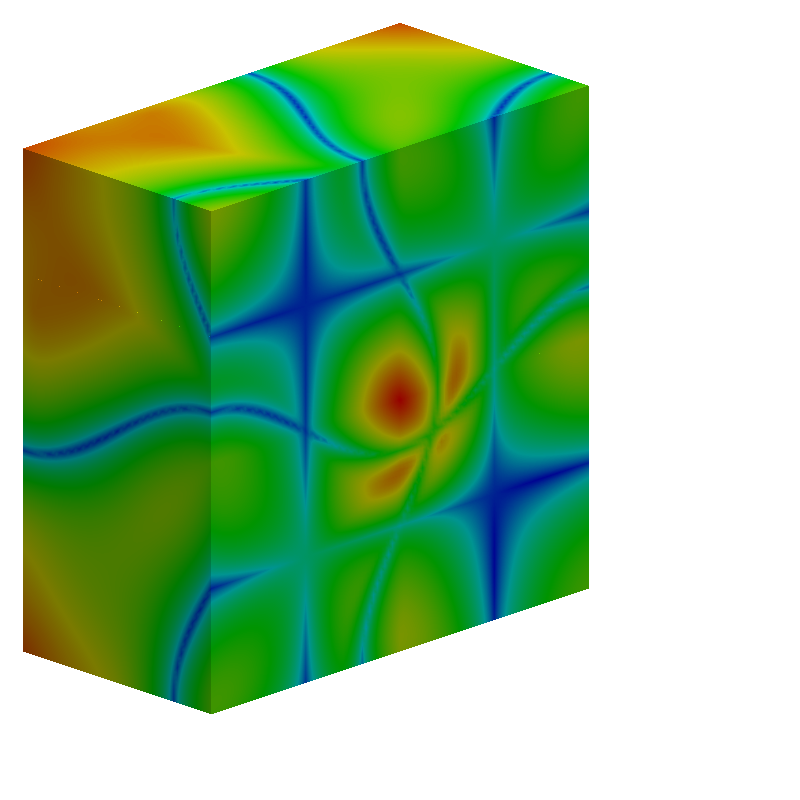}}
			\\
			\stackunder[5pt]{\subfloat{\label{fig:plane}
					\includegraphics[width=0.3\textwidth]{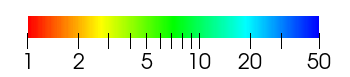}}}{(d)}
			&
			\stackunder[5pt]{\subfloat{\label{fig:surface}
					\includegraphics[width=0.3\textwidth]{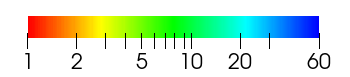}}}{(e)}
			&
			\stackunder[5pt]{\subfloat{\label{fig:surfaces}
					\includegraphics[width=0.3\textwidth]{bar1_60}}}{(f)}
			\\
		\end{tabular}	
		\caption{Anisotropic ratio in logarithmic scale for the different (columns) metric examples and (rows) domain dimensions. The metrics correspond to the ones presented in Table \ref{table:metrics}: Line (a), Curve (b), Curves (c), Plane (d), Surface (e), and Surfaces (f).}
		\label{fig:stretching}	
	\end{figure}
	
	\begin{table}[t!]
		\caption{Metric examples classified in terms of id, name, parameters, anisotropic quantities, and figure. The auxiliar function $g$ is defined in Equation \ref{eq:g}}
		\label{table:metrics}
		\centering
		\small
		\par\medskip
		
		\begin{tabular}{ c l r c r r r c}
			
			\hline\noalign{\smallskip}
			Name & \multicolumn{4}{c}{Parameters} & \multicolumn{2}{c}{Anisotropic} & Fig.\\
			& $\textbf{D}$ & $\varphi$ & $\gamma$ & $h_{\min}^{-1}$ & ratio & quo. & \\
			\noalign{\smallskip}\hline\noalign{\smallskip}
			Line & $\zDline$ & $(x,y)$ & 2 & 100 & 100 & 100 & \ref{fig:stretching}(a) \\
			Curve & $\zDline$ & $\left(x,\frac{g(y,x,1)}{\sqrt{100 + 4\pi^2}}\right)$ & 2 & 100 & 120 & 120 & \ref{fig:stretching}(c) \\
			Curves & $\zDcross$ & $\left( g(x,y,1),g(y,x,1) \right)$ & 2 & 100 & 120 & 120 & \ref{fig:stretching}(e) \\
			\hline
			Plane & $\zDline$ & $(x,y,z)$ & 2 & 50 & 50 & 50 & \ref{fig:stretching}(b) \\
			Surface & $\zDline$ & $\left(x,y,\frac{g(z,y,x)}{\sqrt{100 + 8\pi^2}}\right)$ & 2 & 50 & 60 & 60 & \ref{fig:stretching}(d) \\
			Surfaces & $\zDcross$ & $\left( g(x,y,z),g(y,z,x),g(z,y,x)\right)$ & 2 & 50 & 60 & 3600 & \ref{fig:stretching}(f) \\
			\noalign{\smallskip}\hline\noalign{\smallskip}
		\end{tabular}
	\end{table}
	In Table \ref{table:metrics}, we present six examples of metrics. In the first column, we show the numbering. Then, in the second column we show a descriptive name. Specifically, Line and Plane correspond to the shear layer metrics over a line and a plane, respectively. In contrast, Curve and Surface correspond to the shear layer metrics over a deformed line and a deformed plane, respectively. We use a singular noun for a layer over one entity and a plural noun for an intersection of two layers in 2D and three layers in 3D. In the third column, we present the parameters that characterize the metric, see  \ref{sec:testmetrics}: the shear layer metric $\textbf{D}$, the deformation map $\varphi$ in terms of the function
	\begin{equation}\label{eq:g}
		g(x,y,z) := 10x - \cos(2\pi y)\cos(2\pi z),
	\end{equation}
	the growth factor $\gamma$, and the inverse of the imposed stretching $h_{\min}$, $1/h_{\min}$. Then, in the fourth column we present the approximate anisotropic ratio and quotient defined in Equations \eqref{eq:ratio} and \eqref{eq:quotient}, respectively. Finally, in the last column we include the figure corresponding to the metric.
	
	In Figure \ref{fig:stretching}, we show the anisotropic ratio of the test metrics. We can observe that, it blends between 1 and $1/h_{\min}$ together with a contribution of the deformation $\varphi$. In addition, the maximum anisotropic ratio is attained at the zero-level sets of the last or each component of the deformation map $\varphi$ depending on which shear layer metric is used. That is, according to the ordering presented in Table \ref{table:metrics} at: line $y = 0$; curve $g(y,x,1) = 0$; the curves $g(y,x,1) = 0,\ g(x,y,1) = 0$; plane $z = 0$; surface $g(z,y,x) = 0$, and the surfaces $g(z,y,x) = 0,\ g(y,x,z) = 0,\ g(x,y,z) = 0$, respectively. Finally, note that at the intersection of two entities in 2D and three entities in 3D, the anisotropic ratio attains its minimum value, equal to one. This is because the stretching alignments span all the space, producing a sizing effect without stretching on a particular direction. 
	
	\begin{figure}[t!]
		\centering
		\hspace{-0.35cm}
		\tiny
		\begin{tabular}{cccc}
			\subfloat[]{\label{fig:p1_2_0}
				\includegraphics[width=0.22\textwidth]{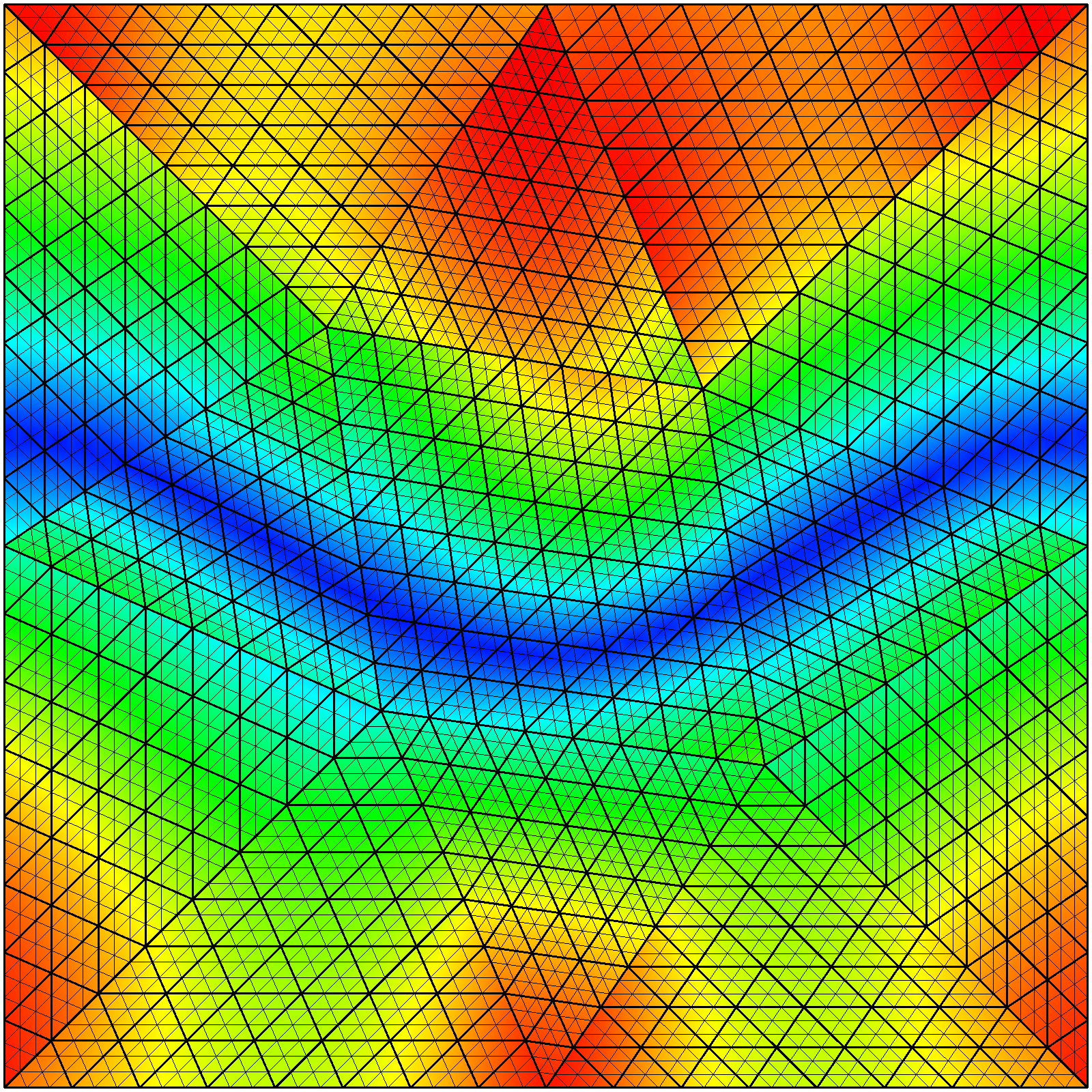}}
			&
			\subfloat[]{\label{fig:p2_2_0}
				\includegraphics[width=0.22\textwidth]{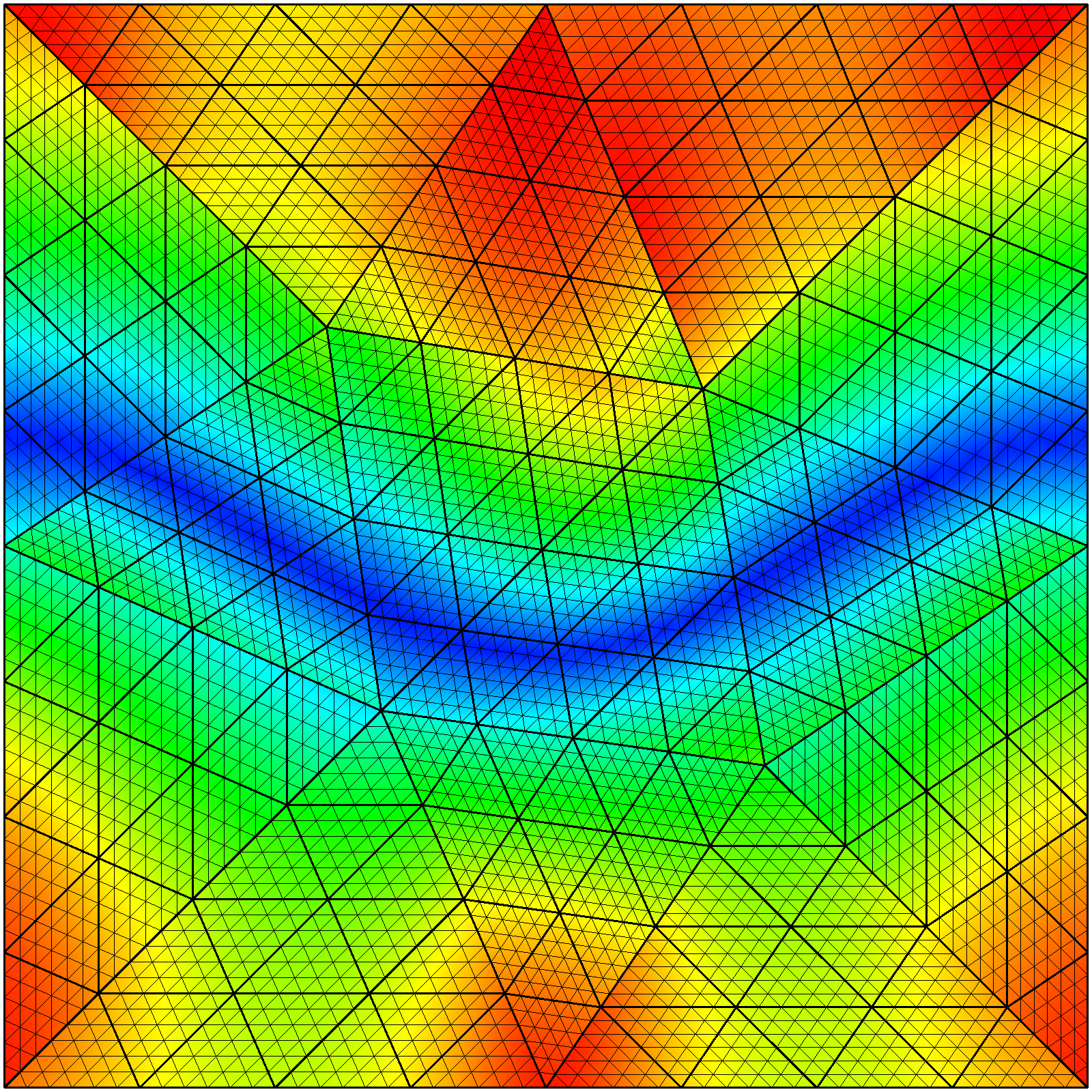}}
			&
			\subfloat[]{\label{fig:p4_2_0}
				\includegraphics[width=0.22\textwidth]{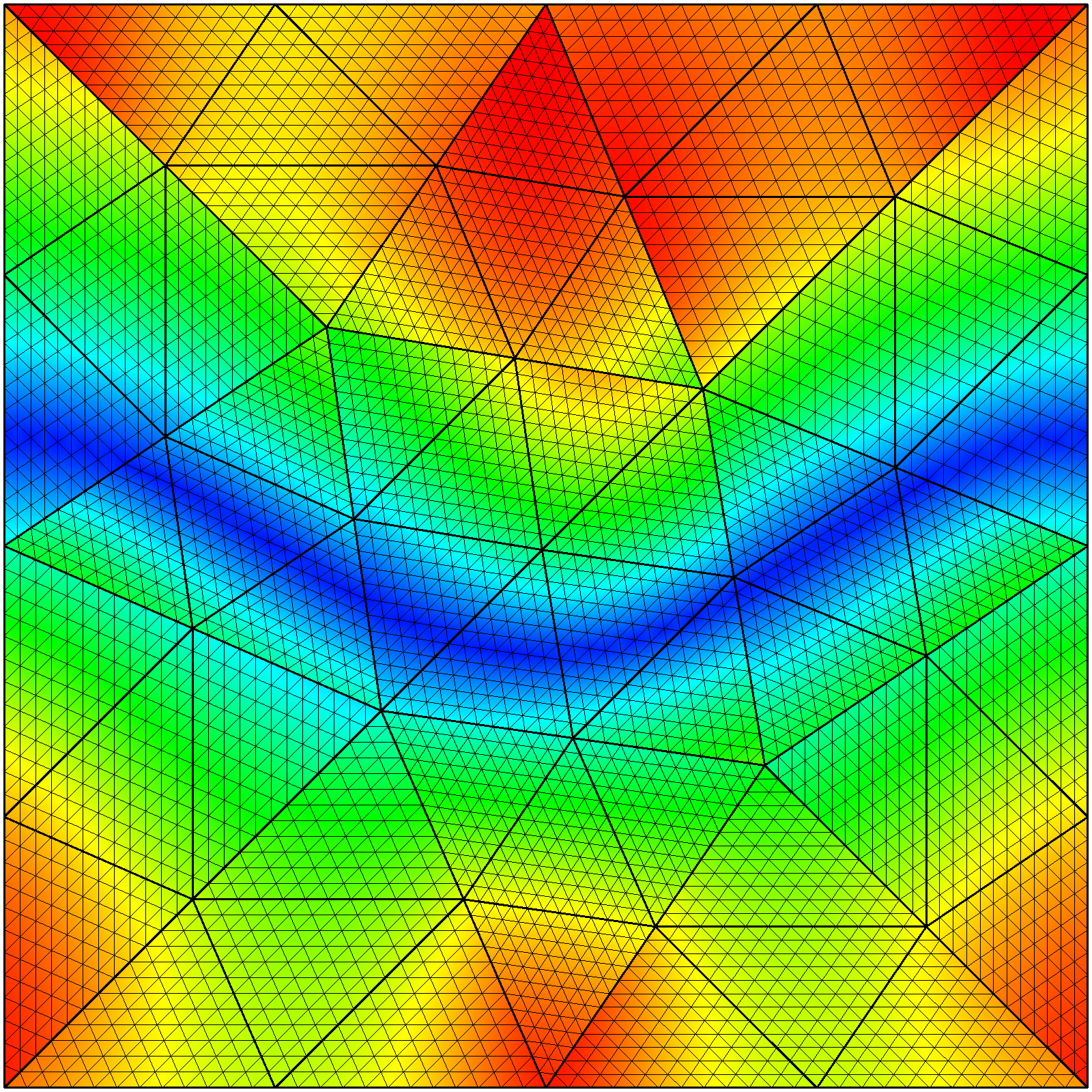}}
			&
			\subfloat[]{\label{fig:p8_2_0}
				\includegraphics[width=0.22\textwidth]{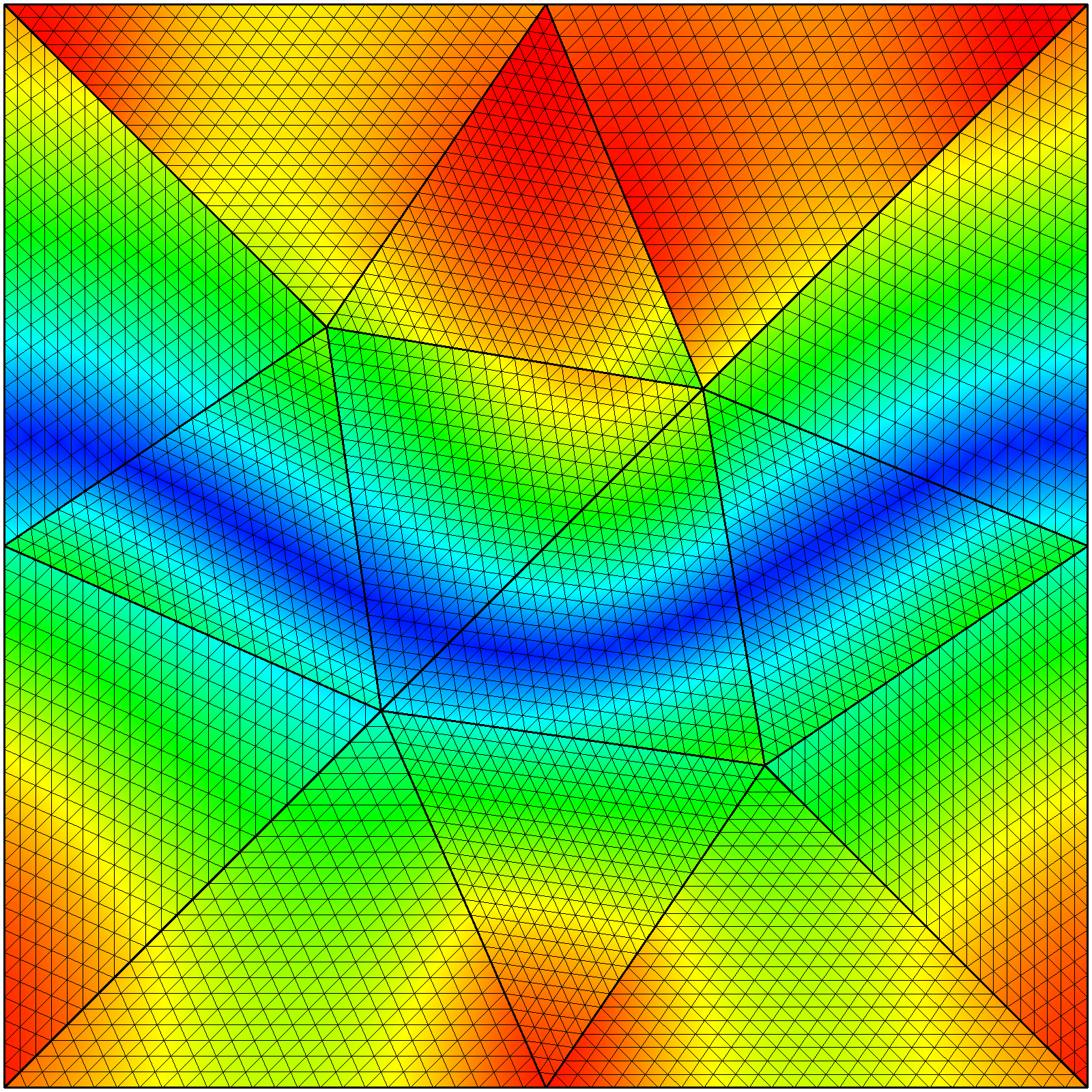}}
			\\
			\subfloat[]{\label{fig:p1_2_1}
				\includegraphics[width=0.22\textwidth]{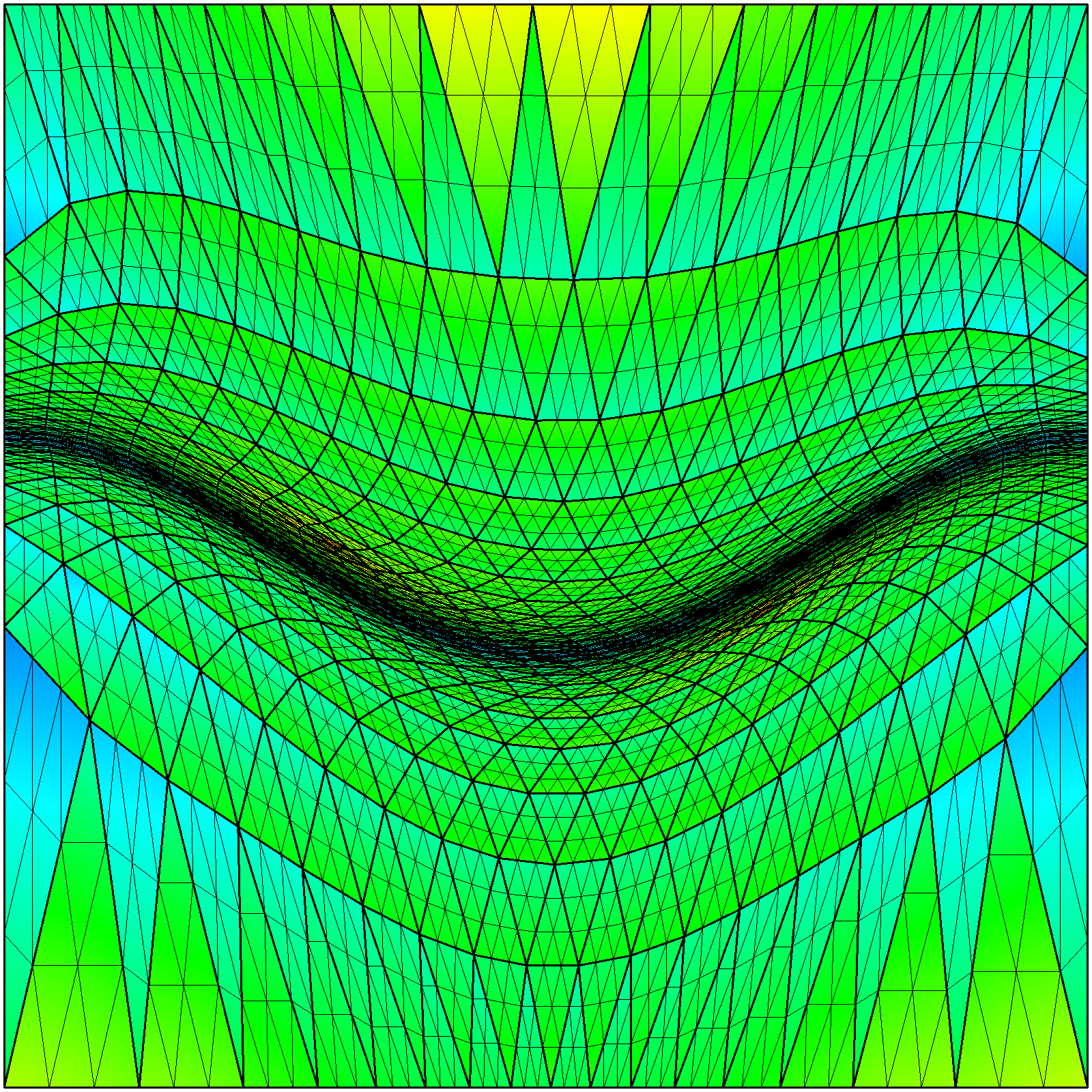}}
			&
			\subfloat[]{\label{fig:p2_2_1}
				\includegraphics[width=0.22\textwidth]{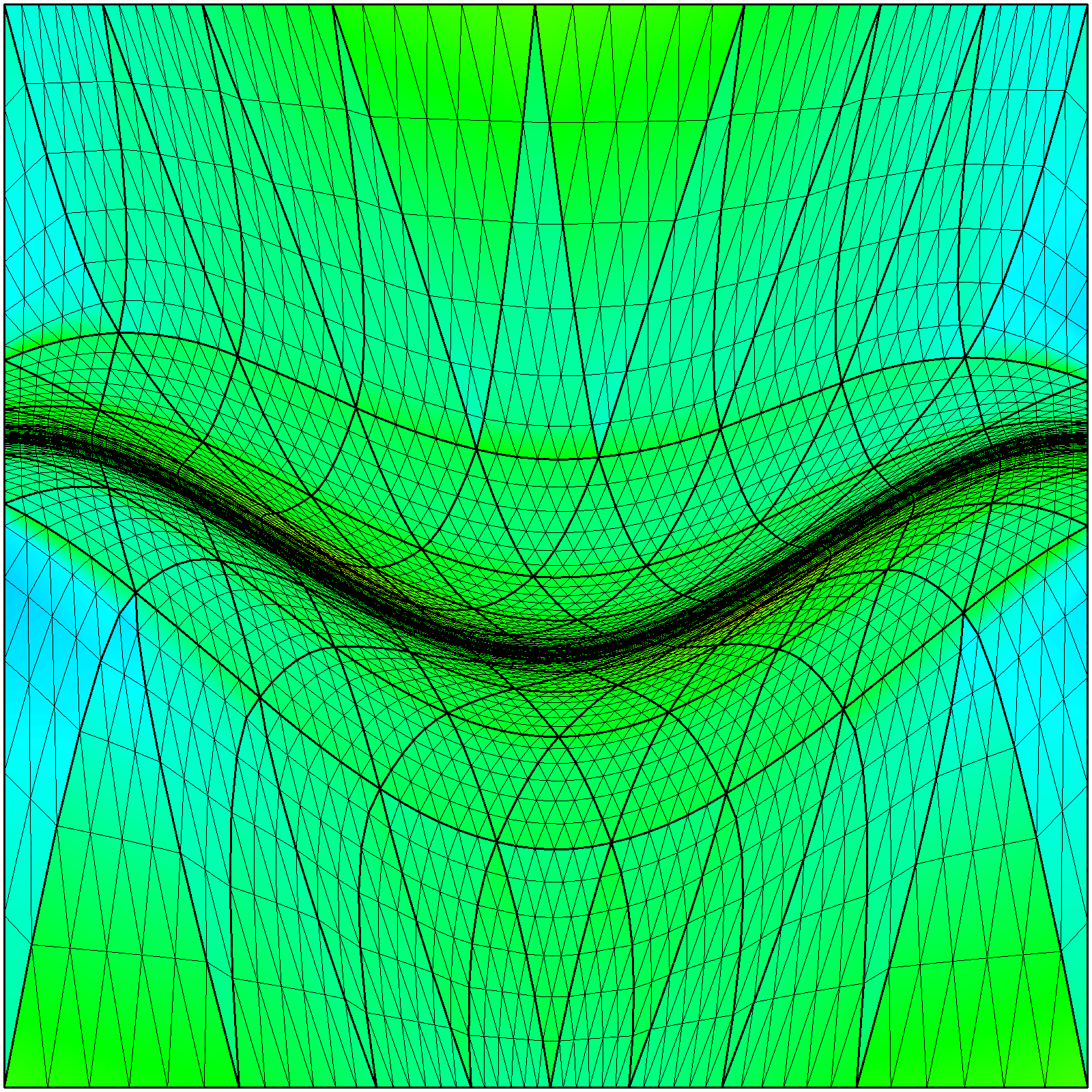}}
			&
			\subfloat[]{\label{fig:p4_2_1}
				\includegraphics[width=0.22\textwidth]{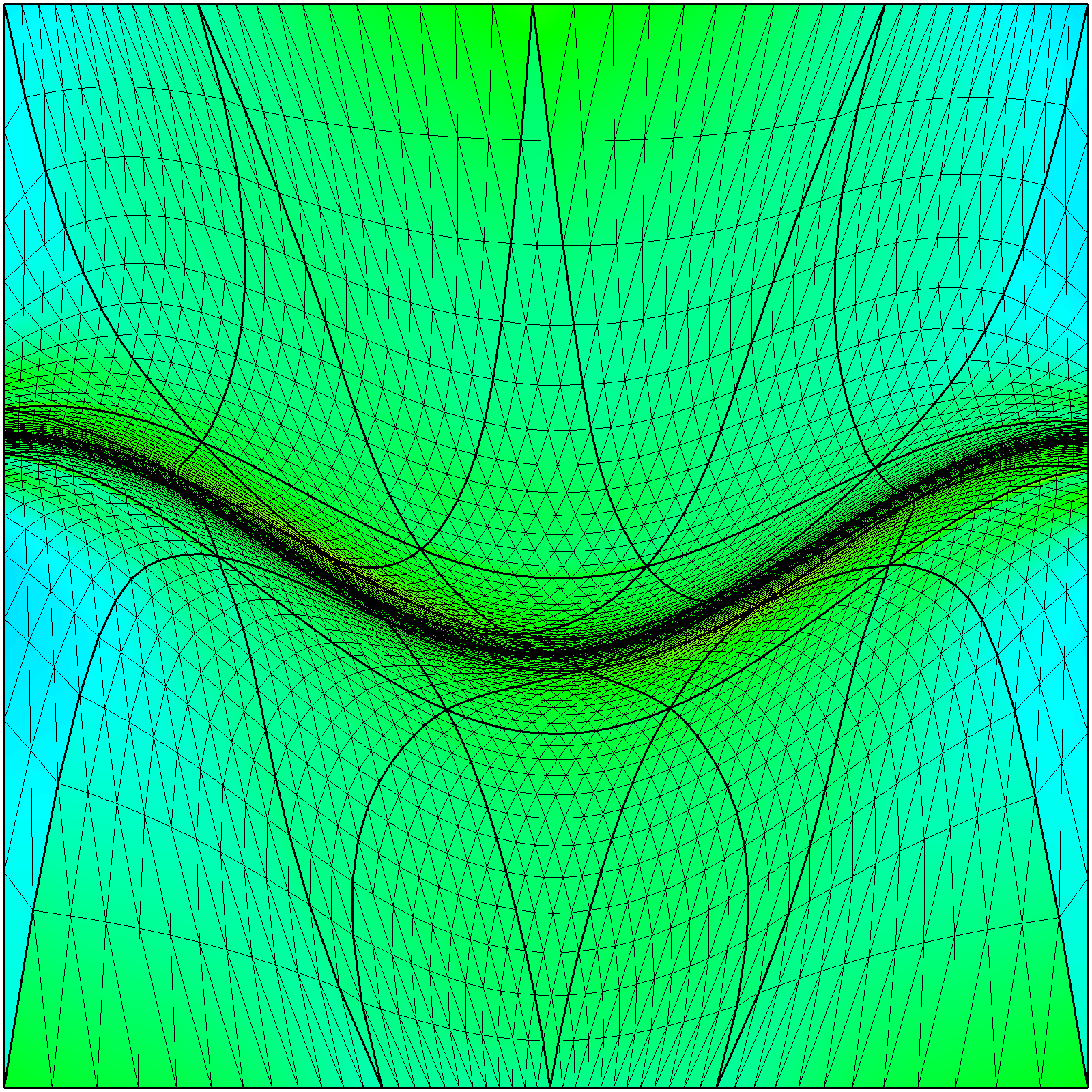}}
			&
			\subfloat[]{\label{fig:p8_2_1}
				\includegraphics[width=0.22\textwidth]{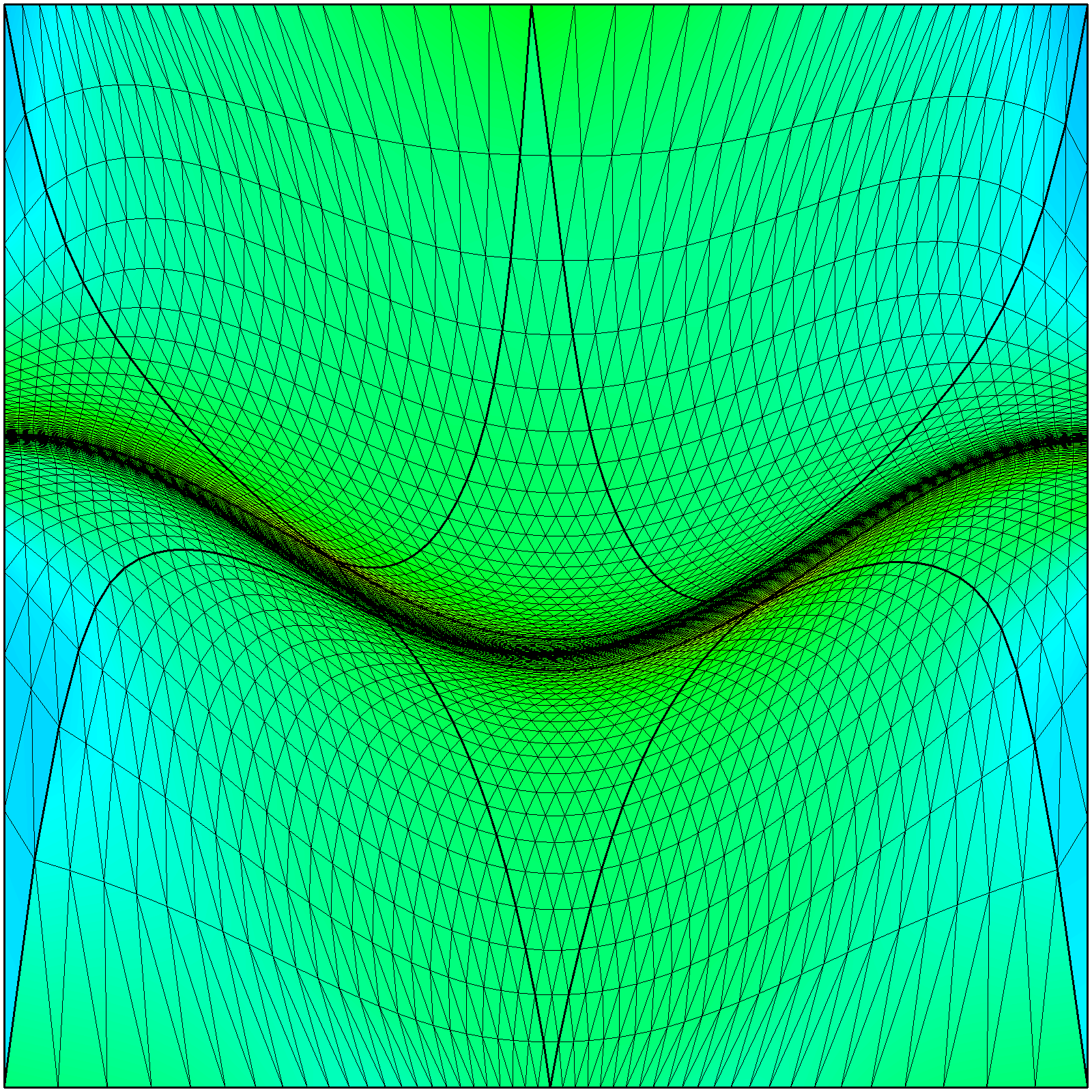}}
			\\
			\subfloat[]{\label{fig:p1_3_0}
				\includegraphics[width=0.22\textwidth]{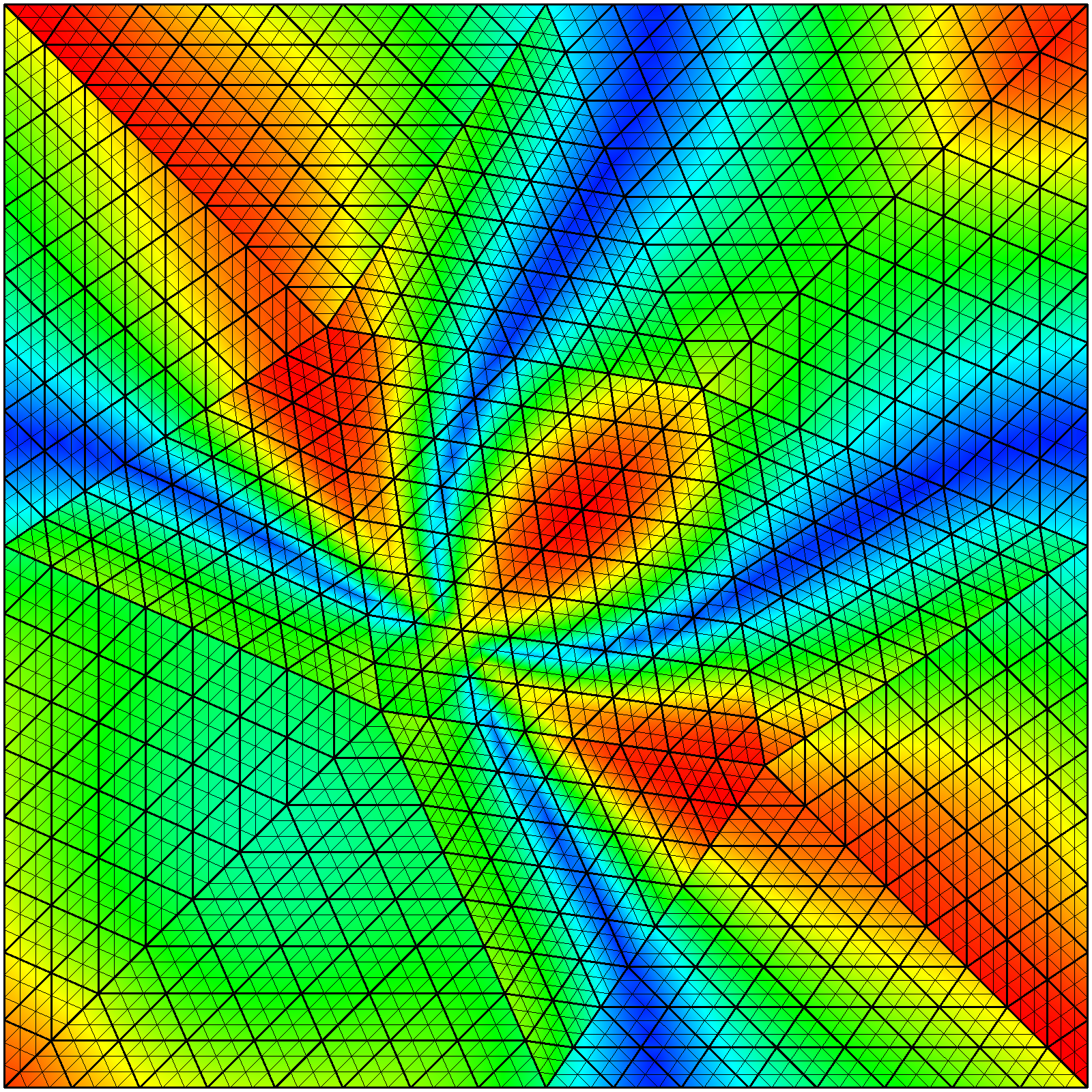}}
			&
			\subfloat[]{\label{fig:p2_3_0}
				\includegraphics[width=0.22\textwidth]{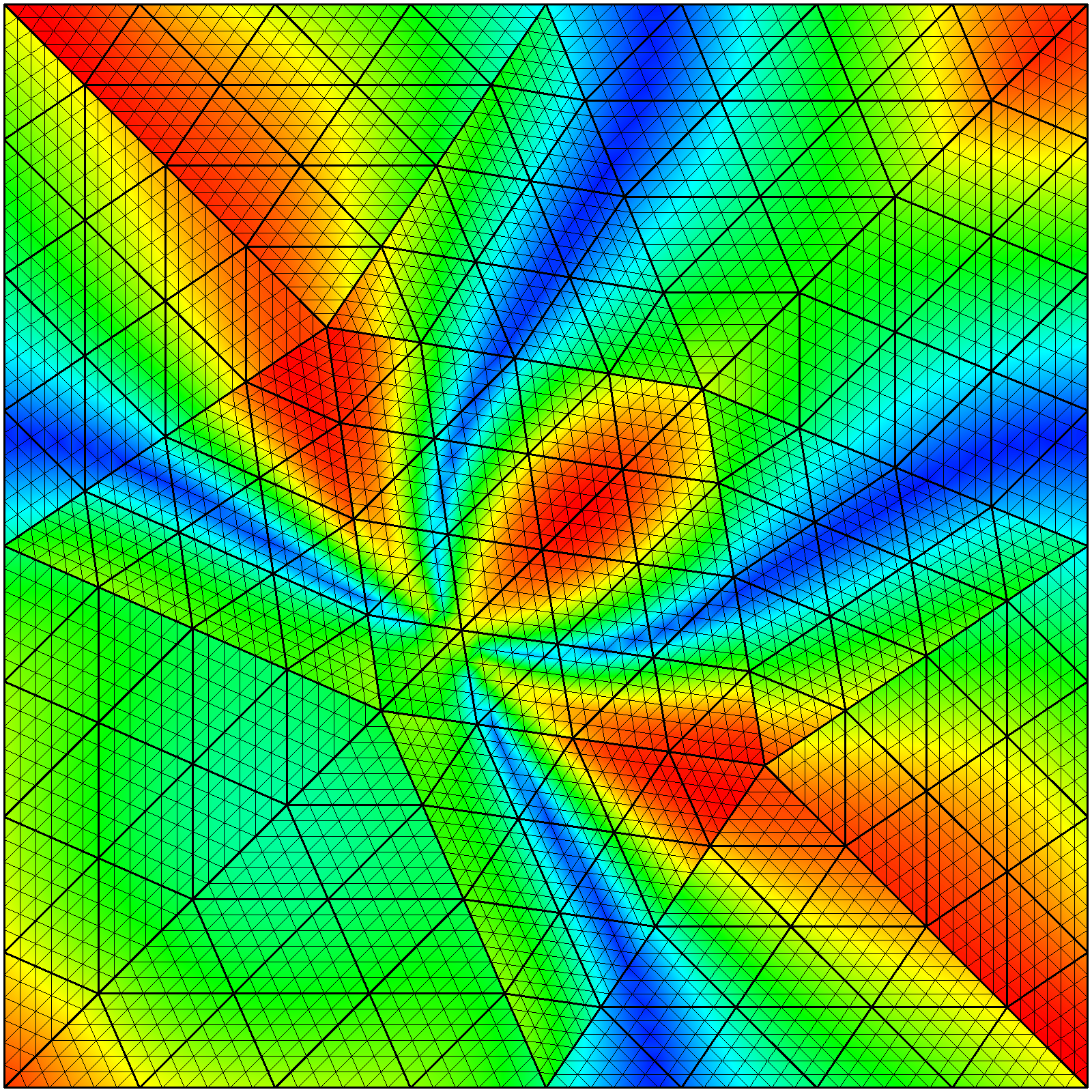}}
			&
			\subfloat[]{\label{fig:p4_3_0}
				\includegraphics[width=0.22\textwidth]{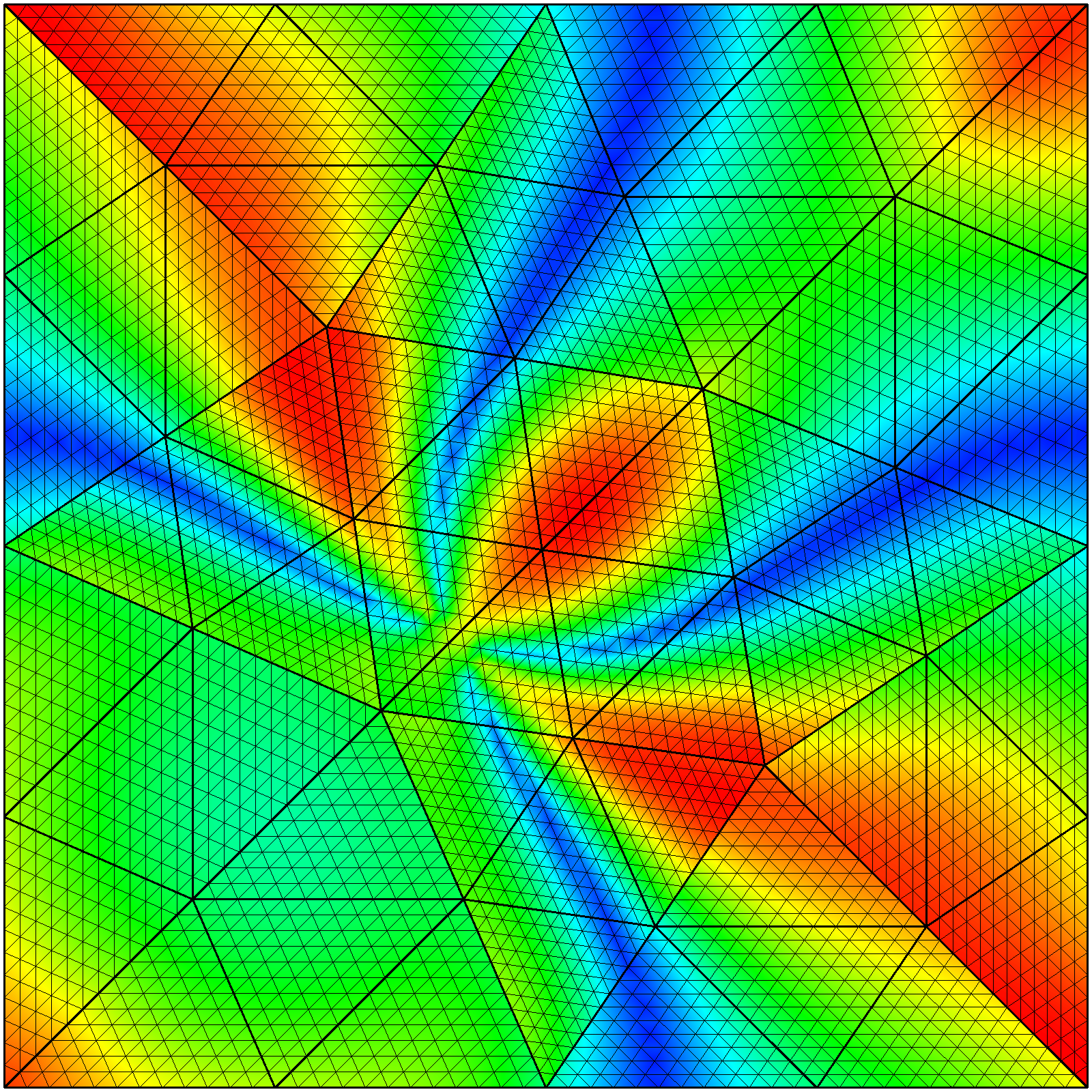}}
			&
			\subfloat[]{\label{fig:p8_3_0}
				\includegraphics[width=0.22\textwidth]{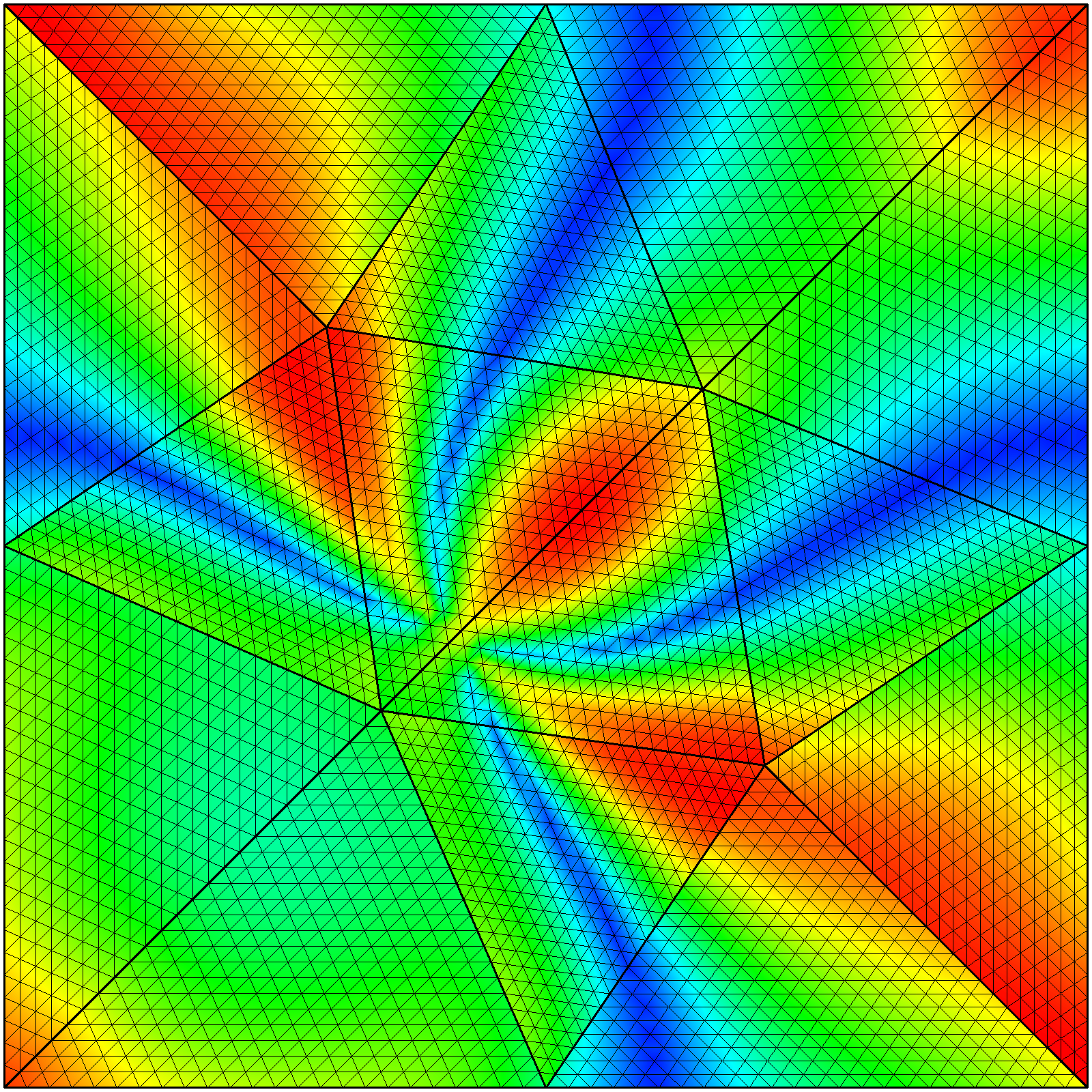}}
			\\
			\subfloat[]{\label{fig:p1_3_1}
				\includegraphics[width=0.22\textwidth]{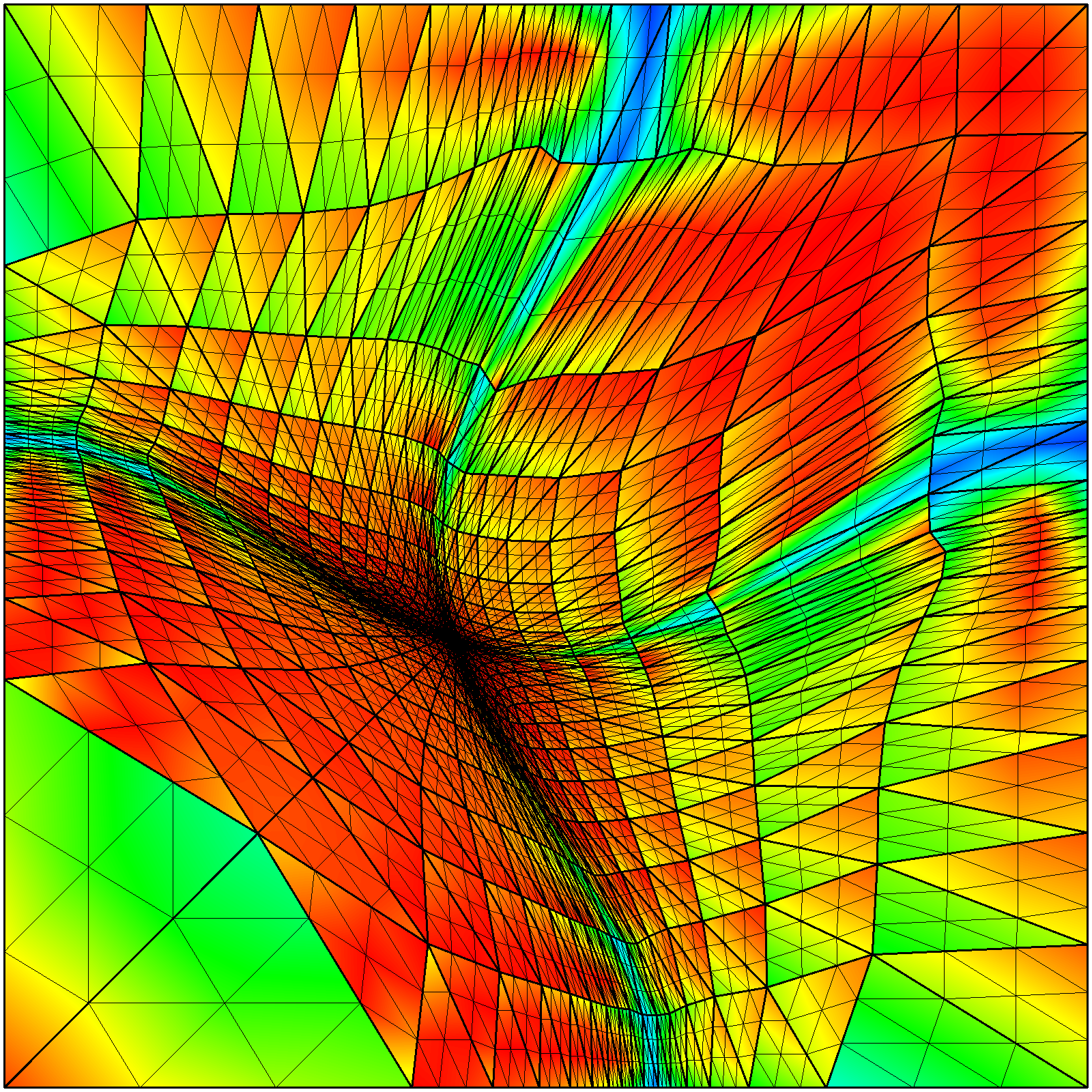}}
			&
			\subfloat[]{\label{fig:p2_3_1}
				\includegraphics[width=0.22\textwidth]{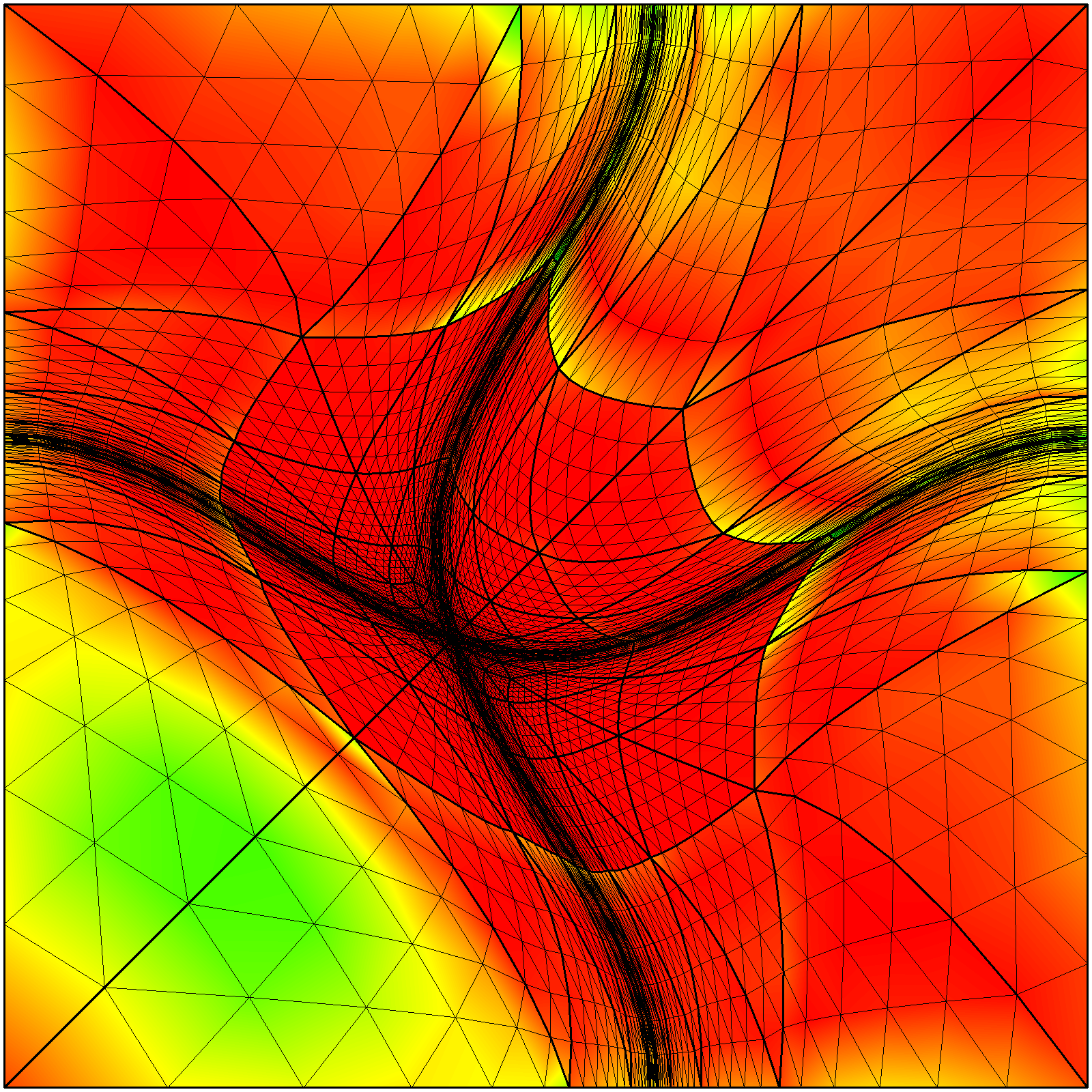}}
			&
			\subfloat[]{\label{fig:p4_3_1}
				\includegraphics[width=0.22\textwidth]{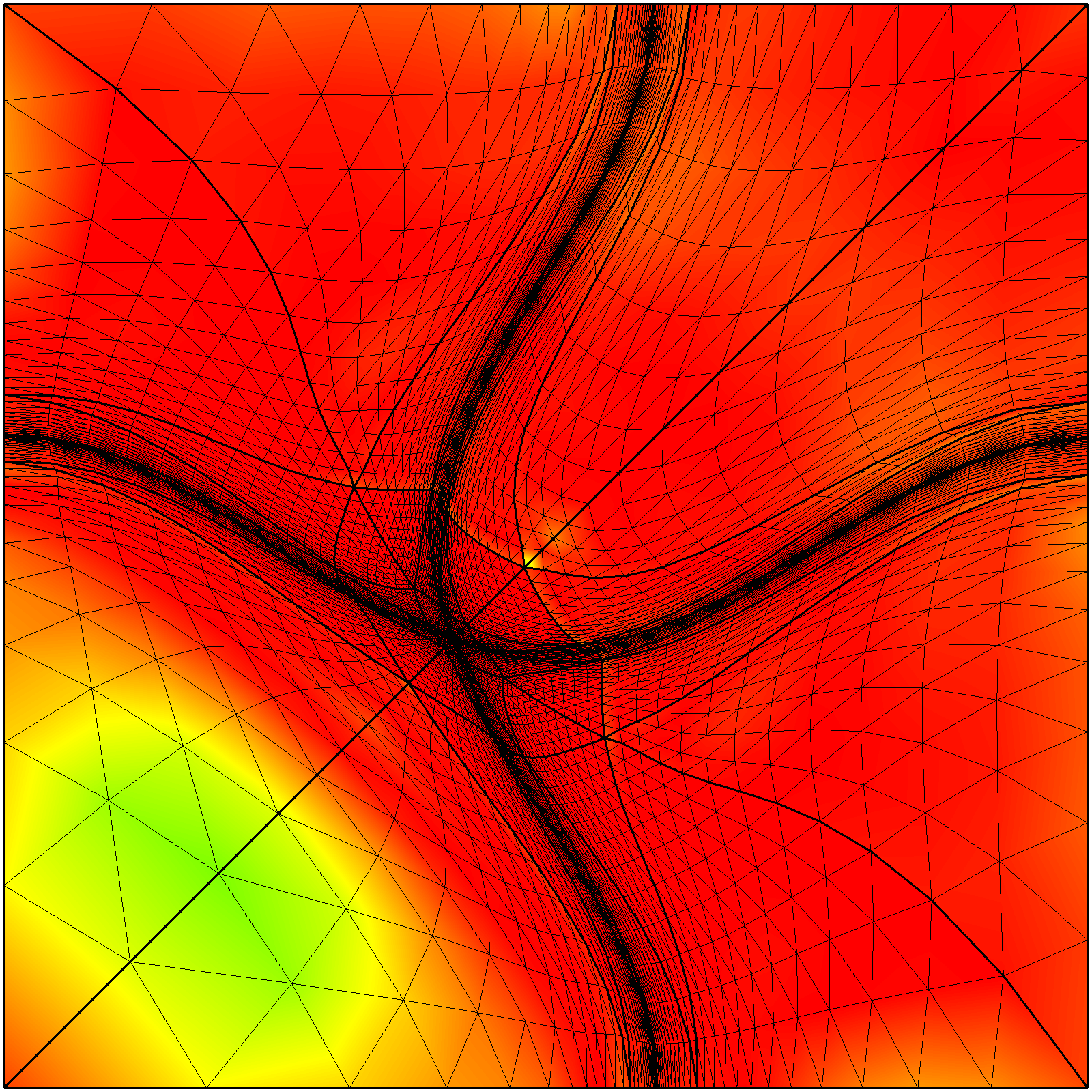}}
			&
			\subfloat[]{\label{fig:p8_3_1}
				\includegraphics[width=0.22\textwidth]{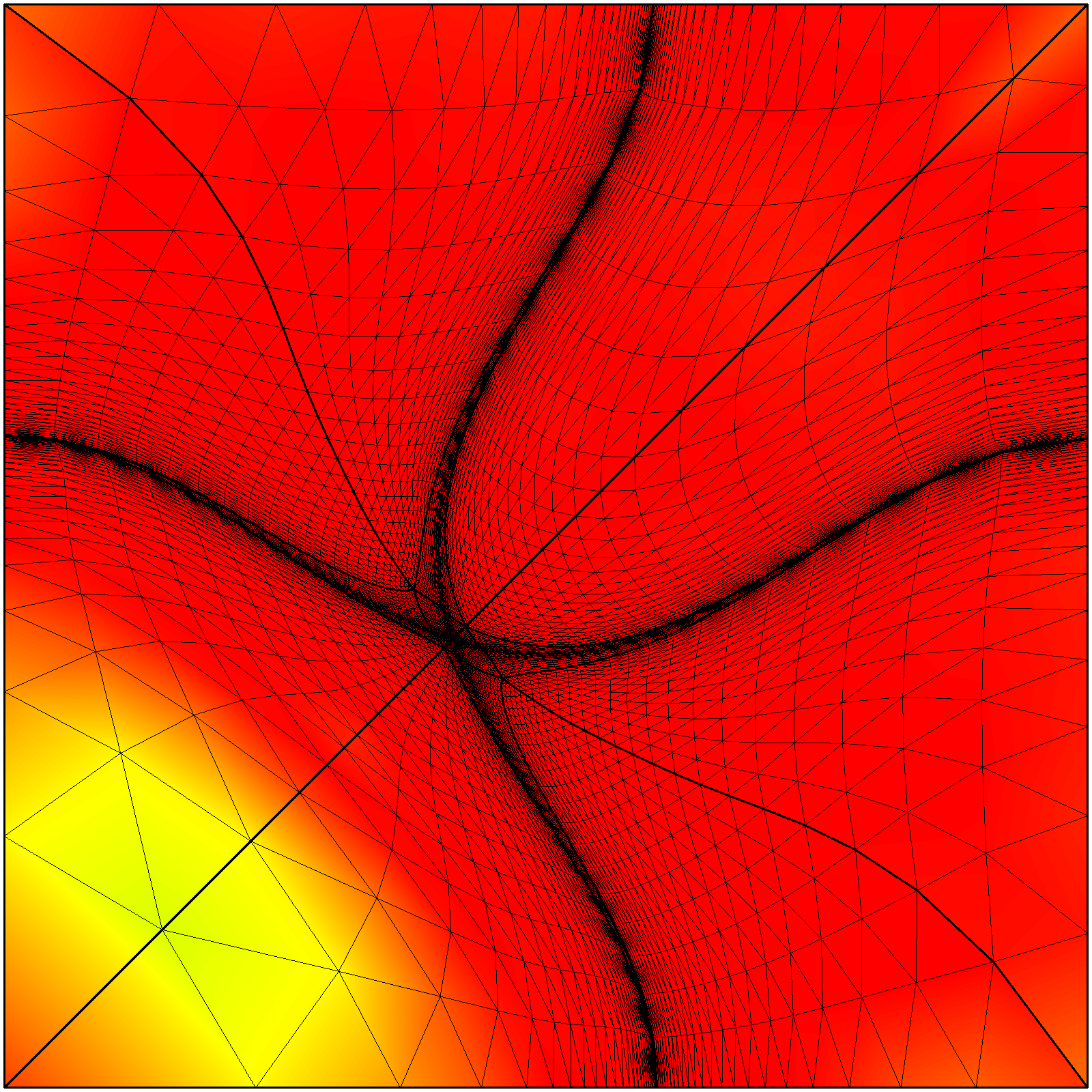}}
			\\
		\end{tabular}
		\\
		\includegraphics[width=0.4\textwidth]{qualBarParaview_color} 
		\caption{Pointwise quality measure for meshes of (columns) polynomial degree 1, 2, 4, and 8 equipped with the (a-h) Curve metric and (i-p) Curves target metric of Table \ref{table:metrics}: (a-d, i-l) initial straight-sided isotropic meshes, and (e-h, m-p) optimized meshes.}
		\label{fig:meshcurves}
	\end{figure}
\begin{figure}[t!]
	\centering
	%		\hspace{0.55cm}
	\begin{tabular}{cccc}
		\subfloat[]{\label{fig:p1_4_0}
			\includegraphics[width=0.3\textwidth]{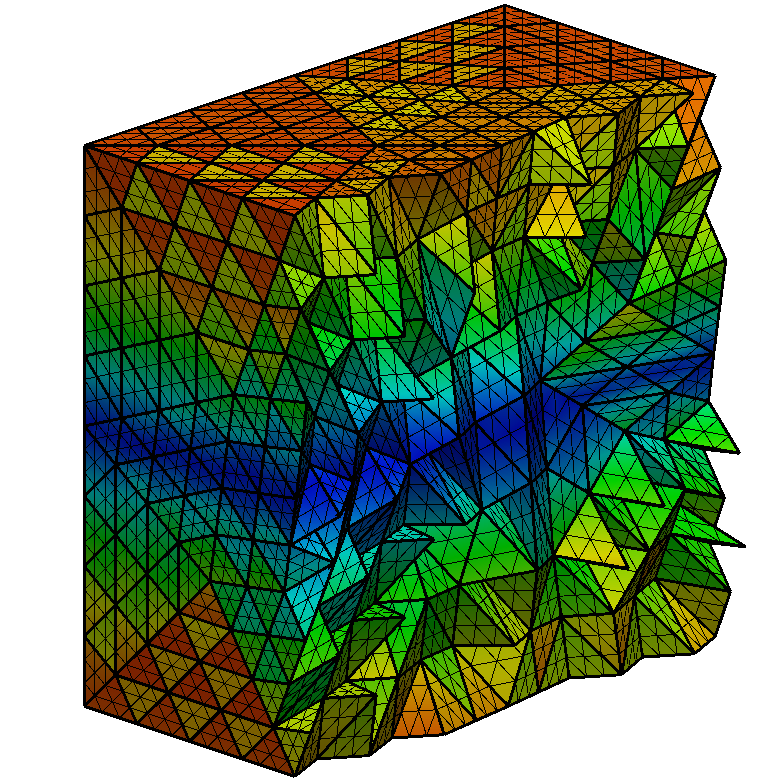}}
		&
		\subfloat[]{\label{fig:p2_4_0}
			\includegraphics[width=0.3\textwidth]{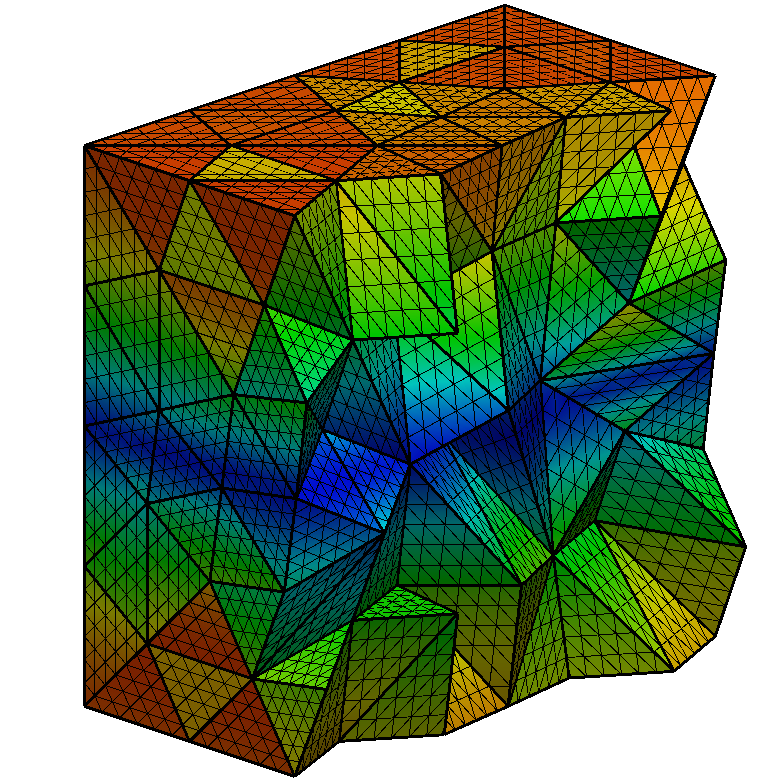}}
		&
		\subfloat[]{\label{fig:p4_4_0}
			\includegraphics[width=0.3\textwidth]{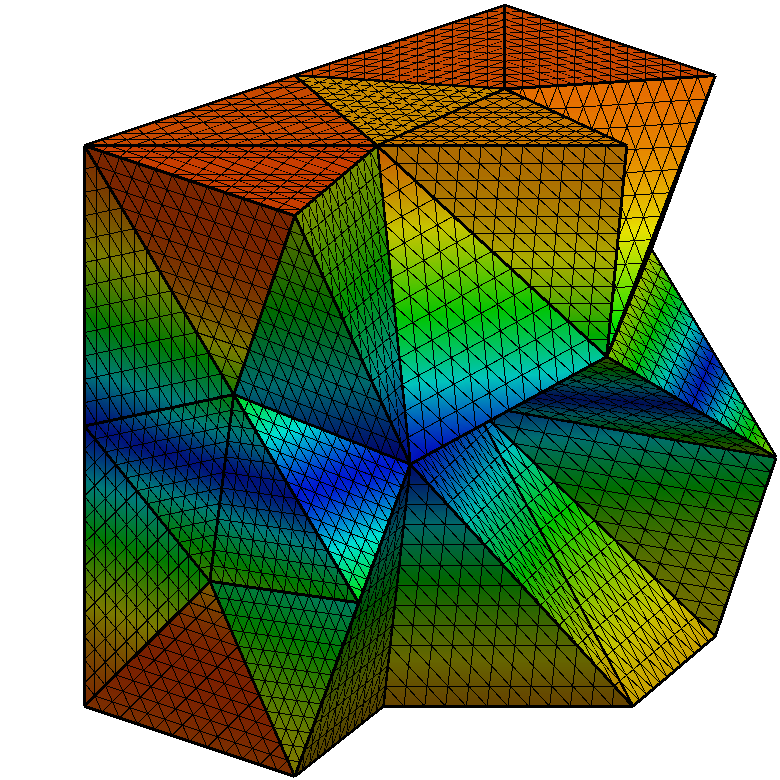}}
		\\
		\subfloat[]{\label{fig:p1_4_1}
			\includegraphics[width=0.3\textwidth]{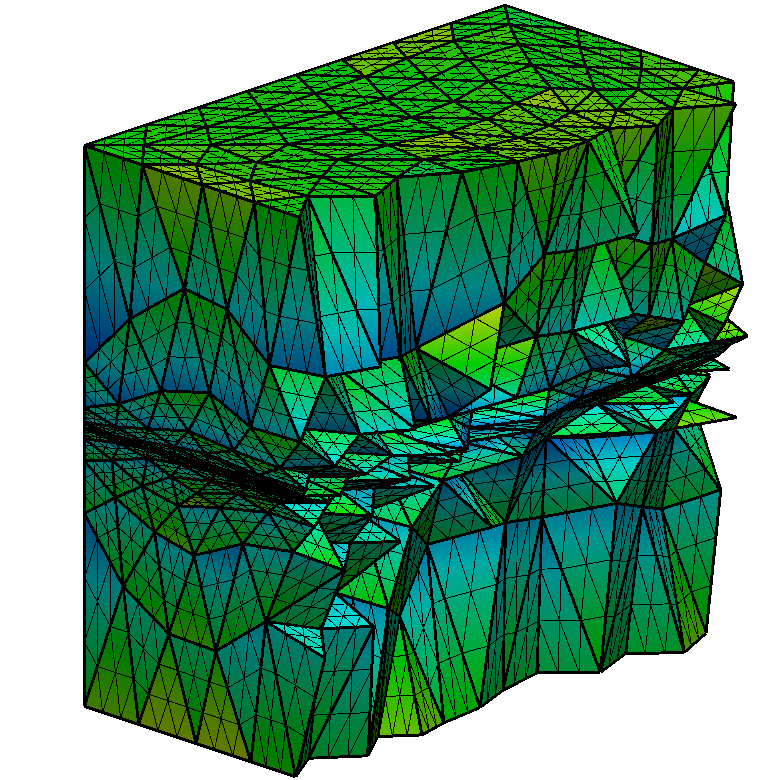}}
		&
		\subfloat[]{\label{fig:p2_4_1}
			\includegraphics[width=0.3\textwidth]{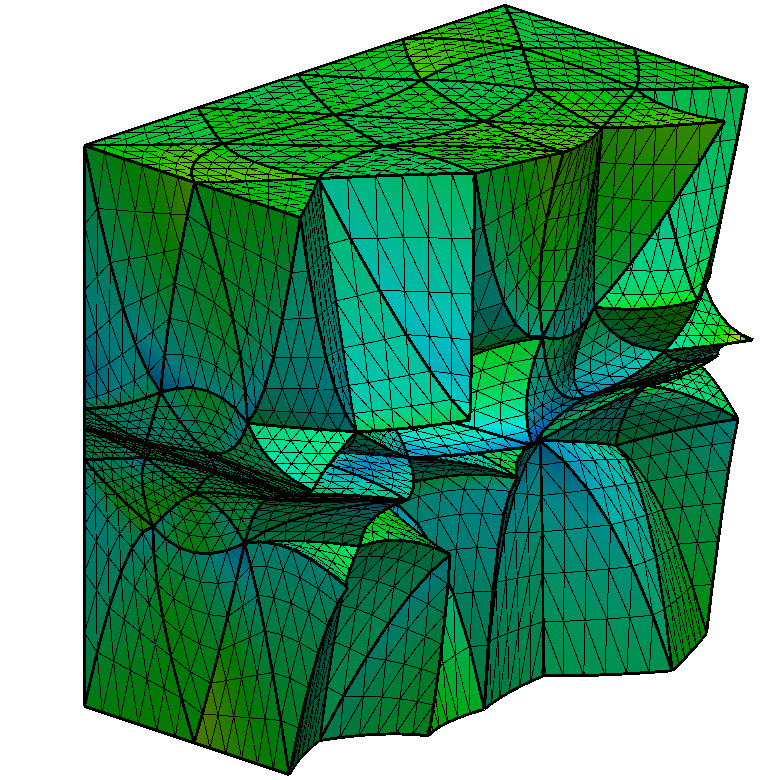}}
		&
		\subfloat[]{\label{fig:p4_4_1}
			\includegraphics[width=0.3\textwidth]{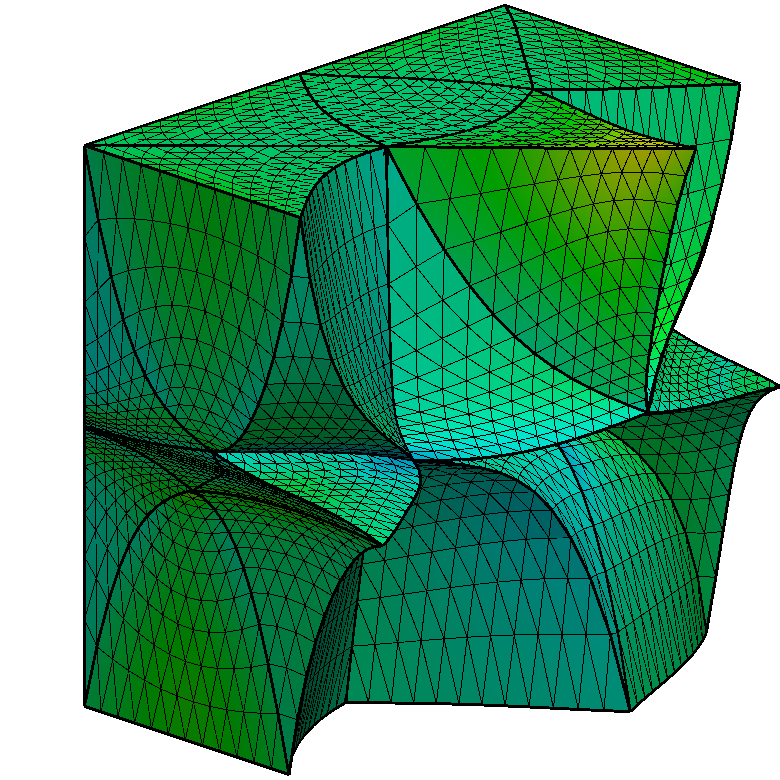}}
	\end{tabular}
	\\
	\includegraphics[width=0.4\textwidth]{qualBarParaview_color} 
	\caption{Pointwise quality measure for meshes of (columns) polynomial degree 1, 2, and 4 equipped with the Plane metric, see Table \ref{table:metrics}. (a-c) initial straight-sided isotropic meshes, and (d-f) optimized meshes.}
	\label{fig:meshsurfaces0}
\end{figure}
\begin{figure}[t!]
	\centering
	%		\hspace{0.55cm}
	\begin{tabular}{cccc}
		\subfloat[]{\label{fig:p1_5_0}
			\includegraphics[width=0.3\textwidth]{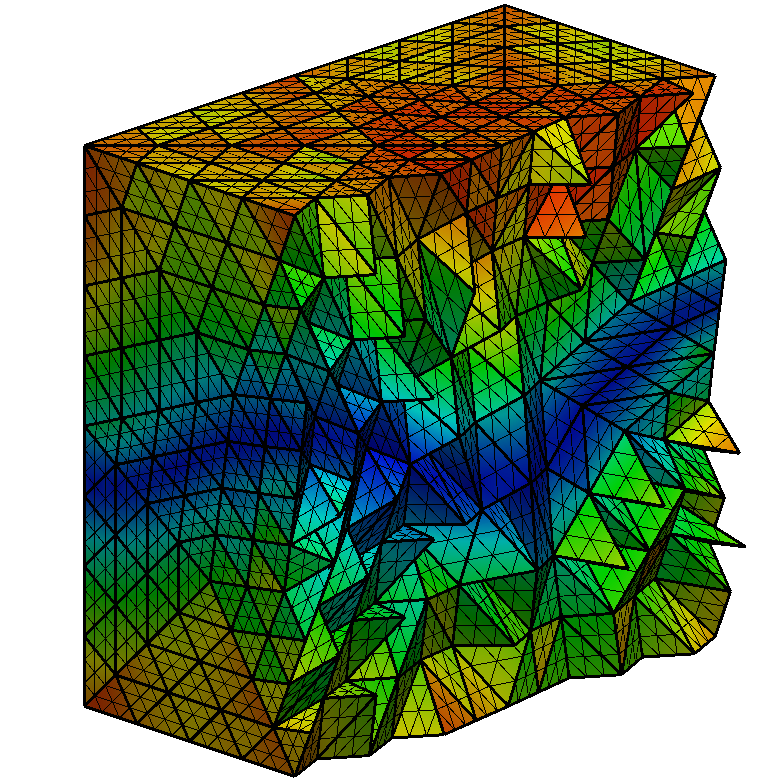}}
		&
		\subfloat[]{\label{fig:p2_5_0}
			\includegraphics[width=0.3\textwidth]{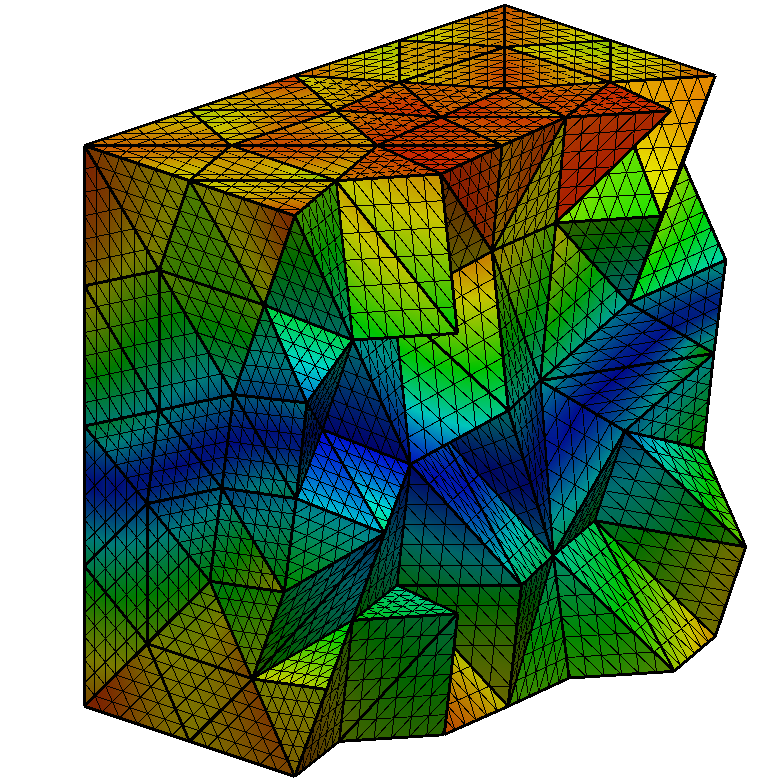}}
		&
		\subfloat[]{\label{fig:p4_5_0}
			\includegraphics[width=0.3\textwidth]{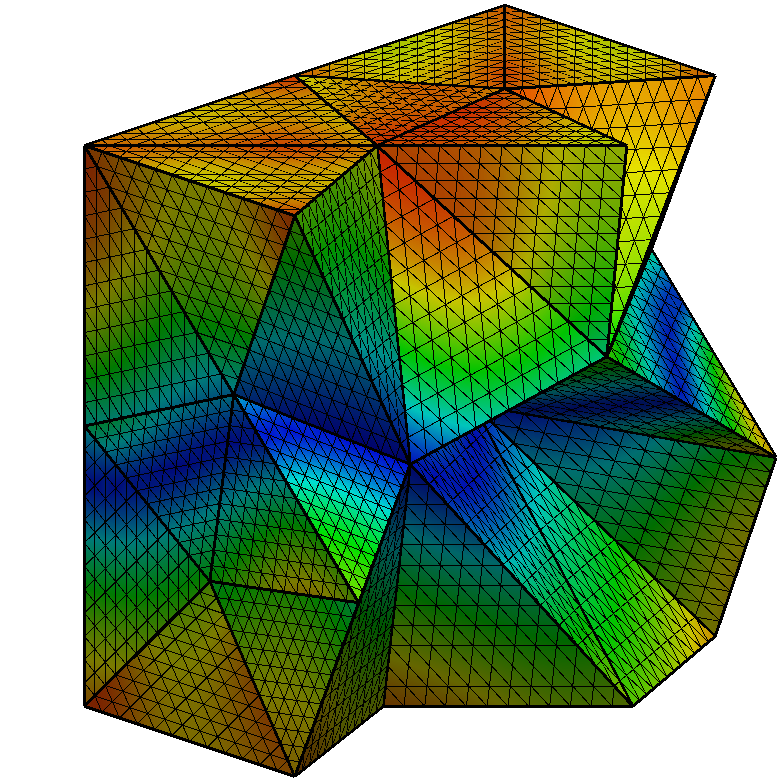}}
		\\
		\subfloat[]{\label{fig:p1_5_1}
			\includegraphics[width=0.3\textwidth]{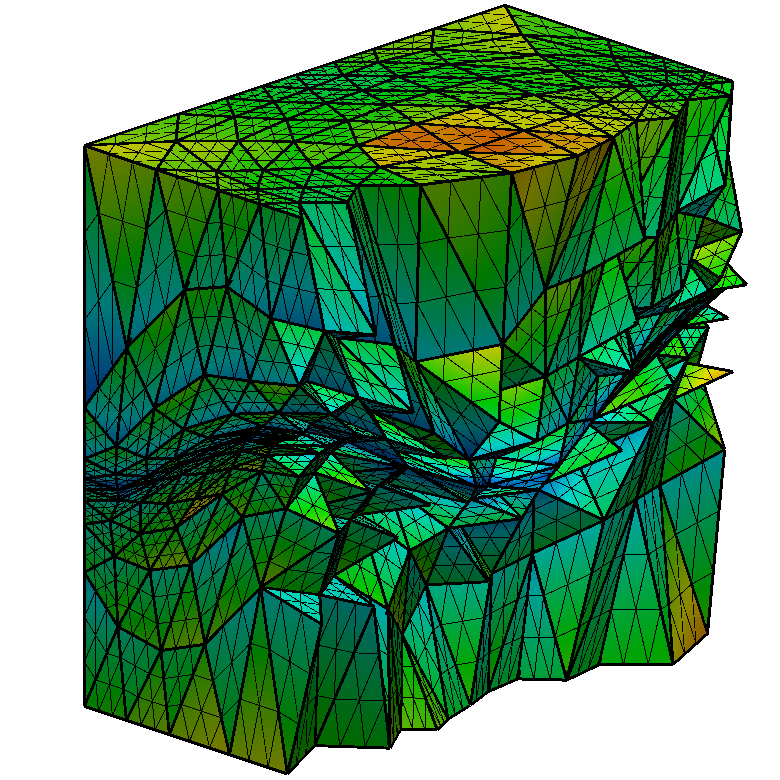}}
		&
		\subfloat[]{\label{fig:p2_5_1}
			\includegraphics[width=0.3\textwidth]{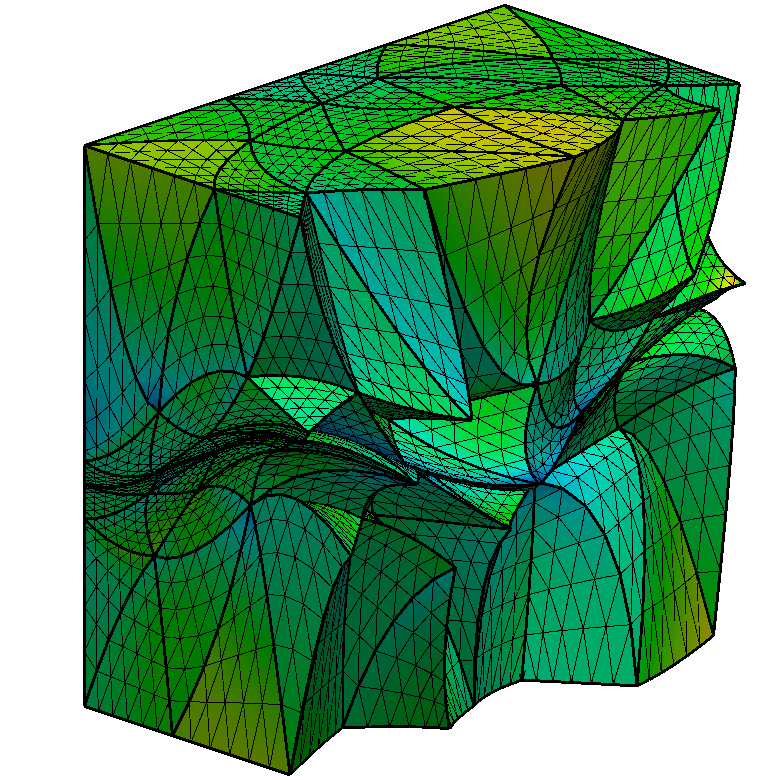}}
		&
		\subfloat[]{\label{fig:p4_5_1}
			\includegraphics[width=0.3\textwidth]{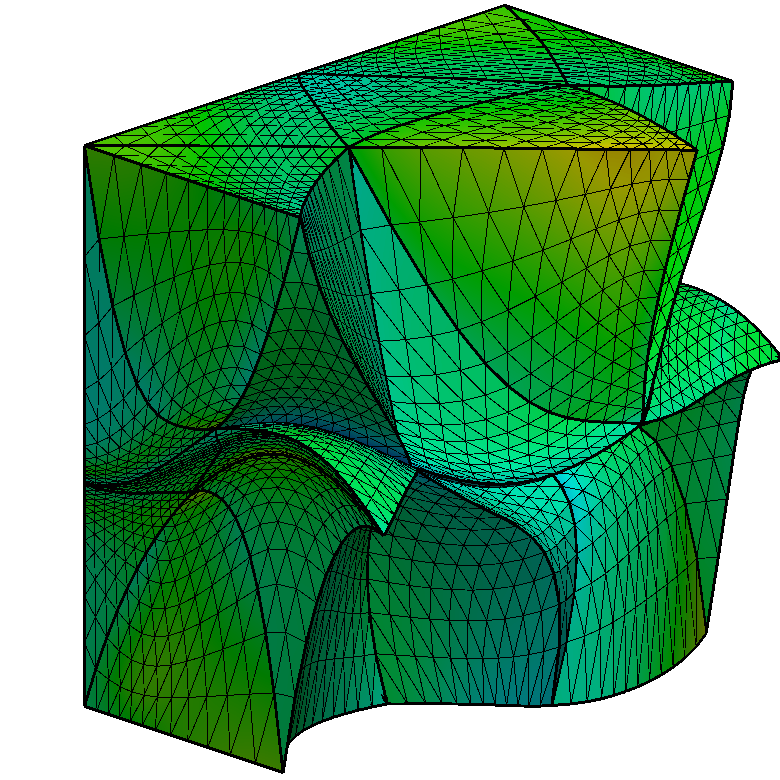}}
	\end{tabular}
	\\
	\includegraphics[width=0.4\textwidth]{qualBarParaview_color} 
	\caption{Pointwise quality measure for meshes of (columns) polynomial degree 1, 2, and 4 equipped with the Surface metric, see Table \ref{table:metrics}. (a-c) initial straight-sided isotropic meshes, and (d-f) optimized meshes.}
	\label{fig:meshsurfaces1}
\end{figure}
\begin{figure}[t!]
	\centering
	%		\hspace{0.55cm}
	\begin{tabular}{cccc}
		\subfloat[]{\label{fig:p1_6_0}
			\includegraphics[width=0.3\textwidth]{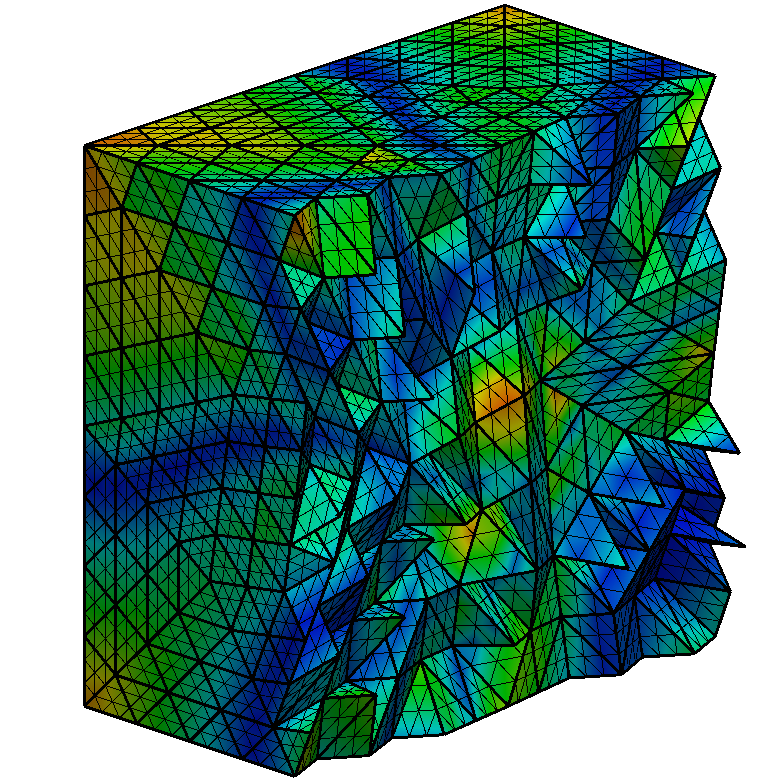}}
		&
		\subfloat[]{\label{fig:p2_6_0}
			\includegraphics[width=0.3\textwidth]{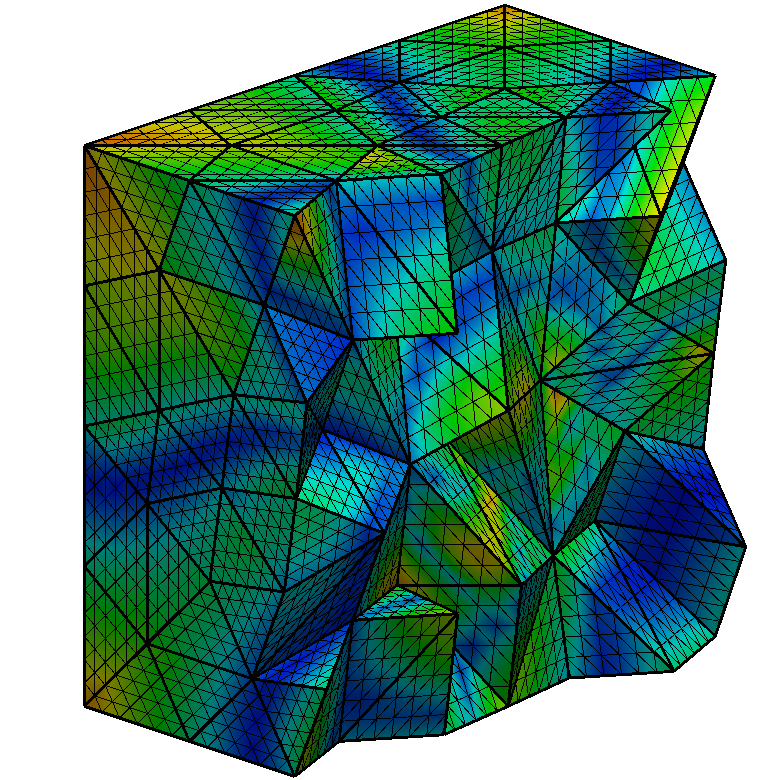}}
		&
		\subfloat[]{\label{fig:p4_6_0}
			\includegraphics[width=0.3\textwidth]{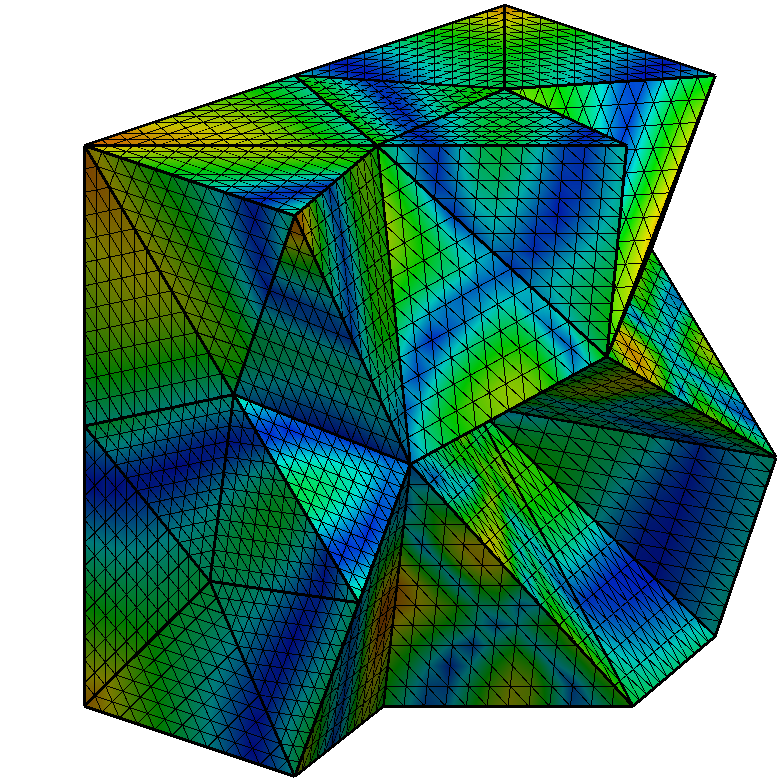}}
		\\
		\subfloat[]{\label{fig:p1_6_1}
			\includegraphics[width=0.3\textwidth]{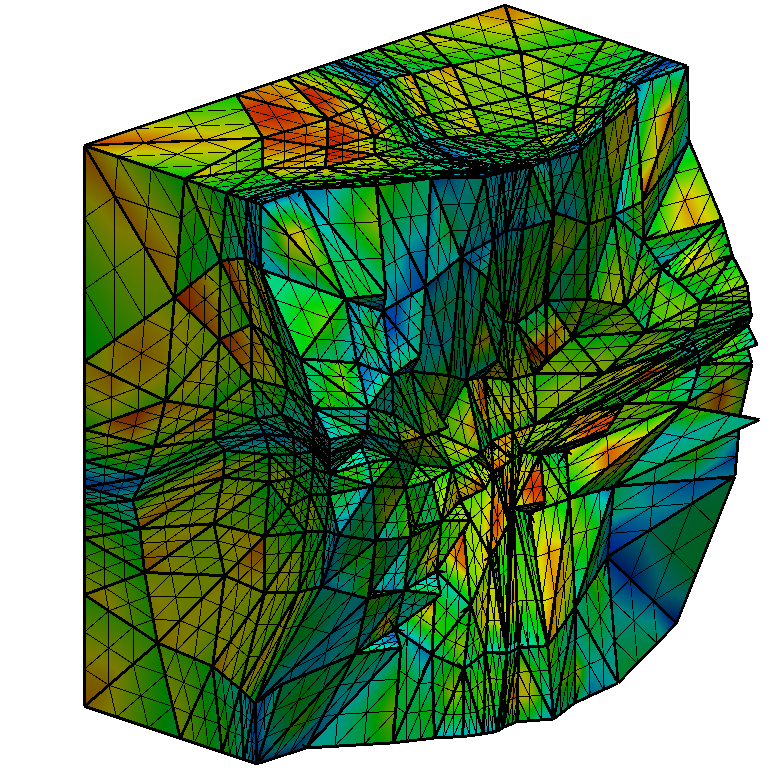}}
		&
		\subfloat[]{\label{fig:p2_6_1}
			\includegraphics[width=0.3\textwidth]{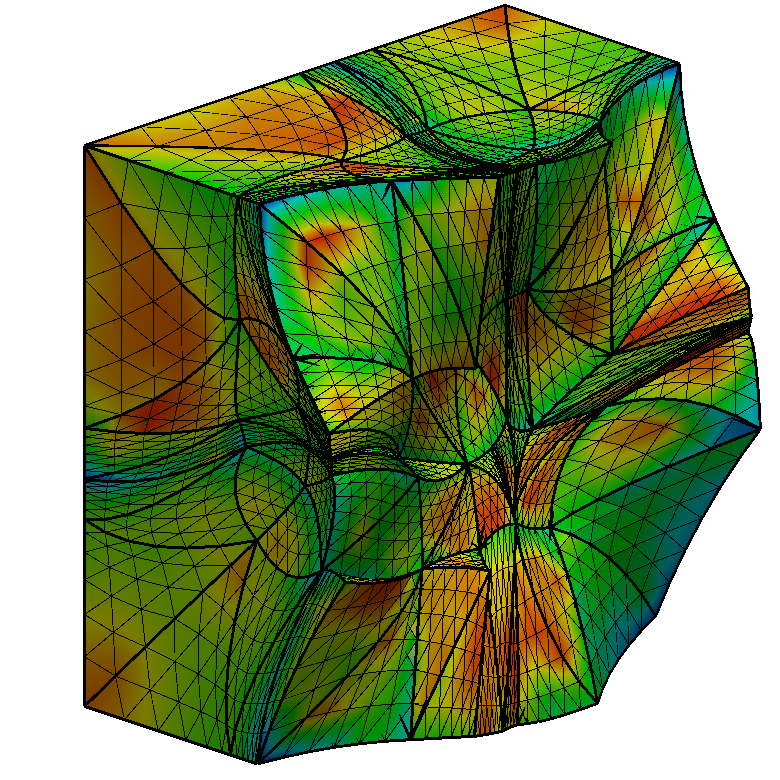}}
		&
		\subfloat[]{\label{fig:p4_6_1}
			\includegraphics[width=0.3\textwidth]{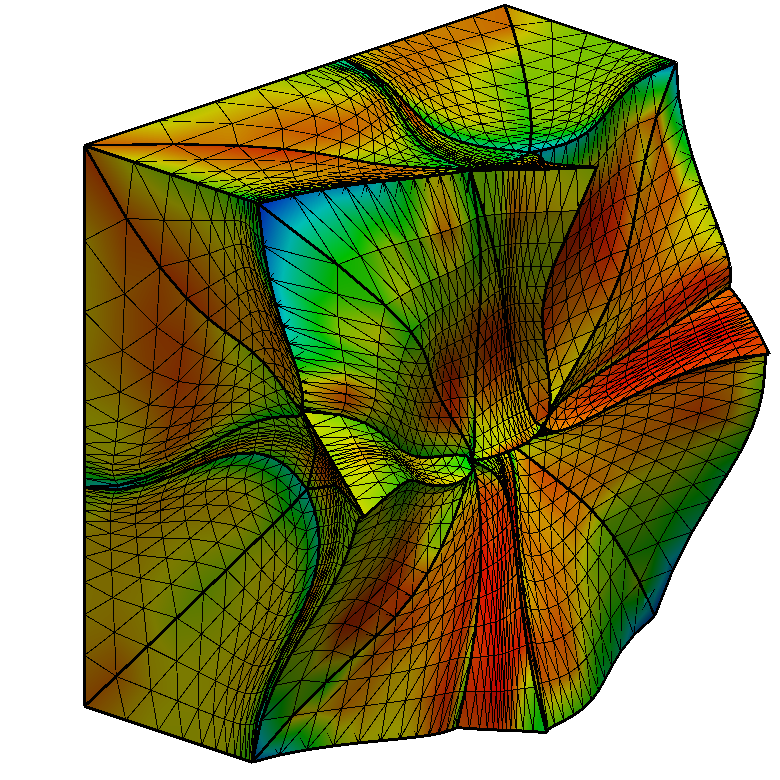}}
		\\
	\end{tabular}
	\\
	\includegraphics[width=0.4\textwidth]{qualBarParaview_color} 
	\caption{Pointwise quality measure for meshes of (columns) polynomial degree 1, 2, and 4 equipped with the Surfaces metric, see Table \ref{table:metrics}. (a-c) initial straight-sided isotropic meshes, and (d-f) optimized meshes.}
	\label{fig:meshsurfaces2}
\end{figure}
	As initial guess of the mesh optimizer we generate isotropic meshes with the MATLAB PDE Toolbox \cite{MATLAB:2017}. The initial isotropic linear unstructured 2D and 3D meshes are presented in Figures \ref{fig:ex}\subref{fig:p8_0} and \ref{fig:meshsurfaces0}\subref{fig:p4_4_0}, respectively. The structured meshes of lower polynomial degree are generated by subdivision.
	
	In 2D, for each considered metric, we generate four meshes of polynomial degree 1, 2, 4, and 8. The meshes feature the same resolution and hence have the same number of nodes, 481 nodes, but a different number of elements, 896, 224, 56, and 14 elements, respectively.The meshes from Figures \ref{fig:ex}, \ref{fig:meshcurves}\subref{fig:p1_2_0}-\ref{fig:meshcurves}\subref{fig:p8_2_1}, and \ref{fig:meshcurves}\subref{fig:p1_3_0}-\ref{fig:meshcurves}\subref{fig:p8_3_1} correspond to the metrics 1, 2, and 3, see Table \ref{table:metrics}.
	
	In 3D, for each considered metric, we consider three meshes of polynomial degree 1, 2, and 4. The meshes feature the same resolution and hence, the same number of nodes, 1577 nodes, but a different number of elements, 7296, 912, and 114 elements, respectively. The meshes from Figures \ref{fig:meshsurfaces0}\subref{fig:p1_4_0}-\ref{fig:meshsurfaces0}\subref{fig:p4_4_1}, \ref{fig:meshsurfaces1}\subref{fig:p1_5_0}-\ref{fig:meshsurfaces1}\subref{fig:p4_5_1}, and \ref{fig:meshsurfaces2}\subref{fig:p1_6_0}-\ref{fig:meshsurfaces2}\subref{fig:p4_6_1} correspond to the metrics 4, 5, and 6, see Table \ref{table:metrics}.
	
	The meshes are colored according to the pointwise stretching and alignment quality measure, proposed in \cite{aparicio2018defining} and detailed in Equation \eqref{eq:point-wise} of Section \ref{sec:minimizationProblem}. As we observe, the elements lying in the region of highest stretching ratio have less quality than the elements lying in the isotropic region.
	
	%by changing the coordinates of all the mesh nodes and
	To obtain an optimal configuration, we minimize the distortion measure by relocating the mesh nodes while preserving their connectivity, see Section \ref{sec:minimizationProblem}. The coordinates of the inner nodes, and the coordinates tangent to the boundary, are the design variables. Thus, the inner nodes are free to move, the vertex nodes are fixed, while the rest of boundary nodes are enforced to slide along the boundary facets of the domain $\Omega$. The total amount of degrees of freedom for the 2D and 3D meshes is 894 and 3957, respectively. The optimized meshes are illustrated in Figures \ref{fig:ex} and \ref{fig:meshcurves} for the 2D cases and in Figures \ref{fig:meshsurfaces0}, \ref{fig:meshsurfaces1}, and \ref{fig:meshsurfaces2} for the 3D cases. We observe that the elements away from the anisotropic region are enlarged vertically whereas the elements lying in the anisotropic region are compressed. Moreover, the minimum quality is improved, and the standard deviation of the element qualities is reduced.
	\subsection{Line-search globalization: standard versus specific-purpose}\label{sec:resultsglob}	
	Following, we compare the line-search globalization strategies, presented in Section \ref{sec:updating}, and their effect in the non-linear optimization method. For this, we apply the Newton method presented in Section \ref{sec:optimizationOverview}, with the corresponding globalization strategy and linear solver, to the first test metric presented in Table \ref{table:metrics}.
	
	The standard and the specific-purpose globalization strategies are presented in Sections \ref{sec:standard} and \ref{sec:final}, respectively. To compare them we use an exact Newton method. Specifically, we consider the optimization method presented in Algorithm \ref{alg:optim1} with a globalization strategy, Line \ref{lst:line:glob1}, and a direct solver, Line \ref{lst:line:newtondirection1}. In particular, the direct solver computes, Line \ref{lst:line:directsolver0} of Algorithm \ref{alg:Newton0}, the exact approximation of the Newton equation presented in Equation \eqref{eq:Newton} using the complete sparse $\text{LDL}^\text{T}$ preconditioner of the CHOLMOD package \cite{chen2008algorithm}.
	\begin{table}[t!]
		\caption{Non-linear and line-search iterations for the exact Newton method composed of standard and specific-purpose LS. Both globalizations are coupled with the direct solver.}
		\label{table:final1}
		\centering
		
		\par\medskip
		
		\begin{tabular}{ c r r r r }
			
			\hline\noalign{\smallskip}
			Mesh & \multicolumn{2}{c}{Non-linear iterations}& \multicolumn{2}{c}{Line-search iterations}\\
			degree & Standard & Specific-purpose& Standard & Specific-purpose\\
			\noalign{\smallskip}\hline\noalign{\smallskip}
			
			1 & 54 & 37 & 159 & 31 \\		
			2 & 82 & 91 & 328 & 155\\
			4 & 88 & 78 & 304 & 92\\
			8 & 203 & 125 & 771 & 171\\
			
			\noalign{\smallskip}\hline\noalign{\smallskip}
		\end{tabular}
	\end{table}	
	
	%Header: \textbf{Comments results}\\
	The results of our numerical experiments allow comparing the standard, Algorithm \ref{alg:LSinitial}, and specific-purpose, Algorithm \ref{alg:LSfinal}, globalization strategies in terms of the required line-search iterations, see Table \ref{table:final1}. For meshes of polynomial degree 1, 2, 4, and 8, we report the number of non-linear and line-search iterations required to optimize the model case. We report these numbers for the exact Newton method equipped with the standard and specific-purpose globalizations.
	
	We conclude that the specific-purpose strategy improves the standard one. The results show that the number of line-search iterations is reduced. Meanwhile, the number of non-linear iterations remain in the same order of magnitude, yet tending to be smaller. We can explain these improvements by highlighting two factors. First, the specific-purpose strategy can enlarge the step length with line-search iterations, a larger advance that promotes to reduce the number of non-linear iterations. Second, for the specific-purpose strategy, by reusing the last step length we promote to reduce the total amount of line-search iterations. In contrast, for the standard line-search strategy each direction has step length at most one, limiting the length of the step.
	
%	We expect a similar behavior for the other metrics of Table \ref{table:metrics}. Specifically, we expect a higher number of required non-linear and line-search iterations for metrics with curved alignment (Curve, Curves, Surface, and Surfaces). In addition, when compared to the standard optimization solver, we expect a higher decrease of the non-linear and line-search iterations in the specific-purpose optimization method. This is because a higher complexity of the metric features stiffens the validity of mesh deformations. This makes the convergence more difficult, especially for standard line-search strategies that only reduce the step-length.
%	designed for initial guests near a minimizer.
	\subsection{Inexact Newton method: standard versus specific-purpose}\label{sec:resultsinexactNewton}
	Next, we compare the inexact Newton methods presented in Section \ref{sec:updatinglinearsystem}. Specifically, we compare the influence of the forcing sequences and of the preconditioner in the non-linear optimization method. For this, we equip the meshes with the first metric presented in Table \ref{table:metrics}. Moreover, we globalize the non-linear solver, Section \ref{sec:optimizationOverview}, with the specific-purpose LS strategy, Section \ref{sec:updating}. Finally, we optimize the meshes using the different approaches to compute the inexact Newton direction.
	
	The standard and the specific-purpose inexact Newton methods are presented in Sections \ref{sec:inexactNewton0} and \ref{sec:inexactNewton}, respectively. We compare them in two steps. In both cases, we compare the standard inexact Newton method, that uses the standard forcing terms, Section \ref{sec:forcingseq0}, with the specific-purpose inexact Newton method, that uses the specific-purpose forcing terms, Section \ref{sec:forcingseq}. In the first case, we use the Jacobi preconditioner presented in Section \ref{sec:preconditioner0}, see Line \ref{lst:line:setprecon0} of Algorithms \ref{alg:Newton0} and \ref{alg:Newton1}. In the second case, we use the Jacobi/$\text{iLDL}^\text{T}(0)$ preconditioner switch presented in Section \ref{sec:preconditioner}, see Line \ref{lst:line:setprecon1} of Algorithms \ref{alg:Newton0} and \ref{alg:Newton2}. Note that the reordering and the curvature inequality checks are also active for the $\text{iLDL}^\text{T}(0)$. In Algorithm \ref{alg:ildl}, the reordering of the degrees of freedom is explicitly provided by the permutation $\sigma$. In addition, the curvature inequality check is explicitly used in Line \ref{lst:line:illconditionedcond} in Algorithm \ref{alg:Newton2}.
%	Note that the $\text{iLDL}^\text{T}(0)$ preconditioner is computed as in Algorithm \ref{alg:ildl}, where the role of the reordering of the degrees of freedom is encapsulated in the permutation $\sigma$. In addition, we account for the curvature inequality check in Line \ref{lst:line:illconditionedcond} in Algorithm \ref{alg:Newton2}.
	
	\begin{table*}[t!]
		\caption{Non-linear iterations and matrix-vector products for the inexact Newton methods with Jacobi preconditioner and Jacobi/$\text{iLDL}^\text{T}(0)$ preconditioners with specific-purpose LS globalization. Inexact Newton methods are distinguished by standard and specific-purpose forcing terms.}
		\label{table:forcingterms}
		\centering
		
		\par\medskip
		\begin{tabular}{ c r r r r l}
			\hline\noalign{\smallskip}
			Mesh & \multicolumn{2}{c}{Non-linear iterations}&\multicolumn{2}{c}{Matrix-vector products}&\\
			degree & Standard & Specific-& Standard & Specific-&Preconditioner\\
			&&purpose&&purpose&\\ 
			\noalign{\smallskip}\hline\noalign{\smallskip}
			
			1 & 24 & 20 & 1856 & 677&Jacobi \\
			& 24 & 17 & 333 & 113&Jacobi/$\text{iLDL}^\text{T}(0)$\\
			\noalign{\smallskip}\hline\noalign{\smallskip}
			2 & 29 & 29 & 1824 & 624&Jacobi \\
			& 30 & 23 & 579 & 155&Jacobi/$\text{iLDL}^\text{T}(0)$\\
			\noalign{\smallskip}\hline\noalign{\smallskip}
			4 & 48 & 40 & 4858 & 1521&Jacobi \\
			& 48 & 31 & 1162 & 223&Jacobi/$\text{iLDL}^\text{T}(0)$\\
			\noalign{\smallskip}\hline\noalign{\smallskip}
			8 & 109 & 73 & 10031 & 2864&Jacobi \\
			& 109 & 83 & 3538 & 1452&Jacobi/$\text{iLDL}^\text{T}(0)$\\
			\noalign{\smallskip}\hline\noalign{\smallskip}
		\end{tabular}
	\end{table*}
	
	The results of our numerical experiments allow comparing between the standard, Line \ref{lst:line:iterative0} of Algorithm \ref{alg:Newton0}, and the specific-purpose, Line \ref{lst:line:iterative2} of Algorithm \ref{alg:Newton2}, inexact Newton methods in terms of the number of required matrix-vector products, see Table \ref{table:forcingterms}. The model case is optimized with the specific-purpose LS strategy for meshes of polynomial degree 1, 2, 4, and 8. For these meshes, we report the number of non-linear iterations and matrix-vector products required by the standard and specific-purpose inexact Newton methods.
	
	We conclude that the specific-purpose preconditioned inexact Newton method significantly improves the standard one. Specifically, the matrix-vector products are reduced by one order of magnitude. On the one hand, the specific-purpose forcing terms feature a total number of matrix-vector products smaller than with the standard one. The number of products is reduced because the forcing terms stop the linear iterations when sufficient accuracy and positive curvature is reached. Accordingly, these conditions ensure that the reduction does not hamper the quality of the inexact Newton direction. Meanwhile, the number of non-linear iterations remain in the same order of magnitude, yet being smaller or equal. On the other hand, the specific-purpose pre-conditioner features a total number of matrix-vector products smaller than with the standard one. This number of products is reduced because the specific-purpose pre-conditioner automatically switches to a more accurate $\text{iLDL}^\text{T}(0)$ decomposition. Specifically, it only improves the accuracy when the Hessian is predicted to be numerically positive. Thus, the solver obtains a highly accurate Newton direction with fewer matrix-vector products. Meanwhile, the number of non-linear iterations remains almost unchanged. Finally, the combination of specific-purpose forcing terms and pre-conditioner features a number of matrix-vector products one order of magnitude smaller than for the standard one. This reduction is because we combine the advantages of the specific-purpose forcing terms and the specific-purpose pre-conditioner. Moreover, the number of non-linear iterations is reduced. In addition, we observe that when augmenting the mesh polynomial degree, the total amount of matrix-vector products and the number of non-linear iterations are increased. This can be explained by highlighting that when the mesh polynomial degree is increased the Hessian $\textbf{H}f$ becomes more ill-conditioned. Hence, the CG method needs more iterations to converge.
	
%	We expect a similar behavior for the other metrics of Table \ref{table:metrics}. Specifically, we expect a higher amount of required matrix-vector products for metrics with curved alignment (Curve, Curves, Surface, and Surfaces). In addition, when compared to the standard optimization solver, we expect a higher decrease of the matrix-vector products in the specific-purpose optimization method. This is because a higher complexity of the metric features stiffens the linear systems of the Newton equation. This demands more matrix-vector products in the CG method, especially for standard forcing sequences and preconditioners that do not account for the accuracy level of the numerical approximation of the Newton direction.
	\subsection{Newton-CG solver: standard versus specific-purpose}\label{sec:optimizationresults}
	%Header: \textbf{Section}\\
	In what follows, we compare both optimization solvers: standard and specific-purpose. To this end, we consider the $r$-adaption problem for the domains, metrics, and meshes presented in Section \ref{sec:domainsmetrics}. Finally, we present the results obtained from the optimization processes.
		\begin{table}[p!]
			\caption{Non-linear and line-search iterations, and matrix-vector products for standard (std.) specific-purpose (spec. purp.) optimization methods.}
			
			\centering
%			\tiny
			\par\medskip
			\begin{tabular}{ c r r r r r r r r}	
				\hline\noalign{\smallskip}
				\multicolumn{2}{c}{Example}& \multicolumn{4}{c}{Iterations} & \multicolumn{3}{c}{Cost indicator}\\ Metric &Mesh&\multicolumn{2}{c}{Non-linear}&\multicolumn{2}{c}{Line-search}& \multicolumn{3}{c}{Matrix-vector products}\\ &deg.&Std.&Spec.&Std.&Spec.&Std.&Spec.&Speedup\\
				&&    &purp.  &    &purp.  &		&purp.&\\
					\noalign{\smallskip}\hline\noalign{\smallskip}
					& 1 & 24 & 17 & 48 & 29 & 1646 & 113 & 14.57 \\
					Line & 2 & 32 & 23 & 81 & 33 & 2309 & 155 & 14.90 \\
					& 4 & 47 & 31 & 197 & 53 & 4284 & 223 & 19.21\\
					& 8 & 74 & 83 & 358 & 196 & 14710 & 1452 & 10.13\\
					\noalign{\smallskip}\hline\noalign{\smallskip}
					& 1 & 30 & 23 & 85 & 29 & 6689 & 145 & 46.13\\
					Curve & 2 & 91 & 37 & 575 & 74 & 29626 & 339 & 87.39\\
					& 4 & 102 & 54 & 582 & 85 & 24497 & 597 & 41.03\\
					& 8 & 374 & 78 & 1721 & 190 & 30725 & 2150 & 14.29 \\
					\noalign{\smallskip}\hline\noalign{\smallskip}
					& 1 & 43 & 28 & 174 & 38 & 4196 & 177 & 23.71\\
					Curves & 2 & 211 & 57 & 1473 & 157 & 35378 & 835 & 42.37\\
					& 4 & 140 & 81 & 768 & 184 & 20535 & 1279 & 16.06\\
					& 8 & 858 & 102 & 3169 & 295 & 43317 & 4056 & 10.68 \\
					\noalign{\smallskip}\hline\noalign{\smallskip}
					& 1 & 45 & 30 & 80 & 27 & 3943 & 378 & 10.43\\
					Plane & 2 & 219 & 89 & 1298 & 91 & 10118 & 1194 & 08.47\\
					& 4 & 229 & 167 & 1534 & 267 & 22437 & 2985 & 07.52\\
					\noalign{\smallskip}\hline\noalign{\smallskip}
					& 1 & 69 & 51 & 87 & 117 & 16859 & 635 & 26.55\\
					Surface & 2 & 287 & 131 & 1968 & 140 & 44547 & 2641 & 16.87\\
					& 4 & 290 & 118 & 2040 & 165 & 67997 & 1980 & 34.34\\
					\noalign{\smallskip}\hline\noalign{\smallskip}
					& 1 & 252 & 58 & 2322 & 64 & 51698 & 554 & 93.32\\
					Surfaces & 2 & 361 & 152 & 2646 & 278 & 72078 & 2250 & 32.03\\
					& 4 & 288 & 164 & 2203 & 272 & 73593 & 2897 & 25.40\\
					\noalign{\smallskip}\hline\noalign{\smallskip}
				\end{tabular}
			\label{table:all}
		\end{table}	
	
	Each optimization solver is composed of a globalization and a linear solver, see Algorithm \ref{alg:optim1}. On the one hand, the standard optimization method is composed of the standard BLS, see Section \ref{sec:standard}, and the standard linear solver with the Jacobi preconditioner, see Section \ref{sec:inexactNewton0}. On the other hand, the specific-purpose optimization method is composed of the specific-purpose LS, see Section \ref{sec:final}, and the specific-purpose linear solver with the Jacobi/$\text{iLDL}^\text{T}(0)$ preconditioner switch, see Section \ref{sec:inexactNewton}.
	
	The results of our numerical experiments allow comparing the specific-purpose optimization method with the standard one, see Table \ref{table:all}. The non-linear and line-search iterations, and the matrix-vector products are reported for 2D meshes of polynomial degree 1, 2, 4, and 8 and for 3D meshes of polynomial degree 1, 2, and 4.
	
%	The results of our numerical experiments allow comparing the specific-purpose optimization method with the standard one, see Table \ref{table:all}. The non-linear and line-search iterations, the matrix-vector products, and the specific-purpose solver speedup are reported for 2D meshes of polynomial degree 1, 2, 4, and 8 and for 3D meshes of polynomial degree 1, 2, and 4. 
	
	We conclude that the specific-purpose optimization method significantly improves the standard one. In particular, the total amount of matrix-vector products is reduced one order of magnitude. Meanwhile, the number of non-linear and line-search iterations is reduced. As detailed in Sections \ref{sec:resultsglob} and \ref{sec:resultsinexactNewton}, these reductions in the number of linear and non-linear iterations arise from combining the specific-purpose inexact Newton solver and the specific-purpose line-search globalization. Moreover, as detailed in Section \ref{sec:resultsinexactNewton}, we observe again that, when augmenting the polynomial degree for each tested case, the total number of matrix-vector products, non-linear iterations, and line-search iterations increases.
	
	\subsection{Application: metric-aware curved high-order optimization of an $h$-adapted mesh}\label{sec:MMG}
	
	To compare the standard and specific-purpose solvers in an adaptive application, we optimize the distortion of an anisotropic mesh previously adapted to match a metric. In practice, adapted meshes are obtained by combining local topological operations that modify the mesh connectivity and local $r$-adaption operations that modify the mesh coordinates \cite{alauzet:AnisotropicMeshAdaptation}. Herein, we optimize the adapted mesh with the standard and specific-purpose optimization methods.
	
	Although we generate meshes adapted to a target metric with MMG \cite{dobrzynski2012mmg3d}, our goal is not to compare the distortion minimization with the MMG package. Actually, we acknowledge MMG because it generates an initial straight-sided mesh that matches the stretching and alignment of the target metric.
	
	We consider the hexahedral domain $\Omega=[-0.5,0.5]^3$ with the Plane metric presented in \cite{ibanez2017first}
	\begin{equation*}\label{eq:MMGmetric}
	\textbf{M} = \frac{1}{h_\text{m}^2}\left(
	\begin{array}{ccc}
	1 & 0 & 0\\
	0 & 1 & 0\\
	0 & 0 & 1/h(z)^2
	\end{array}
	\right),
	\end{equation*}
	where the function $h$ is presented in Equation \eqref{eq:htest} with stretching $h_{\min} = 0.01$, growth factor $\gamma = 2(1 - h_{\min})$, and size $h_\text{m} = 0.1$. Note that the stretching ratio of this metric is similar to the one presented in Figure \ref{fig:stretching}(d).
	\begin{figure}[t!]
		\centering
		\hspace{-0.35cm}
		\begin{tabular}{cc}
			\subfloat[]{\label{fig:mmg0}
				\includegraphics[width=0.5\textwidth]{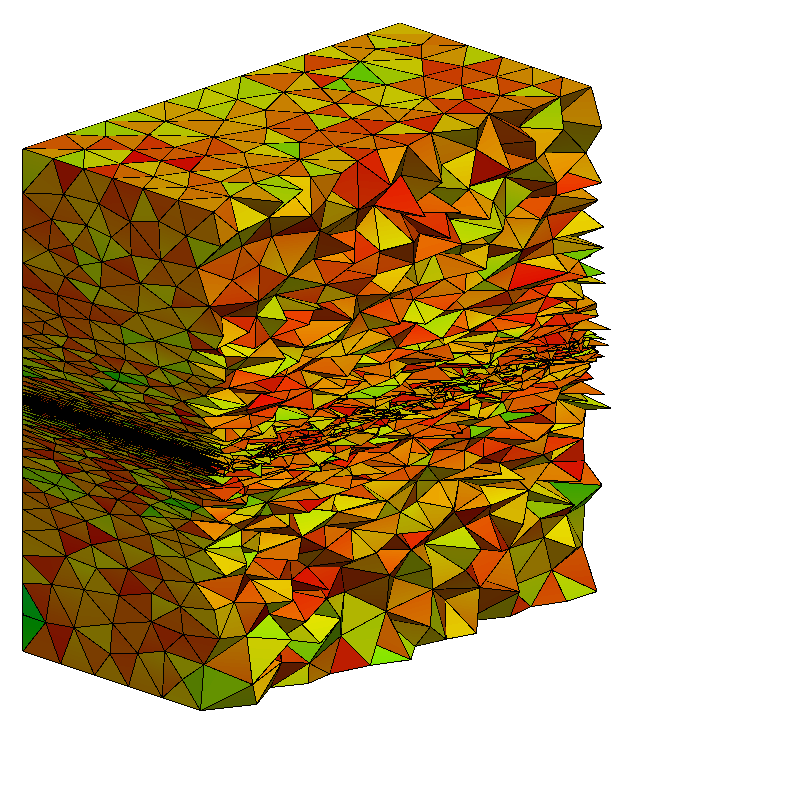}}
			&
			\subfloat[]{\label{fig:mmg1}
				\includegraphics[width=0.5\textwidth]{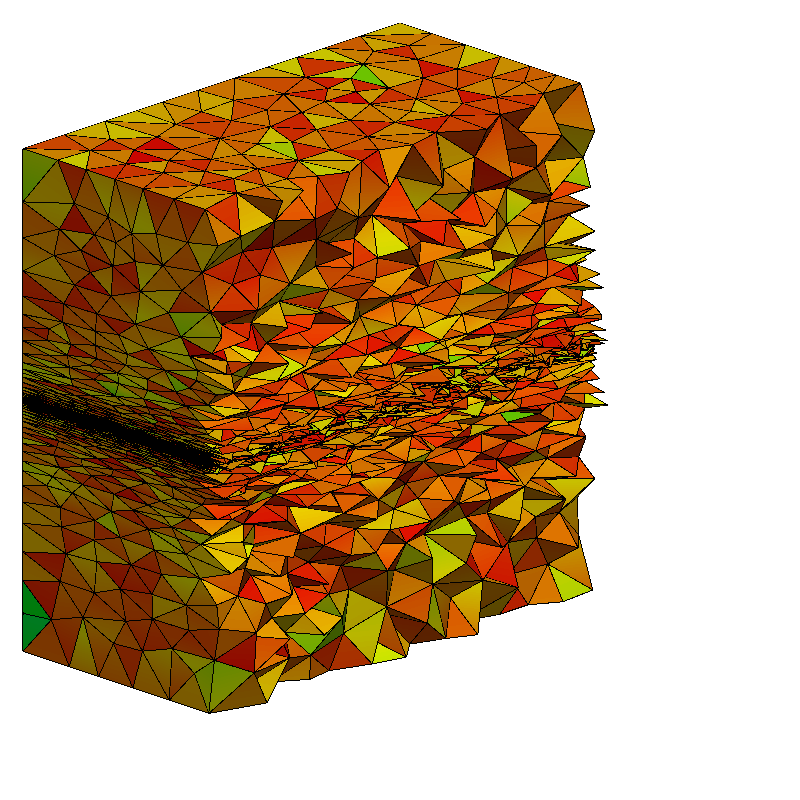}}
			\\
		\end{tabular}
		\\
		\includegraphics[width=0.25\textwidth]{qualBarParaview_color} 
	%	\caption{Linear tetrahedral meshes coupled with the Plane metric. \subref{fig:mmg0} Initial adapted mesh; and \subref{fig:mmg1} the corresponding optimized mesh.}
		\caption{Linear tetrahedral meshes coupled with the Plane metric (where the $z$-axis is the vertical direction). (a) Initial adapted mesh; and (b) the corresponding optimized mesh.}
		\label{fig:mmg}
	\end{figure}
	
	First we generate an initial adapted mesh, see Figure \ref{fig:mmg}\subref{fig:mmg0}. Specifically, we consider an isotropic linear tetrahedral mesh of size $0.01$ with \cite{MATLAB:2017}. We equip such mesh with the target metric evaluated at the mesh vertices and we use the MMG algorithm \cite{dobrzynski2012mmg3d} to obtain an anisotropic mesh. This mesh is composed of 11092 nodes and 57448 tetrahedra.
	We observe that the elements lying in the anisotropy region are stretched and aligned matching the target metric as quantified by the mean (0.8380) and standard deviation (0.3079) of the shape quality, see Table \ref{table:mmg_measures}.
%	We observe that the elements lying in the anisotropy region are stretched and aligned matching the target metric.
	In addition, we observe that far from the anisotropic region the elements are almost isotropic and with an approximate size of $h_{\text{m}} = 0.1$. 
%	Finally, from Table \ref{table:mmg_measures}, we can quantify how the metric is matched for the initial adapted mesh. In particular, from the mean and standard deviation of the shape quality (0.8380 and 0.3079, respectively), we observe that the meshes approximate the target metric.
	%	Points in blue color have low quality and points with red color have high quality.
	%	As in Section \ref{sec:meshes} we obtain an optimal configuration, illustrated in Figure \ref{fig:mmg}\subref{fig:mmg1}, by minimizing the distortion measure presented in Equation \ref{sec:minimizationProblem}.
	%	we obtain an optimal configuration, illustrated in Figure \ref{fig:mmg}\subref{fig:mmg1}, by minimizing the distortion measure presented in Equation \ref{sec:minimizationProblem}.
	
	\begin{table}[t!]
		\caption{Quality and Geometry Statistics of the initial adapted mesh and the corresponding optimized mesh.}
		\label{table:mmg_measures}
		\centering
		\tiny
		\begin{tabular}{ c c c c c c c c c c}
			\hline\noalign{\smallskip}
			Measure & \multicolumn{2}{c}{Minimum}&\multicolumn{2}{c}{Maximum}& \multicolumn{2}{c}{Mean} & \multicolumn{2}{c}{Standard deviation}\\
			&Initial&Optimized&Initial&Optimized&Initial&Optimized&Initial&Optimized\\
			\noalign{\smallskip}\hline\noalign{\smallskip}
			Shape & 0.3565 & 0.3804 & 0.9970 & 0.9972 & 0.8380 & 0.8690 & 0.3079 & 0.2790\\
			Length & 0.2340 & 0.4211 & 1.875 & 1.6481 & 0.9544 & 0.9425 & 0.1814 & 0.1641 \\
			Area & 0.1528 & 0.1806 & 1.8377 & 1.7901 & 0.7265 & 0.7165 & 0.1776 & 0.1591 \\
			Volume & 0.0583 & 0.0819 & 1.7251 & 1.3557 & 0.4865 & 0.4865 & 0.1418 & 0.1317\\			
			\noalign{\smallskip}\hline\noalign{\smallskip}
		\end{tabular}
	\end{table}	
	We obtain an optimal configuration, Figure \ref{fig:mmg}\subref{fig:mmg1}, by minimizing the distortion measure presented in Section \ref{sec:minimizationProblem}. The total amount of degrees of freedom is 30072. To compare both meshes we measure the element quality and geometry statistics, see Table \ref{table:mmg_measures}. The measures are: the elemental shape quality presented in Equation \eqref{eq:elementwise}, the lengths of the edges, the areas of the triangular faces, and the volumes of the tetrahedra. For each measure the minimum is improved, and the maximum and standard deviation are reduced.
	
	\begin{table}[t!]
		\caption{Non-linear iterations, line-search iterations, and matrix-vector products for standard and specific-purpose optimization methods.}
		\centering
		\tiny
		\par\medskip
		\begin{tabular}{ c c c c c c c }
			\hline\noalign{\smallskip}
			\multicolumn{2}{c}{Non-linear iterations}&\multicolumn{2}{c}{Line-search iterations}& \multicolumn{2}{c}{Matrix-vector products}&Speedup\\
			Standard&Specific-purpose&Standard&Specific-purpose&Standard&Specific-purpose&\\
			\noalign{\smallskip}\hline\noalign{\smallskip}
			7&7&0&0&1040&126&8.25\\
			\noalign{\smallskip}\hline\noalign{\smallskip}
		\end{tabular}
		\label{table:mmg_optimization}
	\end{table}
	We compare the specific-purpose optimization method with the standard one in terms of the non-linear and line-search iterations, the matrix-vector products, and the specific-purpose solver speedup, see Table \ref{table:mmg_optimization}. From the results, we conclude that the specific-purpose optimization method significantly improves the standard one. In particular, the total amount of matrix-vector products is reduced almost one order of magnitude. Meanwhile, the total amount of non-linear and line-search iterations stays unchanged. This is because the initial mesh is near to the optimal one and hence, the descent direction of both optimization methods is a faithful approximation of the Newton direction. That is, since the direction of both optimization methods are similar, the same number of non-linear iterations to converge are required. In addition, since the Newton direction has step length equal to one, no line-search iterations are required, see Section \ref{sec:final}.
	
	\section{Discussion}\label{sec:discussion}
	Next, we present a discussion on aspects related to curved boundaries, influence of our specific-purpose methods, non-smooth size variations, and local optimization.
	\paragraph{Influence of curved boundaries}
	To illustrate the solver capabilities, we only considered domains featuring flat boundaries. Nevertheless, as we demonstrated in \cite{aparicio2023combining}, our solver and preconditioner also work with curved boundaries.
	
	\paragraph{Influence of our specific-purpose methodologies}	For the sake of brevity, in Table \ref{table:final1} and Table \ref{table:forcingterms}, we only reported results for the Line metric. Nevertheless, when using our specific-purpose globalization (Table \ref{table:final1}) as well as for the forcing sequences and preconditioner (Table \ref{table:forcingterms}), we obtain similar improvements for all the metrics in Table \ref{table:metrics} It is important to highlight that the influence of the combination of all our improvements is available in Table \ref{table:all}, where we compare the standard and our specific-purpose solver for all the metrics.
	
	\paragraph{Non-smooth size variations} To test the quadratic convergence of the solver, we have used smooth size variations. Note that if non-smooth size variations are used the solver still converges but the quadratic convergence of Newton's method might be hampered.
	
	\paragraph{Local optimization} To devise our solver, we have tested it using examples with all the node coordinates as free variables. The reasoning is that an all-nodes optimization is more stiff and computationally expensive than a local optimization. Because our solver works well for all-nodes optimization, we expect that it is also useful for local optimization. That is, optimization of the free nodes of a mesh cavity while the rest of mesh nodes remain fixed.
	\section{Concluding remarks}\label{sec:conclusions}
	
	We have presented a specific-purpose non-linear solver for curved high-order metric-aware mesh optimization. To this end, the solver combines a specific-purpose line search with a specific-purpose preconditioned Newton-CG solver with dynamic forcing terms.
	
	The proposed specific-purpose line-search globalization continues the step length by ensuring sufficient decrease and progress of the objective function. Compared with a standard backtracking line search, it reduces the number of line search iterations because it reuses the step length. Meanwhile, the number of non-linear iterations remains in the same order of magnitude, yet it tends to be smaller because the line search can enlarge the step length.
	
	Regarding the proposed specific-purpose preconditioned Newton-CG solver with dynamic forcing terms, it reduces the total number of matrix-vector products by one order of magnitude. It also reduces the number of non-linear iterations. Compared with standard solvers, the proposed solver significantly improves the performance indicators (matrix-vector products) by one order of magnitude because it combines the advantages of the proposed forcing terms and pre-conditioner.  
	
	When we combine both ingredients, we also conclude that the specific-purpose non-linear solver reduces the total number of matrix-vector products by one order of magnitude. Moreover, it also reduces the number of non-linear and line-search iterations. Compared with standard solvers, all these iteration numbers are reduced because it combines the advantages of the specific-purpose inexact Newton solver and the specific-purpose line-search globalization.
	
	For the standard and the specific-purpose solvers, we observe that higher polynomial degrees and stretching ratios lead to higher total numbers of matrix-vector products, non-linear iterations, and line-search iterations. Still, these iteration numbers are smaller for the specific-purpose solver.
	
	In perspective, the proposed solver might help to demonstrate the advantages of curved high-order $r$-adaption driven by a target metric. Thanks to the solver, these advantages might be shown for high polynomial degrees and metrics featuring non-uniform sizing, high stretching ratios, and curved alignment --- specifically the requirements fulfilled by the solver. Because the problem is similar, it might be interesting to study its application when the $r$-adaptation is driven by a target deformation matrix. More broadly, it might be interesting to apply the solver in minimization problems where the non-linear objective is indefinite for initial approximations out of the positive definite region surrounding the local minima --- exactly the issues addressed by our solver for curved $r$-adaptation.

	\section*{Acknowledgements}
	This project has received funding from the European Research Council (ERC) under the European Union's Horizon 2020 research and innovation programme under grant agreement No 715546. This work has also received funding from the Generalitat de Catalunya under grant number 2017 SGR 1731. The work of X. Roca has been partially supported by the Spanish Ministerio de Econom\'ia y Competitividad under the personal grant agreement RYC-2015-01633.
%	The work of the second author has been partially supported by Grant IJC2020-045140-I from MCIN/AEI/10.13039/501100011 033 and “European Union NextGenerationEU/PRTR”
	%
	%%Vancouver style references.
%	\bibliographystyle{model1-num-names}
	\bibliography{references}
	
	\appendix
	\section{Standard CG}\label{sec:CG}
	%		\label{sec:linearsystem}
	
	%%Header: \textbf{Overview CG method}\\
	The CG method is an iterative procedure to solve linear systems of equations. As many other iterative solvers, it starts from an initial guess. Then, it generates a sequence of approximate solutions, derived from the previous ones, which in the limit are supposed to converge to an analytical solution. In this section, we present the algorithmic details of the standard preconditioned CG method \cite{saad:IterativeMethods,Dembo1983}.
	
	%	In what follows, we describe the algorithmic details of the standard inexact method presented in Algorithm \ref{alg:CG} with $\epsilon = 0$, for further details see	\cite{Dembo1983,nocedal2006numerical}.
	
	%%Header: \textbf{Algorithm description}\\
	\begin{algorithm}[t!]
		\caption{Conjugate Gradients \cite{Dembo1983}}\label{alg:CG}
		\textbf{Input:} $H,\ g,\ p,\ \text{preconfun},\ i_{\max},\ \eta,\ \epsilon$\\
		\textbf{Output:} $p^*,\ d^*$
		\begin{multicols}{2}
			\begin{algorithmic}[1]
				\Procedure{CG}{}
				\State $d \gets 0,\ \beta \gets 0$\label{lst:line:CGsetup0}
				\State $r \gets -g - Hp,\ i \gets 1$\label{lst:line:CGsetup1}
				\State $z \gets \text{preconfun}(r)$\label{lst:line:CGprecon1}
				\While{$i \leq i_{\max}$} \label{lst:line:CGiter}
				%				\While{Iteration condition} \label{lst:line:CGiter}
				\State $\tilde{z} \gets z,\ \tilde{r} \gets r$
				\State $d \gets z + \beta d$\label{lst:line:CGstep}
				\If{$d^T H d < \epsilon d^T d$} \label{lst:line:CGnegcurv}
				%				\If{Curvature condition} \label{lst:line:CGnegcurv}
				\State $p^*\gets p,\ d^*\gets d$\label{lst:line:CGnegcurvpointstep}
				\State \textbf{return}
				\Else
				\State $\alpha \gets \frac{r^Tz}{d^T H d}$
				\State $p \gets p + \alpha d,\ r \gets r - \alpha Hd$\label{lst:line:CGpointupdate}
				\EndIf
				\State $z \gets \text{preconfun}(r)$\label{lst:line:CGprecon2}
				\If{$\| r\| < \eta \|g\|$} \label{lst:line:CGresidual}
				%				\If{Residual condition} \label{lst:line:CGresidual}
				\State $p^*\gets p$
				\State \textbf{return}
				\EndIf
				\State $\beta\gets \frac{r^Tz}{\tilde{r}^T\tilde{z}}$
				\State $i \gets i+1$
				\EndWhile
				\EndProcedure
			\end{algorithmic}
		\end{multicols}
	\end{algorithm}
	Consider the CG method, presented in Algorithm \ref{alg:CG}, at the $k$-th non-linear iterate $x_k$, applied to the preconditioned version of the linear system of Newton Equation \eqref{eq:Newton} with a preconditioner $M_k$ of the Hessian matrix $\text{H}f(x_k)$
	\begin{equation*}\label{eq:Newtonpreconditioned}
	M_k\text{H} f(x_{k})p_{k} = -M_k\nabla f(x_{k}).
	\end{equation*}
	We denote as $p_{k}^i$ the direction corresponding to the $i$-th iteration of the CG method and similarly for the CG-residual $r$ and the CG-step $d$.
	The input arguments are the Hessian matrix $H_k = \text{H}f(x_k)$, the gradient vector $g_k = \nabla f(x_k)$, an initial guess $p^0_k$ which in this work is set to be equal to the zero vector $\textbf{0}$, the maximum number of iterations $i_{\max}$, the residual forcing value $\eta_k$, the curvature forcing value $\tau_k$ and the preconditioner function $z = \text{preconfun}(r)$ which solves the linear system $M_k z = r$. In Lines \ref{lst:line:CGsetup0} and \ref{lst:line:CGsetup1}, we setup the main variables: the CG-step $d$, the CG-step multiplier $\beta$, the residual $r^0_k := - \nabla f(x_k) - \text{H}(x_k) p^{0}_{k}$ and, the current iteration value $i$. In Lines \ref{lst:line:CGprecon1} and \ref{lst:line:CGprecon2}, the function $\text{preconfun}$ is applied to a vector $r$. Then, in Line \ref{lst:line:CGiter}, we proceed to the main loop. We compute a step $d_k^{i}$ at each CG-iteration $i \geq 1$, Line \ref{lst:line:CGstep}, providing a new direction $p_{k}^{i} = p_{k}^{i-1} + \alpha_k^i d^{i}_k$, Line \ref{lst:line:CGpointupdate}, and a residual,
	\begin{equation*}\label{eq:defresidual}
	r^{i}_{k}:=r\left(p^{i}_k ;x_k\right) := - \nabla f(x_k) - \text{H}(x_k) p^{i}_{k}.
	\end{equation*}
	The main loop iterates while $d^i_k$ is a negative curvature direction that is, $d^\text{T}Hd < 0$, Line \ref{lst:line:CGnegcurv}, the imposed tolerance is achieved by the residual $r^i_k$, Line \ref{lst:line:CGresidual}, or the iteration has exceeded the limit permitted, Line \ref{lst:line:CGiter}. Finally, the outputs of the algorithm are the CG-point $p^*$ and the CG-step $d^*$.
	
	%%Header: \textbf{Negative curvature case}\\
	The CG method is designed for positive definite systems $\text{H}f(x)$, which usually appear at points $x$ near an optimum $x^*$. However, for points far from an optimum the Hessian $\text{H}f$ may not be positive definite. Then, as in the standard CG-algorithm, we terminate the CG-iteration, in Line \ref{lst:line:CGnegcurv} with $\epsilon = 0$, whenever a CG-step $d_k^{i}$ of negative curvature is encountered. In this case, we provide the last direction $p_k^{i-1}$, see \cite{nocedal2006numerical}.
	%	pag. 169
	
	%%Header: \textbf{Scaled steepest-descent at first cg-iteration}\\
	It could happen that at the first CG-iteration the algorithm stops because a CG-step of negative curvature is encountered. In such case, the CG method returns the \emph{scaled steepest-descent direction} presented in \cite{bellavia2007globalization} given by
	\begin{equation*}\label{eq:scaledSteepestDecent}
	M_k p_k = - \nabla f(x_k).
	\end{equation*}
	\section{Setting $c_{max}$: quadratic convergence without line-search iterations}\label{sec:set_c_max}
	
	In what follows, we propose the value of the non-constant parameter $c_{\max}$. It is required for the sufficient-progress condition of the specific-purpose LS globalization strategy, see Section \ref{sec:final}. For this reason, we choose a value that preserves the main features of a second-order Newton method.
	
	%%Header: \textbf{Justification cmax: ensure quadratic convergence without additional steps} \\
	We propose to set $c_{\max} = 0.25$ to obtain quadratic convergence near a minimizer $x^*$ without additional line-search iterations. To this end, two conditions are required. First, it is required that the current point $x$ is sufficiently near a minimizer $x^*$ so the second order model presented in Equation \eqref{eq:quadraticmodel} is a faithful approximation. Second, it is required that step is an exact approximation to the Newton direction so it satisfies the Newton Equation presented in Equation \eqref{eq:Newton}. Notice that, the Newton direction has step length equal to one in the region of quadratic convergence. Then, by applying the Newton Equation in the second order term of the quadratic model and, assuming that the step length of the Newton direction is one as desired, we obtain the equation
	\begin{align*}
	f(x + s) &\approx f(x) + s^\text{T} \nabla f(x) + \frac{1}{2}s^\text{T} \text{H}f(x) s\\
	&\overset{s=-\text{H}f(x)^{-1}\nabla f(x)}{=} f(x) + s^\text{T} \nabla f(x) - \frac{1}{2}s^\text{T} \nabla f(x) = f(x) + \frac{1}{2}s^\text{T} \nabla f(x).
	\end{align*}
	Now, in terms of the predictor this equation can be reduced to
	\begin{equation*}
	\rho(s;x) = \frac{f(x) - f(x + s)}{-s^\text{T} \nabla f(x)} \approx \frac{1}{2}.
	\end{equation*}
	This shows that, as the current point $x$ tends to a minimizer $x^*$ with the Newton direction, the predictor tends to the value $\frac{1}{2}$. Since $\rho(s;x) > c_{\min}$, no reducing iterations are performed to the step length. Moreover, even if $\rho(s;x) > c_{\max}$, no amplifying iterations will be performed because the step length equal to one is the optimal for the Newton direction near a minimizer and hence, the second condition $f(x + \gamma s) < f(x + s)$ will not be fulfilled. That is, we do not amplify the step when $\rho(s;x) = 0.5$ and $x$ is in the convergence region. To prevent a modification of the step length we need to avoid, in Line \ref{lst:line:updatein2LS} of Algorithm \ref{alg:LSfinal}, the reducing update which will require an additional amplifying iteration in the next non-linear iteration. Thus, $c_{\max}$ must be smaller or equal to $0.5$. In summary, to preserve unit step-length during convergence, it is sufficient to choose a constant $c_{\max}$ satisfying $c_{\min} < c_{\max} \leq 0.5$.
	
	To obtain quadratic convergence near a minimizer without line-search iterations, we choose $c_{\min} = 10^{-4}$ and $c_{\max} = 0.25$. First, we choose $c_{\min} = 10^{-4}$ because our $c_{\min}$ corresponds to the constant $c$ of successful backtracking line-search implementations \cite{nocedal2006numerical}. Second, we choose an equispaced value of $c_{\max}$ between the limits $0$ and $0.5$ because we want to promote amplifications of directions that match the predictor model. Note that we could choose $c_{\max} = 0.5$, but it would promote less amplifications than $c_{\max} = 0.25$ because a smaller $c_{\max}$ promotes more amplifications. Note also that we could choose $c_{\max}$ just over $c_{\min}$, but we might amplify undesired directions because a smaller $c_{\max}$ corresponds to a smaller match to the predictor model.

%	This shows that, as the current point $x$ tends to a minimizer $x^*$ with the Newton direction, the predictor tends to the value $\frac{1}{2}$. Since $\rho(s;x) > c_{\min}$, no reducing iterations are performed to the step length. Moreover, even if $\rho(s;x) > c_{\max}$, no amplifying iterations will be performed because the step length equal to one is the optimal for the Newton direction near a minimizer and hence, the second condition $f(x + \gamma s) < f(x + s)$ will not be fulfilled. That is, we do not amplify the step when $\rho(s;x) = 0.5$ and $x$ is in the convergence region.
%	To prevent a modification of the step length we need to avoid, in Line \ref{lst:line:updatein2LS} of Algorithm \ref{alg:LSfinal}, the reducing update which will require an additional amplifying iteration in the next non-linear iteration. Hence, it is sufficient to choose a constant $c_{\max}$ satisfying $c_{\min} < c_{\max} \leq 0.5$. The constant $c_{\max} = 0.25$ has been chosen since it is equally spaced from its limits $0$ and $0.5$. This permits to obtain quadratic convergence near a minimizer without line-search iterations.
	%perque alpha es 1 en zona de convergencia
	%	 As it is mentioned before, the accepted step-lengths have a predictor between $c_{\min}$ and $c_{\max}$. 
	
	%%Header\textbf{Mirar Nocedal, implementacio wolfe complexitat}\\
	\section{Normalized curvature}\label{sec:NormalizedCurvature}
	%by Equation \eqref{eq:forcingcurv0}.
	%\begin{equation}\label{eq:defcurvatureStandard}
	%{d^{i}_k}^{\text{T}}  \text{H}f(x_k)  d^{i}_k > 0.
	%\end{equation}
	%	\begin{equation}\label{eq:defcurvatureSteihaug}
	%	{d^{i}_k}^{\text{T}}  \text{H}f(x_k)  d^{i}_k > \epsilon\ {d^{i}_k}^{\text{T}} d^{i}_k,
	%	\end{equation}
	%	where $\epsilon > 0$ is a constant and sufficiently small parameter.
	%Following the definition presented in Equation \eqref{eq:forcingcurv0}.
	In this section, we propose a criterion to check the curvature sign. On the one hand, it is standard to check the positiveness of curvature by means of the scalar product. On the other hand, a curvature constraint to limit the number of CG iterations, guarantee stability and sufficient positive curvature it is proposed \cite{Dembo1983}. For this reason, we propose to define the normalized curvature of a direction $p$ at a point $x$ as
	\begin{equation*}\label{eq:defcurvature}
	\kappa(p;x) := \frac{p^{\text{T}}  \text{H}f(x)  p}{p^{\text{T}}  p}.
	\end{equation*}
	Thus, the constraint $\kappa^i_k > 0$ remains unchanged, while the constraint $\kappa^i_k > \epsilon$ becomes
	\begin{equation*}\label{eq:defcurvature2}
	\kappa^i_k := \kappa(d^{i}_k;x_k) = \frac{{d^{i}_k}^{\text{T}}  \text{H}f(x_k)  d^{i}_k}{{d^{i}_k}^{\text{T}} d^{i}_k} > \epsilon.
	\end{equation*}
	This motivates us to propose a dynamic curvature forcing sequence $\{\tau_{k}\}$ detailed in Equation \eqref{eq:forcingseqs}. 
	%\section{Appendix}\label{sec:appendix}
	\section{Ordering of the mesh nodes}\label{sec:ordering}
	
	%	\textbf{Ordering of the mesh nodes}\\
	%For the development of the optimization method we have experienced some sensitivity of the optimization process depending on the ordering of the mesh nodes. Furthermore, the optimization problem may not converge for some mesh node orderings leading to steps of relative length less than $10^{-16}$ at early non-linear iterations. This sensitivity may be due to numerical instabilities of the matrix-vector products and to the assembly of the elemental contributions.
	%Header: \textbf{Introduction Node ordering}\\
	Herein we fix an arrangement for the degrees of freedom to obtain results independent on the node ordering. This arrangement is performed in two steps. First, we perform a mesh node ordering. It is involved in the mesh distortion evaluation, see Section \ref{sec:minimizationProblem}. Second, we perform an arrangement for the degrees of freedom over the mesh node ordering. It determines an arrangement for the optimization method, see Section \ref{sec:optimizationOverview}. Note that, both arrangements may perturb the numerical conditioning of the Hessian and hence, the total number of matrix-vector products performed in the optimization problem.
	
	We propose a node ordering that aims to concentrate the contributions of comparable magnitudes according to the stretching and alignment of the target metric. Our mesh node ordering relies in the class of \emph{spectral} orderings \cite{clift1995spectral,paulino1994nodeI,paulino1994nodeII}. However, it slightly differs from the methods presented in the literature since it is focused to take into account information about the anisotropy of the target metric instead of the mesh connectivity only. We remark that the presented node ordering is used to couple the matrix elements according to their magnitude independently to the chosen preconditioner.
	%and it is not used for a complete factorization so it does not pretend to reduce the matrix bandwidth or to produce a high quality preconditioner.
	
	%Header: \textbf{Laplace-Beltrami eigenfunction ordering}\\
	The node ordering is given by the partial ordering relation of an eigenfunction with lowest non-zero eigenvalue of the Laplace-Beltrami operator. That is, for a piece-wise polynomial mesh $\mathcal{M}$ of a bounded domain $\Omega$ equipped with a metric $\textbf{M}$ and with Lipschitz boundary $\partial \Omega$, the ordering of the mesh nodes that we propose is computed from an eigenfunction with the lowest non-zero eigenvalue $\lambda_1 > 0$ of the Laplacian eigenproblem with Neumann boundary conditions (see \cite{chavel1984eigenvalues})
	\begin{equation*}
	\left\lbrace\begin{array}{cc}
	-\Delta_{\text{\textbf{G}}} u = \lambda_1 u & \text{in } \Omega \\
	\frac{\partial u}{\partial \textbf{n}} = 0 & \text{on } \partial\Omega
	\end{array}\right.,
	\end{equation*}
	where, in our case, we set $\text{\textbf{G}} = \textbf{D}\zphysicalmap^{\text{T}}\ \zmetric\ \textbf{D}\zphysicalmap$ which is the embedded or extrinsic mesh metric in the metric space $\left( \mathds{R}^m,\textbf{M} \right)$, $\textbf{D}\zphysicalmap$ is the Jacobian of the mapping $\zphysicalmap$ between the reference element and the physical element of the mesh and $\textbf{n}$ is the outward normal of the boundary $\partial \Omega$. Then, the partial ordering relation $u(\textbf{x}_i) < u(\textbf{x}_j)$ (where $\textbf{x}_i$ are the nodes of the mesh for any ordering) determines an ordering of the mesh nodes. In this work an eigenfunction $u$ is computed using a continuous Galerkin finite element method over the mesh $\mathcal{M}$. Moreover, this reordering algorithm is used only one time, before the non-linear optimization.
	%However, we suggest to use it at each non-linear iteration since the metric $\text{\textbf{G}}$ changes as the mesh nodes move.
	
	%Header: \textbf{Ordering of the components}\\
	Once we have defined the mesh node ordering, we prescribe an arrangement for the degrees of freedom. Note that, each node $\textbf{x}\in \mathds{R}^m$ contains $m$ degrees of freedom. In the 2D case $(m = 2)$, each node contains 2 degrees of freedom, the $x$-component and the $y$-component which we locate contiguous on the corresponding global mesh node $\zx_i\in \mathds{R}^m$. For example, if the mesh nodes are denoted by $\textbf{x}_i = \left( x_i,y_i \right),\ i = 1,...,k$ then the corresponding variable representing the mesh is given by $(x_1,y_1,x_2,y_2,...,x_k,y_k)$. The gradient is given by $\nabla f = \left( \frac{\partial f}{\partial x_1},\frac{\partial f}{\partial y_1},\frac{\partial f}{\partial x_2},\frac{\partial f}{\partial y_2},...,\frac{\partial f}{\partial x_k},\frac{\partial f}{\partial y_k} \right)$ and the components of the Hessian $\text{H} f$ are then straightforward. Analogously, we apply this procedure for 3D meshes. In our case, the nodes lying at the boundary of the domain are permitted to slide on the boundary where they belong.
	%	Herein, the target domains are the quadrilateral $\left[-\frac{1}{2}, \frac{1}{2}\right]^2$ and the hexahedron $\left[-\frac{1}{2}, \frac{1}{2}\right]^3$. Thus, the parameterization of the boundary nodes is replaced by directly reducing the corresponding degrees of freedom of the fixed coordinates. For example, lets assume that the node 1 is on the bottom boundary: $x_1\in\left[-\frac{1}{2}, \frac{1}{2}\right],\ y_1 = -\frac{1}{2}$, with the variable $y_1$ fixed. Then the degrees of freedom of the mesh correspond to $(x_1,\hat{y}_1,x_2,y_2,...,x_k,y_k) = (x_1,x_2,y_2,...,x_k,y_k)$.
	
	\section{Test metrics}\label{sec:testmetrics}
	
	In this section, we detail the shear layer metric $\textbf{D}$ of the target metric $\zmetric$, see Equation \eqref{eq:metricDeformation} and Table \ref{table:metrics}. Specifically, we propose two choices for the metric $\textbf{D}$, a shear layer over a curve (surface) $\zDline$ or a shear layer over two intersecting curves (three intersecting surfaces) $\zDcross$.
	%Header: \textbf{Alignment shear layer: $\textbf{D}(h_1,...,h_d)$}\\
	%	The considered metric $\textbf{M}$ attains the highest level of anisotropy, close to the curve described by the points $(x,y)\in\Omega$ such that $\varphi(x,y)=(x,0)$ in 2D, and close the surface described by the points $(x,y,z)\in\Omega$ such that $\varphi(x,y,z)=(x,y,0)$ in 3D.
	
	On the one hand, the shear layer $\zDline$ aligns with the $x$-axis ($xy$-plane) in the 2D case (3D case). It requires a constant unit element size along the $x$-direction ($xy$-directions), and a non-constant element size along the $y$-direction ($z$-direction). This vertical element size grows linearly with the distance to the $x$-axis ($xy$-plane), with a factor $\gamma = 2$, and starts with the minimal value $h_{\min} = 10^{-2}$ ($h_{\min} = 2\cdot 10^{-2}$). Thus, for the 2D example illustrated in Figure \ref{fig:stretching}(a), between $y=-0.5$ and $y=0.5$ the stretching ratio blends from $1:100$ to $1:1$. For the 3D case, illustrated in Figure \ref{fig:stretching}(b), between $z=-0.5$ and $z=0.5$ the stretching ratio blends from $1:50$ to $1:1$. To match the shear layer, we define the metric as:
	%	\begin{equation}\label{eq:BL-}
	%	\textbf{D}(h_1,...,h_d) = \text{diag}
	%	\end{equation}	
	\begin{equation}\label{eq:BL-}
	\zDline := \left(
	\begin{array}{cc}
	1 & 0\\
	0 & 1/h(y)^2
	\end{array}
	\right),\quad
	\zDline := \left(
	\begin{array}{ccc}
	1 & 0 & 0\\
	0 & 1 & 0\\
	0 & 0 & 1/h(z)^2
	\end{array}
	\right).
	\end{equation}	
	where the function $h$ is defined by $h(x):= h_{\min} + \gamma |x|$, see Equation \eqref{eq:htest}.
	The metric of Equation \eqref{eq:BL-} is the metric induced by the following deformation
	\begin{equation*}\label{eq:BL2D-}
	\zmapline(x,y) = \left( x,H(y)\right),\quad \zmapline(x,y,z) = \left( x,y,H(z)\right),
	\end{equation*}
	that is $\zDline = \nabla \zmapline^\mathrm{T} \cdot \nabla \zmapline$ and being $H$ the function given by
	\begin{equation}\label{eq:H}
	H(x):= \frac{1}{\gamma}\log\left(\frac{h(x)}{h_{\min}}\right).
	\end{equation}
	On the other hand, we consider a metric $\zDcross$ consisting in the intersection of shear layers with a stretching in each axis direction at the corresponding orthogonal hyperplane: in the $x$-direction at the line $x = 0$ and in the $y$-direction at the line $y = 0$ in 2D and with a stretching in the $x$-direction at the plane $x = 0$, in the $y$-direction at the plane $y = 0$ and in the $z$-direction at the plane $z = 0$ in 3D, that is
	\begin{equation}\label{eq:BL+}
	\zDcross := \left(
	\begin{array}{cc}
	1/h(x)^2 & 0\\
	0 & 1/h(y)^2
	\end{array}
	\right),\quad
	\zDcross := \left(
	\begin{array}{ccc}
	1/h(x)^2 & 0 & 0\\
	0 & 1/h(y)^2 & 0\\
	0 & 0 & 1/h(z)^2
	\end{array}
	\right).
	\end{equation}
	The metric of Equation \eqref{eq:BL+} is the metric induced by the deformation
	\begin{equation*}\label{eq:BL2D+}
	\zmapcross(x,y) = \left( H(x), H(y)\right),\quad \zmapcross(x,y,z) = \left( H(x),H(y),H(z)\right),
	\end{equation*}
	that is $\zDcross = \nabla \zmapcross^\mathrm{T} \cdot \nabla \zmapcross$ and being $H$ the function presented in Equation \eqref{eq:H}.
	%Header: \textbf{Stretching shear layer}\\
	As expected, the 2D intersection shear layer metric presented in Equation \eqref{eq:BL+} aligns with the $x$-axis at the line $y = 0$ and with the $y$-axis at the line $x=0$, requires a constant unit element size along the diagonals of the square $x + y = 0$ and $x - y = 0$. Locally, the element size grows linearly along each axis with the distance to the orthogonal line, with a factor $\gamma= 2$, and starts with the minimal value $h_{\min}=10^{-2}$. Thus, between $y=-0.5$ and $y=0.5$ the stretching ratio blends from $1:100$ to $1:1$.
	%	Thus, as illustrated in Figure \ref{fig:metric+}, between $y=-0.5$ and $y=0.5$ the stretching ratio blends from $1:100$ to $1:1$.
	
	Analogously, the 3D intersection shear layer metric presented in Equation \eqref{eq:BL+} aligns with the $xy$-axis at the plane $z = 0$, with the $zx$-axis at the plane $y=0$ and with the $yz$-axis at the plane $x = 0$, requires a constant unit element size along the 4 diagonal lines of the cube. Locally, the element size grows linearly along each axis with the distance to the corresponding orthogonal plane, with a factor $\gamma= 2$, and starts with the minimal value $h_{\min} = 2 \cdot 10^{-2}$. Thus, between $z=-0.5$ and $z=0.5$ the stretching ratio blends from $1:2500$ to $1:1$. Note that in this case, the maximum stretching ratio is given by $h_{\min}^{2}$ and it is attained at the intersection of each plane $x = 0$, $y = 0$ and $z = 0$ with the boundary of the hexahedron $\Omega$.
	
	%Header: \textbf{Stretching deformation map}\\
	For $\textbf{D} = \zDline$ the metric $\textbf{M}$ attains the highest level of stretching ratio, close to the curve described by the points $(x,y)\in\Omega$ such that $\varphi(x,y)=(x,0)$ in 2D and close the surface described by the points $(x,y,z)\in\Omega$ such that $\varphi(x,y,z)=(x,y,0)$ in 3D. Note that the metric $\textbf{M}$ is induced by the map ${\psi} := \zmapline\circ {\varphi}$. Analogously, for $\textbf{D} = \zDcross$ the metric $\textbf{M}$ attains the highest level of stretching ratio, close the intersection of curves described by the points $(x,y)\in\Omega$ such that $\varphi(x,y) = (x,0)$ or $\varphi(x,y) = (0,y)$ in 2D and close the surface described by the points $(x,y,z)\in\Omega$ such that $\varphi(x,y,z)=(x,y,0)$ or $\varphi(x,y,z)=(x,0,z)$ or $\varphi(x,y,z)=(0,y,z)$ in 3D. Note that the metric $\textbf{M}$ is induced by the map ${\psi} := \zmapcross\circ {\varphi}$.
\end{document}